\documentclass[aps, prd, fleqn, nofootinbib, 12pt]{revtex4-2}

\usepackage{
	amsmath, 
	amssymb, 
	graphicx,
	hyperref, 
	float, 
	enumerate, 
	pgffor,
    longtable,
    xcolor,
	multirow,
	mathrsfs,
	bm,
	slashed,
    lineno,
    soul,
    pdfpages,
    breqn,
    courier,
    xparse,
    fvextra,
    array,
    pdflscape,
    adjustbox
}

\hypersetup{
	colorlinks=true
}

\usepackage{tikz}

\def\beq{\begin{equation}}
\def\eeq{\end{equation}}
\def\beqa{\begin{eqnarray}}
\def\eeqa{\end{eqnarray}}

\def\pp#1{\left(#1\right)}

\def\cc#1{\left\{#1\right\}}
\def\vv#1{\left\vert#1\right\vert}

\def\CC#1{\Big\{#1\Big\}}

\def\nn{\nonumber \\ &}

\def\d#1{\mathop{{\rm d}#1}}

\def\vec#1{\bm{#1}}

\allowdisplaybreaks

\def\GeV{{\rm GeV}}
\def\TeV{{\rm TeV}}
\def\fb{{\rm fb}}

\DeclareMathAlphabet{\mymathbb}{U}{BOONDOX-ds}{m}{n}

\def\bar{\overline}
\def\tilde{\widetilde}

\DefineVerbatimEnvironment{verbatim}{Verbatim}{breaklines=true}

\def\smeftatnlo{{\small \sc SMEFT@NLO}}
\def\madgraph{{\small \sc MadGraph5\_aMC@NLO}}
\def\feynarts{{\small \sc FeynArts}}
\def\feyncalc{{\small \sc FeynCalc}}
\def\lhapdf{{\small \sc LHAPDF}}
\def\vegas{{\small \sc Vegas}}
\def\ninja{{\small \sc Ninja}}
\def\collier{{\small \sc Collier}}

\usepackage{listings}
\lstdefinelanguage{mypython}{
  language=Python,
  keywords={def, return, if, elif, else, for, while, in, import, from, as, with, try, except, raise, class, self, None, True, False},
  keywordstyle=\color{blue}\bfseries,
  ndkeywords={int, float, str, list, dict, set, tuple},
  ndkeywordstyle=\color{magenta},
  identifierstyle=\color{black},
  comment=[l]{\#},
  commentstyle=\color{gray}\ttfamily,
  stringstyle=\color{orange},
  sensitive=true
}
\makeatletter
\AtBeginDocument{\let\LS@rot\@undefined}
\makeatother 

\begin{document}

\title{Constraining dimension-6 SMEFT with higher-order predictions for $p p \to t W$}
\author{Nikolaos Kidonakis and Kaan \c{S}im\c{s}ek \\
\textsl{Department of Physics, Kennesaw State University, Kennesaw, Georgia 30144, USA}}

\begin{abstract}
	We study single-top production in association with a $W$ boson at the Large Hadron Collider (LHC) as a probe of dimension-6 Standard Model effective field theory (SMEFT) at leading order, next-to-leading order, and approximate next-to-next-to-leading order accuracy in quantum chromodynamics (QCD). The process is sensitive to operators that modify the top-quark weak and chromomagnetic dipole interactions, and we perform three-parameter linear and quadratic SMEFT fits using doubly differential top-quark distributions in transverse momentum and rapidity for the Run II and Run III configurations at the LHC. We provide a detailed account of the uncertainties and quantify the impact of the different uncertainty components across bins and perturbative orders. We find that effective scales up to 2 TeV can be probed in nonmarginalized fits, while in marginalized fits the corresponding scales are around 0.5 and 1.5 TeV for linear and quadratic fits, respectively.
\end{abstract}

\maketitle

\section{Introduction\label{sec:intro}} 
In the Large Hadron Collider (LHC) era, the Standard Model effective field theory (SMEFT) has become a standard and systematic framework for describing possible heavy new physics through higher-dimensional operators. A complete nonredundant dimension-6 basis, the Warsaw basis, was established in 2010~\cite{Grzadkowski:2010es}, and equivalent alternative organizations such as the Higgs basis and the strongly interacting light Higgs basis were developed and related back to it by field redefinitions~\cite{Pomarol:2013zra, Falkowski:2015fla}. The operator program has since been pushed to higher orders, including complete dimension-8 bases and progress toward even higher dimensions~\cite{Murphy:2020rsh, Li:2020gnx, Harlander:2023psl}, supported by systematic enumeration tools such as Hilbert-series methods~\cite{Henning:2015daa, Henning:2015alf, Graf:2020yxt, Graf:2022rco}. In parallel, the SMEFT toolkit has matured through the derivation of complete Feynman rules and public implementations enabling automated predictions~\cite{Dedes:2017zog, Dedes:2019uzs, Dedes:2023zws, Brivio:2017btx, Brivio:2020onw}, together with extensive studies of renormalization, operator mixing, and fundamental consistency conditions, including positivity bounds that restrict parts of parameter space under assumptions like a causal and unitary ultraviolet (UV) completion~\cite{Baker:2019sli, Zhang:2020jyn, deRham:2022hpx}. Recently, the complete set of one-loop threshold corrections in the dimension-6 SMEFT, relating broken-phase electroweak inputs to $\bar{\rm MS}$ symmetric-phase parameters and providing a process-independent ingredient for RG running and matching, was computed in analytical form~\cite{Biekotter:2026dlb}. A broad overview can be found in~\cite{Brivio:2017vri}.
\par 
Phenomenologically, the SMEFT has been used to interpret data and constrain new physics from early sector-by-sector studies, such as electroweak precision tests, four-fermion interactions, Higgs and gauge-boson observables, and complementary low-energy and flavor measurements, to increasingly comprehensive global analyses~\cite{Han:2004az, Cirigliano:2012ab, Chen:2013kfa, Ellis:2014dva, Wells:2014pga, Hartmann:2016pil, Cirigliano:2016nyn, Falkowski:2017pss}. As LHC measurements improved, the program shifted toward global fits combining Higgs, electroweak, and top-quark observables with correlations and growing scope across many Wilson coefficients~\cite{deBlas:2016ojx, Hartland:2019bjb, Biekotter:2018ohn, Grojean:2018dqj, Baglio:2020oqu, Boughezal:2020uwq, Boughezal:2020klp, Ethier:2021ydt, Carrazza:2019sec}, sometimes supplemented by heavy-flavor and low-energy inputs or controlled by flavor assumptions, and in some cases incorporating uncertainties deriving from parton distribution functions (PDFs) directly into the effective field theory (EFT) treatment~\cite{Ellis:2018gqa, Dawson:2018dxp}. More recently, there has also been rising attention to effects beyond leading dimension-6 order, including explicit dimension-8 contributions and more general higher-order terms in SMEFT parameters, with studies exploring their impact on diboson, Drell-Yan, and Higgs observables~\cite{Alioli:2020kez, Boughezal:2021tih, Cohen:2020qvb, Li:2022rag, Dawson:2022cmu}. The SMEFT approach is now also routinely used by the experimental community, for example in recent CMS EFT interpretations~\cite{CMS:2022ubq}.
\par 
For phenomenological applications, the limiting factor is often theoretical accuracy; namely, turning precision measurements into meaningful constraints on Wilson coefficients requires predictions that are computed consistently in the EFT expansion and at sufficiently high perturbative order. This has motivated substantial progress on SMEFT calculations beyond leading order (LO), with a range of next-to-leading order (NLO) studies now available~\cite{Grober:2015cwa,Zhang:2016omx,Passarino:2016pzb,Englert:2019rga, Dawson:2024pft, Bellafronte:2025jbk, Bellafronte:2026mhp}. On the practical side, automated NLO QCD computations in the SMEFT can be performed in \madgraph~\cite{Alwall:2014hca} through the \smeftatnlo~package~\cite{Degrande:2020evl}.
\par 
Top-quark production processes provide opportunities for the study of the SMEFT and its effective higher-dimensional operators. In Ref.~\cite{Kidonakis:2023htm}, calculations at approximate next-to-next-to-leading order (aNNLO) were performed for the chromomagnetic dipole operator in top-antitop pair production. In this study, we use higher-order calculations for the associated production of a top quark with a $W$ boson ($tW$ production) to constrain dimension-6 SMEFT operators. Rather than aiming for the most competitive global bounds, our focus is on establishing perturbative-QCD control in this channel and quantifying how higher-order corrections and a realistic QCD uncertainty budget affect EFT sensitivity and the perturbative stability of the extracted constraints.
\par 
The $tW$ process has the second largest cross section among single-top production processes at LHC energies, and many cross-section measurements have been taken by ATLAS and CMS \cite{ATLAS:2012bqt,CMS:2012pxd,CMS:2014fut,ATLAS:2015igu,ATLAS:2016ofl,ATLAS:2017quy,CMS:2018amb,ATLAS:2019hhu,ATLAS:2020cwj,CMS:2021vqm,CMS:2022ytw,ATLAS:2024ppp,CMS:2024okz}. Results for $tW$ production and decays appeared at LO in Refs.~\cite{Ladinsky:1990ut,Heinson:1996zm,Moretti:1997ng,Belyaev:1998dn}, and with some additional corrections in~\cite{Tait:1999cf,Belyaev:2000me}. The full NLO cross section for this process was presented in Refs.~\cite{Zhu:2002uj,Campbell:2005bb,Cao:2008af}. NLO calculations with decays of the top quark and the $W$ boson were made in Ref. \cite{Campbell:2005bb}, and with parton showers in \cite{Frixione:2008yi,Re:2010bp}. Some progress towards 2-loop QCD amplitudes was made in \cite{Chen:2021gjv,Long:2021vse,Chen:2022ntw,Wang:2022enl,Chen:2022yni,Chen:2022pdw}.
\par 
The calculation of soft-gluon corrections beyond NLO from QCD threshold resummation was studied for $tW$ production in a series of papers in Refs. \cite{Kidonakis:2006bu,Kidonakis:2007ej,Kidonakis:2010ux,Kidonakis:2012rm,Kidonakis:2013zqa,Kidonakis:2015wva,Kidonakis:2016sjf,Kidonakis:2018ybz,Kidonakis:2019nqa,Kidonakis:2021vob}, where cross sections and differential distributions at aNNLO and higher were presented (see also \cite{Kidonakis:2006bu,Kidonakis:2007ej,Kidonakis:2012rm,Kidonakis:2013zqa,Kidonakis:2015wva,Kidonakis:2018ybz,Kidonakis:2019nqa,Kidonakis:2010tc,Kidonakis:2011wy,Kidonakis:2013yoa,Kidonakis:2025eia} for other single-top processes). The soft-gluon corrections are large and numerically dominant, and thus it is important to calculate these contributions beyond NLO. Higher-order results based on resummation in another formalism, namely Soft-Collinear Effective Theory, have appeared in \cite{Li:2019dhg,Ding:2025xhc,Ding:2025onw}.
\par 
This paper is organized as follows. In Sec.~\ref{sec:analytical}, we describe $tW^-$ production at the LHC within the SMEFT framework and define the set of dimension-6 operators and Wilson coefficients relevant to our analysis. In Sec.~\ref{sec:resum}, we briefly describe the formalism for the calculation of the soft-gluon corrections used to construct the aNNLO predictions. In Sec.~\ref{sec:numerical}, we describe the numerical setup, present the Standard Model (SM) and SMEFT distributions, and detail our statistical treatment and uncertainty modeling. In Sec.~\ref{sec:results}, we present the SMEFT fit results. We conclude in Sec.~\ref{sec:conclusion}.

\section{$tW$ production at the LHC within the SMEFT\label{sec:analytical}} 
The SMEFT is a model-independent extension of the SM in which one builds operators $O_k^{(n)}$ of mass dimension $n>4$ at a UV scale $\Lambda$ beyond accessible collider energy using the existing SM spectrum and introduces Wilson coefficients $C_k^{(n)}$ as effective couplings. The SMEFT Lagrangian is schematically given by 
\begin{gather} 
    \mathcal L = \mathcal L_{\rm SM} + \sum_{n>4} {1 \over \Lambda^{n-4}} \sum_k C_k^{(n)} O_k^{(n)}. 
\end{gather} 
The SMEFT Lagrangian shifts the SM vertices in a gauge-invariant manner and respects all underlying symmetries. Here, we focus on the case $n = 6$. 
\par 
In this work, we adopt the $\{G_F,m_Z,m_W\}$ scheme, as it is the natural choice for LHC processes with external $W$ bosons. The choice necessarily affects the operators involved. In order to be consistent in our calculations at various orders, we borrow the Feynman rules from the \smeftatnlo~model file~\cite{Degrande:2020evl}. We adopt the five-flavor scheme and set the Cabibbo-Kobayashi-Maskawa (CKM) matrix element $V_{tb}=1$, while $V_{ts}$ and $V_{td}$ are taken to vanish. This choice is standard in single-top analyses, as off-diagonal contributions are suppressed by both the CKM hierarchy and the small parton luminosities for $d$ and $s$ quarks. 
\par
\begin{figure} 
    [H] 
    \centering 
    \includegraphics[width=.5\linewidth]{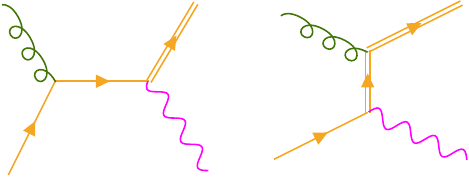} 
    \caption{The Feynman diagrams for the partonic process contributing to $p p \to t W^-$ at leading order.} 
    \label{fig:lo_diagrams} 
\end{figure}

 
We consider the hadronic process $p p \to t W^-$. The underlying partonic process at leading order is $g b \to t W^-$.
The Feynman diagrams at leading order are presented in Fig.~\ref{fig:lo_diagrams}. The double lines with an arrow indicate the top quark.
There are two classes of dimension-6 SMEFT operators of interest from the Warsaw basis~\cite{Grzadkowski:2010es} presented in Table~\ref{tab:ops}. First, we have the operators that directly modify the $qqg$ and $tbW$ vertices at leading order in SMEFT parameters. These are weak and chromomagnetic couplings, as well as the couplings of the quark current to operators of the form $[\varphi D \varphi]$. The second class of operators is the ones that enter through the input parameters. 
\begin{table}
    [H]
    \centering
    \caption{The dimension-6 SMEFT operators from the Warsaw basis~\cite{Grzadkowski:2010es} that affect the observable at leading order in Wilson coefficients.}
    \label{tab:ops}
    {\renewcommand{\arraystretch}{1}\begin{tabular}{|c|c|}
        \hline
        Couplings & Input parameters  \\
        \hline
        $O_{dG} = [\bar q \sigma^{\mu\nu} T^A d] \varphi G^A_{\mu\nu}$ & $O_{\ell\ell} = [\bar \ell \gamma_\mu \ell] [\bar \ell \gamma^\mu \ell] $ \\
        $O_{uG} = [\bar q \sigma^{\mu\nu} T^A u] \tilde \varphi G^A_{\mu\nu}$ & $O_{\varphi\ell}^{(3)} = [\varphi^\dagger i \stackrel{\leftrightarrow}{D}{}_\mu^I \varphi] [\bar \ell \tau^I \gamma^\mu \ell] $ \\
        $O_{dW} = [\bar q \sigma^{\mu\nu} d] \tau^I \varphi W_{\mu\nu}^I $ & \\
        $O_{uW} = [\bar q \sigma^{\mu\nu} u] \tau^I \tilde \varphi W_{\mu\nu}^I $ & \\ 
        $O_{\varphi q}^{(3)} = [\varphi^\dagger i \stackrel{\leftrightarrow}{D}{}_\mu^I \varphi] [\bar q \tau^I \gamma^\mu q] $ & \\ 
        $O_{\varphi u d} = i [\tilde \varphi^\dagger \stackrel{\leftrightarrow}{D}{}_\mu \varphi] [\bar u \gamma^\mu d] $ & \\ 
        \hline
    \end{tabular}}
\end{table}
In Table~\ref{tab:ops}, $\ell$ and $q$ denote the left-handed SU(2)${}_L$ lepton and quark doublets, while $u$ and $d$ are the corresponding right-handed SU(2)${}_L$ singlets, with generation indices suppressed; $\bar\psi \equiv \psi^\dagger \gamma^0$. $\varphi$ is the SU(2)${}_L$ scalar doublet, with $\tilde\varphi \equiv i\tau^2 \varphi^\ast$ its charge-conjugate. $D_\mu$ is the SM covariant derivative, and the $\tau^I$ are the Pauli matrices acting on SU(2)${}_L$ indices, with the left-right derivative defined such that 
\begin{gather}
    \varphi^\dagger i \stackrel{\leftrightarrow}{D}{}_\mu \varphi = i \varphi^\dagger (D_\mu - \stackrel{\leftarrow}{D}{}_\mu) \varphi, \\ 
    \varphi^\dagger i \stackrel{\leftrightarrow}{D}{}_\mu^I \varphi = i \varphi^\dagger (\tau^I D_\mu - \stackrel{\leftarrow}{D}{}_\mu \tau^I) \varphi, 
\end{gather}
with $\varphi^\dagger \stackrel{\leftarrow}{D}{}_\mu \varphi = (D_\mu \varphi^\dagger) \varphi$, while the $T^A$ are the SU(3)${}_c$ generators acting on color indices. $W_{\mu\nu}^I$ and $G_{\mu\nu}^A$ are the gauge field strength tensors for SU(2)${}_L$ and for SU(3)${}_c$, respectively. The Dirac matrices $\gamma^\mu$ and $\sigma^{\mu\nu} \equiv \tfrac{i}{2}[\gamma^\mu,\gamma^\nu]$ appear in the fermion bilinears; square brackets indicate Lorentz and gauge-invariant contractions within each bilinear/current, with implicit contraction of spinor, SU(2)${}_L$, and color indices as appropriate.
\par
\begin{figure}
    [H]
    \centering
    \includegraphics[width=0.5\linewidth]{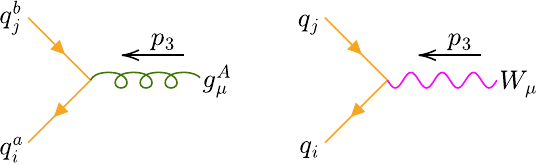}
    \caption{The relevant SMEFT Feynman rules for the partonic process of interest, adopted from~\cite{Degrande:2020evl}.}
    \label{fig:smeft_feynman_rules}
\end{figure}

 
With a massless bottom quark, the operators $O_{dG}$, $O_{dW}$, and $O_{\varphi ud}$ do not contribute because they involve chirality flip. With that, we are left with only five operators, now explicitly written in the \textit{top basis}~\cite{Degrande:2020evl}: $O_{tG}$, $O_{tW}$, $O_{\varphi Q}^{(3)}$, $O_{\ell\ell}$, and $O_{\varphi \ell}^{(3)}$. With real Wilson coefficients (since we are not immediately interested in $CP$-violating effects), the relevant Feynman rules from~\cite{Degrande:2020evl} can be summarized as follows:
\begin{gather}
    V_{ttg}^\mu = 2 i \sqrt{\pi \alpha_s} T^A_{ab} \cc{\gamma^\mu - {2^{1/4} C_{tG} \over \sqrt{G_F} \Lambda^2} i \sigma^{\mu\nu} p_{3\nu}}, \\ 
    V_{bbg}^\mu = 2 i \sqrt{\pi \alpha_s} T^A_{ab} \gamma^\mu, \\ 
    V_{tbW}^\mu = 2^{3/4} i \sqrt{G_F} m_W \cc{\gamma^\mu P_L - {C_{tW} \over G_F m_W \Lambda^2} i \sigma^{\mu\nu} p_{3\nu} P_L + {C_p \over 2 \sqrt2 G_F \Lambda^2} \gamma^\mu P_L},
\end{gather}
where we have defined $C_p = 2 C_{\varphi Q}^{(3)} + C_{\ell\ell}^{[1221]} - C_{\varphi \ell}^{(3)[11]} - C_{\varphi \ell}^{(3)[22]}$, with the numbers in square brackets indicating the generation indices; the Feynman rules are illustrated in Fig.~\ref{fig:smeft_feynman_rules}. Since three of the operators of interest appear only in a certain linear combination, at the end of the day we have only three Wilson coefficients of interest: $C_{tG}$, $C_{tW}$, and $C_p$.
The hadronic cross section at the $pp$ center-of-mass (c.m.) energy $s$ is computed as
\begin{gather}
    \sigma(s) = \sum_{i=-N_f}^{N_f} \sum_{j=-N_f}^{N_f} \int \d{x_1} \d{x_2} f_i(x_1, \mu_F) f_j(x_2, \mu_F) \sigma_{ij}(\hat s),
\end{gather}
where $N_f$ is the number of active flavors, $i$ and $j$ are the flavor indices with the negative values indicating antiquarks and 0 the gluon, $x_1$ and $x_2$ are the Bjorken-$x$ parameters for the incoming partons, $f_i$ and $f_j$ are the parton distribution functions (PDFs) defined at the factorization scale $\mu_F$, and $\sigma_{ij}$ is the partonic cross section corresponding to the partonic process $f_i f_j \to t W^-$ given by
\begin{gather}
    \sigma_{ij} (\hat s) = F \int \bar{|\mathcal A_{ij}|^2} \d{\rm LIPS}, 
\end{gather}
where $\hat s$ is the c.m. energy for the partonic process, $F = 1/(2\hat s)$ is the flux factor for massless initial states and $\d{\rm LIPS}$ is the two-particle Lorentz-invariant phase space. Here, $\bar{|\mathcal A_{ij}|^2}$ denotes the spin and color-averaged squared amplitude for the said partonic process and the only contributions at leading order are the $bg$ and $gb$ processes.  
\par 
We parametrize the cross section as
\begin{gather}
    \sigma = \sigma_{\rm SM} + \sum_k C_k \sigma_k + \sum_{k\leq k'} C_k C_{k'} \sigma_{kk'}.
\end{gather}
at all orders. For the leading-order amplitudes, we use \feynarts~\cite{Hahn:2000kx} and \feyncalc~\cite{Shtabovenko:2016sxi, Shtabovenko:2020gxv, Shtabovenko:2023idz, Mertig:1990an}. 
For the partonic process
\begin{gather}
    b(p_1) + g(p_2) \to t(p_3) + W^-(p_4),
\end{gather}
the squared SM amplitude is obtained to be 
\begin{align}
    \bar{|\mathcal A_{bg}|^2}_{\rm SM} &= \frac{2 \sqrt{2} \pi  G_F \alpha _s}{3 \hat{s}
   \left(m_t{}^2-\hat{u}\right){}^2}
    \CC{
        m_t{}^4 \left(-2 \hat{u} m_W{}^2-2
   m_W{}^4+\left(\hat{s}+\hat{u}\right)^2\right) \nn 
        +m_t{}^2 \left(2 m_W{}^2 \left(-\hat{s}
   \hat{u}+\hat{s}^2+2 \hat{u}^2\right)-2 \hat{u} m_W{}^4+4 m_W{}^6-\hat{u}
   \left(\hat{s}+\hat{u}\right)^2\right) \nn 
        -m_t{}^6 \left(2 \hat{s}+\hat{u}\right)-2
   \hat{u} m_W{}^2 \left(-2 m_W{}^2 \left(\hat{s}+\hat{u}\right)+2
   m_W{}^4+\hat{s}^2+\hat{u}^2\right)+m_t{}^8}.
\end{align}
Here, we have defined $\hat s = (p_1 + p_2)^2$, $\hat t = (p_1 - p_3)^2$, and $\hat u = (p_1 - p_4)^2$, which satisfy $\hat s + \hat t + \hat u = m_t{}^2 + m_W{}^2$ at leading order. 
The shifts to the squared amplitude for the $bg$ process characterized by $C_k$ and $C_k C_{k'}$ are presented in Tables~\ref{tab:lo-ampsq-shifts-c} and \ref{tab:lo-ampsq-shifts-cc}. The $gb$ channel is given by $\hat u \to \hat t$. 
{\renewcommand{\arraystretch}{1}
\begin{table}
[H]
\centering
\caption{Linear corrections to the squared SMEFT amplitude. The expressions in the second column are to be multiplied by $\frac{2 \pi  \alpha _s}{3 \Lambda ^2 \hat{s}
   \left(m_t{}^2-\hat{u}\right){}^2}$.}
 \label{tab:lo-ampsq-shifts-c}
\begin{tabular}{|c|l|}
 \hline 
 $C_k$ & $\bar{|\mathcal A_{bg}|^2}_k$ \\ 
 \hline 
 $C_{tG}$ & $2\ 2^{3/4} \hat{s} \sqrt{G_F} m_t m_W{}^2
   \left(m_t{}^2-\hat{u}\right) \CC{m_t{}^2+2 \hat{s}-\hat{u}}$ \\
    \hline 
 $C_{tW}$ & $\begin{array}{l}2 \sqrt{2} m_t m_W \CC{-m_t{}^4 \left(6 m_W{}^2+2 \hat{s}+3
   \hat{u}\right)+m_t{}^2 \left(6 m_W{}^4-\left(\hat{s}-\hat{u}\right)
   \left(\hat{s}+3 \hat{u}\right)\right) \\ +\hat{u} \left(6 m_W{}^2
   \left(\hat{s}+\hat{u}\right)-6 m_W{}^4+\left(\hat{s}-3 \hat{u}\right)
   \left(\hat{s}+\hat{u}\right)\right)+3 m_t{}^6}\end{array}$ \\
    \hline
 $C_p$ & $\begin{array}{l}m_t{}^4 \left(-2
   \hat{u} m_W{}^2-2 m_W{}^4+\left(\hat{s}+\hat{u}\right)^2\right) \\ +m_t{}^2
   \left(2 m_W{}^2 \left(-\hat{s} \hat{u}+\hat{s}^2+2 \hat{u}^2\right)-2
   \hat{u} m_W{}^4+4 m_W{}^6-\hat{u}
   \left(\hat{s}+\hat{u}\right)^2\right) \\ -m_t{}^6 \left(2
   \hat{s}+\hat{u}\right)-2 \hat{u} m_W{}^2 \left(-2 m_W{}^2
   \left(\hat{s}+\hat{u}\right)+2
   m_W{}^4+\hat{s}^2+\hat{u}^2\right)+m_t{}^8\end{array}$ \\
   \hline 
\end{tabular}
\end{table}
}
{\renewcommand{\arraystretch}{1}
\begin{table}
[H]
\centering 
\caption{Quadratic corrections to the squared SMEFT amplitude. The expressions in the second column are to be multiplied by $\frac{\pi  \alpha _s}{6 \sqrt{2} \Lambda ^4 \hat{s} G_F
   \left(m_t{}^2-\hat{u}\right){}^2}$.}
 \label{tab:lo-ampsq-shifts-cc}
\begin{tabular}{|c|l|}
 \hline 
 $C_k C_{k'}$ & $\bar{|\mathcal A_{bg}|^2}_{kk'}$  \\ 
 \hline 
 $C_{tG}{}^2$ & $-8 \sqrt{2} \hat{s} G_F \left(m_t{}^2-\hat{u}\right){}^2 \CC{-2
   m_W{}^2 \left(\hat{s}+\hat{u}\right)+m_t{}^2
   \left(m_W{}^2-\hat{u}\right)+2 m_W{}^4+\hat{s} \hat{u}}$ \\
    \hline
 $C_{tG} C_{tW}$ & $-16
   \sqrt[4]{2} \hat{s} \sqrt{G_F} m_W \left(m_t{}^2-\hat{u}\right){}^2
   \CC{m_W{}^2-2 \hat{u}}$ \\
    \hline
 $C_{tG} C_p$ & $4\ 2^{3/4} \hat{s} \sqrt{G_F} m_t m_W{}^2
   \left(m_t{}^2-\hat{u}\right) \CC{m_t{}^2+2 \hat{s}-\hat{u}}$ \\
    \hline
 $C_{tW}{}^2$ & $\begin{array}{l}8
   \CC{-m_t{}^6 \left(3 m_W{}^2+4 \hat{s}+2 \hat{u}\right)+m_t{}^4
   \left(-m_W{}^2 \left(2 \hat{s}+\hat{u}\right)+2 m_W{}^4+2 \left(4 \hat{s}
   \hat{u}+\hat{s}^2+\hat{u}^2\right)\right) \\ +m_t{}^2 \left(m_W{}^2
   \left(\hat{s}+\hat{u}\right) \left(\hat{s}+5 \hat{u}\right)-4 \hat{u}
   m_W{}^4+2 m_W{}^6-2 \hat{u} \left(6 \hat{s} \hat{u}+3
   \hat{s}^2+\hat{u}^2\right)\right) \\ -\hat{u} \left(-2 m_W{}^4
   \left(\hat{s}+\hat{u}\right)+m_W{}^2 \left(\hat{s}+\hat{u}\right)^2+2
   m_W{}^6-4 \hat{s} \hat{u} \left(\hat{s}+\hat{u}\right)\right)+2
   m_t{}^8}\end{array}$ \\
    \hline
 $C_{tW} C_p$ & $\begin{array}{l}4 \sqrt{2} m_t m_W \CC{-m_t{}^4 \left(6 m_W{}^2+2
   \hat{s}+3 \hat{u}\right)+m_t{}^2 \left(6
   m_W{}^4-\left(\hat{s}-\hat{u}\right) \left(\hat{s}+3
   \hat{u}\right)\right) \\ +\hat{u} \left(6 m_W{}^2
   \left(\hat{s}+\hat{u}\right)-6 m_W{}^4+\left(\hat{s}-3 \hat{u}\right)
   \left(\hat{s}+\hat{u}\right)\right)+3 m_t{}^6}\end{array}$ \\
    \hline
 $C_p{}^2$ & $\begin{array}{l}m_t{}^4 \left(-2
   \hat{u} m_W{}^2-2 m_W{}^4+\left(\hat{s}+\hat{u}\right)^2\right) \\ +m_t{}^2
   \left(2 m_W{}^2 \left(-\hat{s} \hat{u}+\hat{s}^2+2 \hat{u}^2\right)-2
   \hat{u} m_W{}^4+4 m_W{}^6-\hat{u}
   \left(\hat{s}+\hat{u}\right)^2\right) \\ -m_t{}^6 \left(2
   \hat{s}+\hat{u}\right)-2 \hat{u} m_W{}^2 \left(-2 m_W{}^2
   \left(\hat{s}+\hat{u}\right)+2
   m_W{}^4+\hat{s}^2+\hat{u}^2\right)+m_t{}^8\end{array}$ \\
 \hline 
\end{tabular}
\end{table}
}
At NLO QCD, the real-emission contributions to $pp\to tW^-$ contain diagrams with an intermediate resonant $\bar t$, corresponding to $pp\to t\bar t$ followed by $\bar t\to W^- \bar b$, and thus overlap with the $tW^-$ signal definition. To avoid double counting with the $t\bar t$ channel, we adopt the diagram-removal prescription and exclude the resonant $\bar t$ contributions in our NLO predictions, following the implementation within \smeftatnlo~\cite{Degrande:2020evl} in \madgraph~\cite{Alwall:2014hca}.

\section{Soft-gluon corrections\label{sec:resum}}

In this section, we review the soft-gluon resummation formalism that we use for the calculation of soft-gluon corrections through NNLO for $tW$ production \cite{Kidonakis:2006bu,Kidonakis:2007ej,Kidonakis:2010ux,Kidonakis:2012rm,Kidonakis:2013zqa,Kidonakis:2015wva,Kidonakis:2016sjf,Kidonakis:2018ybz,Kidonakis:2019nqa,Kidonakis:2021vob}. In addition to the kinematical variables ${\hat s}$, ${\hat t}$, and ${\hat u}$ that we defined previously for the $bg \to tW^-$ partonic process, we also introduce the variable $s_4={\hat s}+{\hat t}+{\hat u}-m_t^2-m_W^2$ which vanishes at partonic threshold where there is no energy available for additional radiation.

The soft-gluon corrections appear in the perturbative series for the cross
section via plus distributions involving $s_4$, specifically
$[(\ln^k(s_4/m_t^2))/s_4]_+$, where $k$ takes values from 0 to $2n-1$ in the
$n$th order corrections in the strong coupling.

The resummation of soft-gluon contributions follows as a consequence of the 
factorization of the cross section into functions that describe soft and collinear
emission in the partonic process. We take Laplace transforms of the partonic
scattering cross section,
${\hat \sigma}(N)=\int (ds_4/s) \;  e^{-N s_4/s} {\hat \sigma}(s_4)$, with $N$ the
transform variable, and write a factorized expression:
\beq
\sigma_{bg \to tW}(N)=
H_{bg \to tW} \; S_{bg \to tW}\left(\frac{m_t}{N \mu}\right)\;
\psi_b\left (N_b \right) \;
\psi_g\left (N_g \right) \; ,
\label{factorized}
\eeq
where $\mu$ is the scale, $H_{bg \to tW}$ is the hard-scattering function,
$S_{bg \to tW}$ is the soft-gluon function for non-collinear soft-gluon emission,
and $\psi_b$ and $\psi_g$ are functions which describe soft and collinear emission from, respectively, the incoming bottom quark and gluon.
The soft function $S_{bg \to tW}$ requires renormalization and its $N$-dependence can be resummed via renormalization group evolution.

The resummed partonic cross section takes the form
\beqa
\sigma_{bg \to tW}^{res}(N) &=&   
\exp\left[ \sum_{i=b,g} E_i(N_i)\right] \, 
H_{bg \to tW} \left(\alpha_s(\sqrt{s})\right) \, 
S_{bg \to tW} \left(\alpha_s(\sqrt{s}/{\tilde N}) \right) 
\nonumber \\ && \times 
\exp \left[2 \int_{\sqrt{s}}^{{\sqrt{s}}/{\tilde N}} \frac{d\mu}{\mu}\, \Gamma_{S \, bg \to tW}
\left(\alpha_s(\mu)\right)\right]
\label{rgeres}
\eeqa
where the exponents $E_i$ resum universal collinear and soft emissions from the incoming partons, and $\Gamma_{S \, bg \to tW}$ is the soft anomalous dimension which resums noncollinear soft-gluon contributions in this process. More details and explicit expressions for these quantities can be found in Refs. \cite{Kidonakis:2006bu,Kidonakis:2007ej,Kidonakis:2010ux,Kidonakis:2012rm,Kidonakis:2013zqa,Kidonakis:2015wva,Kidonakis:2016sjf,Kidonakis:2018ybz,Kidonakis:2019nqa,Kidonakis:2021vob}.

By expanding the resummed cross section to fixed order and, then, inverting back to momentum space, we derive results for soft-gluon corrections through NNLO without any need for a prescription. Thus, we calculate aNNLO cross sections that are matched to NLO as described in the next section.

\section{Numerical analysis\label{sec:numerical}}
In this section, we detail our calculations for the numerical analysis. We carry out the calculations at leading order (LO), approximate next-to-leading order (aNLO), and unmatched approximate next-to-next-to-leading order (uaNNLO) using in-house codes. We use \lhapdf~\cite{Buckley:2014ana} and \vegas~\cite{Lepage:2020tgj, Lepage:vegas} for Python, and at next-to-leading order (NLO) we use \madgraph~\cite{Alwall:2014hca} in conjunction with \ninja~\cite{Mastrolia:2012bu, Peraro:2014cba, Hirschi:2016mdz} and \collier~\cite{Denner:2016kdg}. The matched approximate next-to-next-to-leading (aNNLO) order is then obtained by using the relation
\begin{gather}
    \text{aNNLO} = \text{NLO} - \text{aNLO} + \text{uaNNLO},
\end{gather}
where aNLO is given by the LO calculations plus the first-order soft-gluon corrections.
\par
\begin{figure}
    [H]
    \centering
    \includegraphics[width=0.32\linewidth]{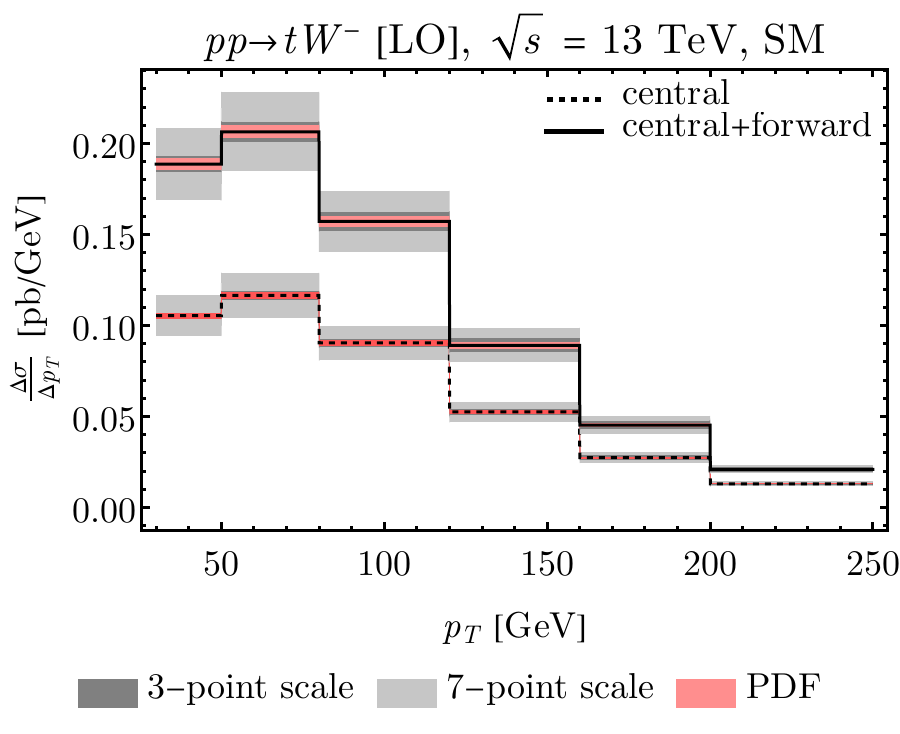}
    \includegraphics[width=0.32\linewidth]{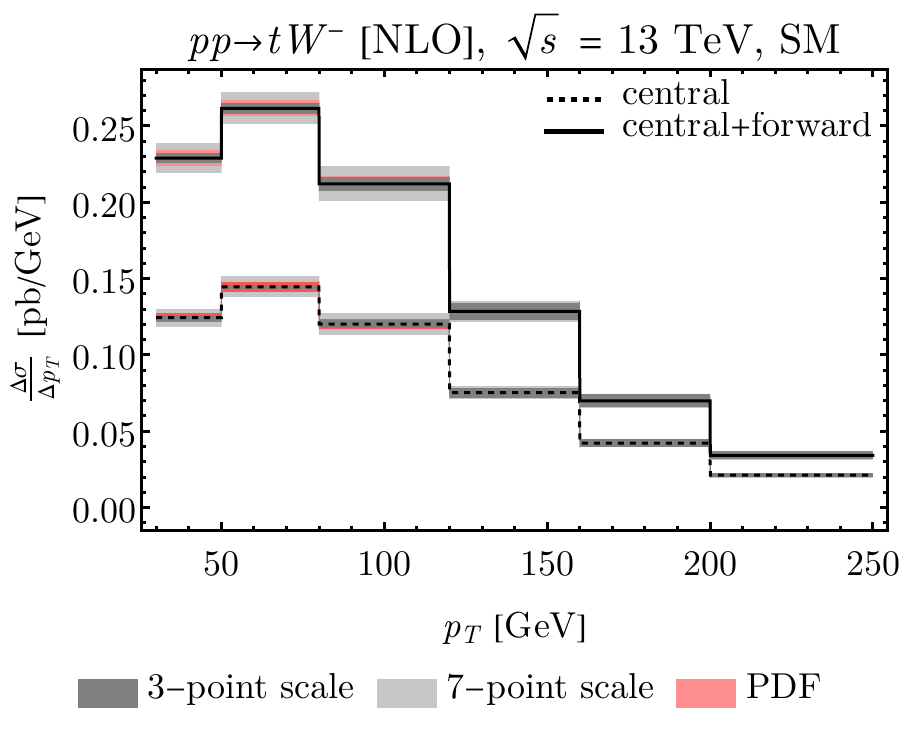}
    \includegraphics[width=0.32\linewidth]{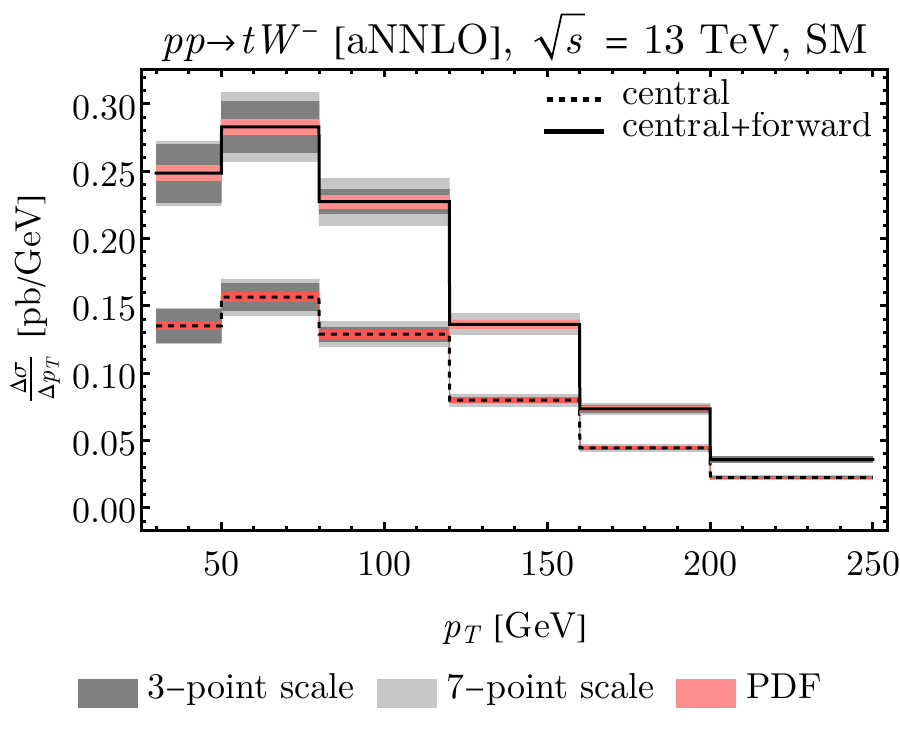}
    \caption{The top-quark distributions in $tW^-$ production in the SM across the assumed bins at LO (left), NLO (center), and aNNLO (right) at 13 TeV.}
    \label{fig:distributions_sm}
\end{figure}

 
The numerical values of the parameters are given by
\begin{gather}
    G_F = 1.1663788 \times 10^{-5} \ \GeV^{-2}, \\
    m_t = 172.5 \ \GeV, \\
    m_Z = 91.1880 \ \GeV, \\
    m_W = 80.3692 \ \GeV.
\end{gather}
The renormalization and factorization scales are fixed at $\mu_R = \mu_F = \mu_0 = m_t$ and we consider variations $\mu_R = f_R \mu_0$ and $\mu_F = f_F \mu_0$ with $f_R, f_F \in \cc{0.5, 1, 2}$. We use the MSHT20 NLO and NNLO PDFs~\cite{Bailey:2020ooq}. We assume $N_f=5$, and the running of the strong coupling $\alpha_s$ is provided from the PDF set. We set $\Lambda = 1 \ \TeV$. We impose basic fiducial acceptance requirements on the reconstructed top quark,
\begin{gather}
    |y|<3, \quad p_T > 30 \ \GeV,
\end{gather}
where the transverse momentum threshold serves to exclude the near-threshold region and ensures infrared-safe and numerically stable predictions in the binned analysis. We consider the binning
\begin{gather}
    p_T \in [30,\,50,\,80,\,120,\,160,\,200,\,250] \ \GeV, \\
    0 < |y| < 1, \quad 1 < |y| < 3.
\end{gather}
We refer to the first $y$ region as \textit{central} and the second as \textit{forward}. Thus, in total, we have 12 bins presented in Table~\ref{tab:bins}. We assume an integrated luminosity of $\mathcal{L}=139~\fb^{-1}$ for 13 TeV (Run II) and $\mathcal{L} = 300~\fb^{-1}$ for 13.6 TeV (Run III). Experimental systematics are modeled by a 5\% per-bin uncorrelated efficiency/selection uncertainty as a proxy for residual detector, efficiency, unfolding, and background-modeling effects, motivated by the typical size of experimental systematics reported in unfolded differential top-quark measurements at 13~TeV~\cite{CMS:2017xio}, and a 1.7\% fully correlated luminosity uncertainty consistent with the standard ATLAS Run~II luminosity calibration~\cite{ATLAS:2021xhq, ATLAS:2022hro}. We introduce fully correlated PDF uncertainties and uncorrelated scale variations as theoretical uncertainties.
{\renewcommand{\arraystretch}{1}
\begin{table}
    [H]
    \centering
    \caption{The top-quark transverse momentum and rapidity bins assumed in our analysis.}
    \label{tab:bins}
    \begin{tabular}{|c|c|c||c|c|c|}
        \hline 
        Bin & $p_T$ [GeV] & $y$ & Bin & $p_T$ [GeV] & $y$ \\
        \hline 
        1 & [30, 50] & \textit{central} & 7 & [120, 160] & \textit{central} \\ 
        2 & [30, 50] & \textit{forward}  & 8 & [120, 160] & \textit{forward} \\
        3 & [50, 80] & \textit{central} & 9 & [160, 200] & \textit{central} \\ 
        4 & [50, 80] & \textit{forward} & 10 & [160, 200] & \textit{forward} \\
        5 & [80, 120] & \textit{central} & 11 & [200, 250] & \textit{central} \\ 
        6 & [80, 120] & \textit{forward} & 12 & [200, 250] & \textit{forward} \\  
        \hline 
    \end{tabular}
\end{table}
}
The top-quark distributions across the bins in $tW^-$ production in the SM together with 3-point and 7-point scale variations, as well as PDF uncertainties, at LO, NLO, and aNNLO at 13 TeV are given in Fig.~\ref{fig:distributions_sm}. In this figure, the dashed line indicates the \textit{central} rapidity contribution at a given $p_T$ bin, whereas the solid line is the \textit{central} and \textit{forward} contributions combined. The dark gray shade indicates the 3-point scale variation, light gray the 7-point scale variation, and red the PDF variations. In the aNNLO distribution, we propagate PDF uncertainties by applying the same relative PDF variation observed at NLO. We have verified that this prescription reproduces the aNNLO PDF dependence to excellent accuracy for the distributions considered. The pictured behavior is similar for the 13.6 TeV distributions.
For the Run II (III) configuration, at LO, the PDF uncertainties range from 1.6\% (1.6\%) to 2.5\% (2.4\%) and the 3-point scale variations from 1.5\% (1.3\%) to 6.0\% (5.7\%), and the 7-point scale variations are about 11\% (11\%) and do not vary considerably. At NLO, the PDF uncertainties are about 2.5\% (2.5\%), and the 3-point scale uncertainties vary between 0.94\% (0.24\%) and 7.3\% (7.8\%), whereas the 7-point scale variations range from 4.0\% (4.4\%) to 8.1\% (8.3\%). At aNNLO, the PDF percentage uncertainties are the same as NLO, and the 3-point scale variations range from 1.5\% (1.6\%) to 9.2\% (8.2\%) and the 7-point scale uncertainties from 5.9\% (5.1\%) to 9.7\% (10\%). We choose 3-point variations as the basis for scale variations in the rest of the work. 
\par
\begin{figure}
    [H]
    \centering
    \includegraphics[width=0.4875\linewidth]{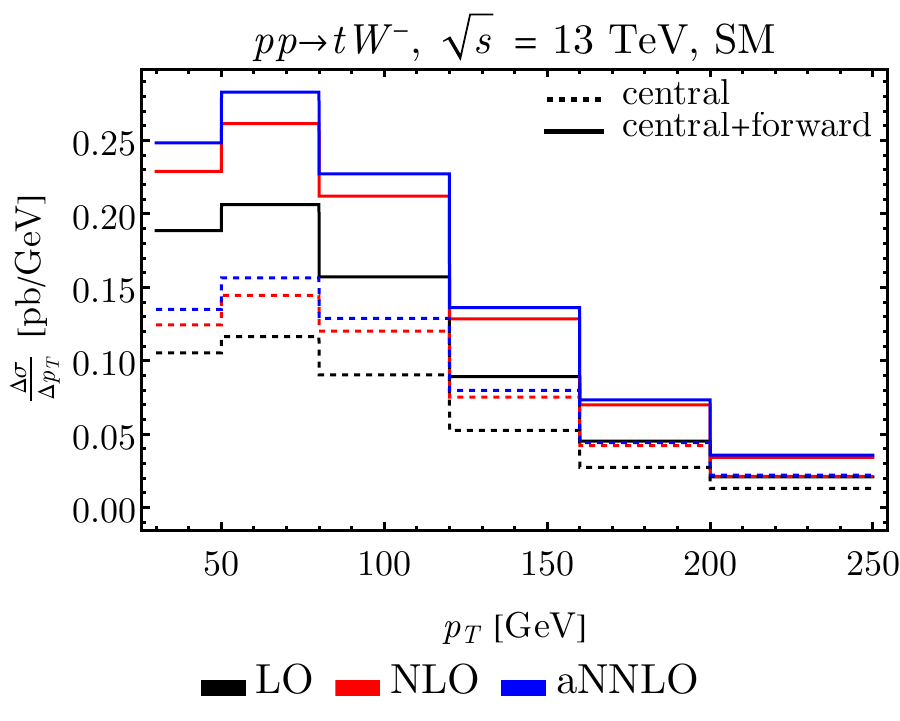}
    \includegraphics[width=0.4875\linewidth]{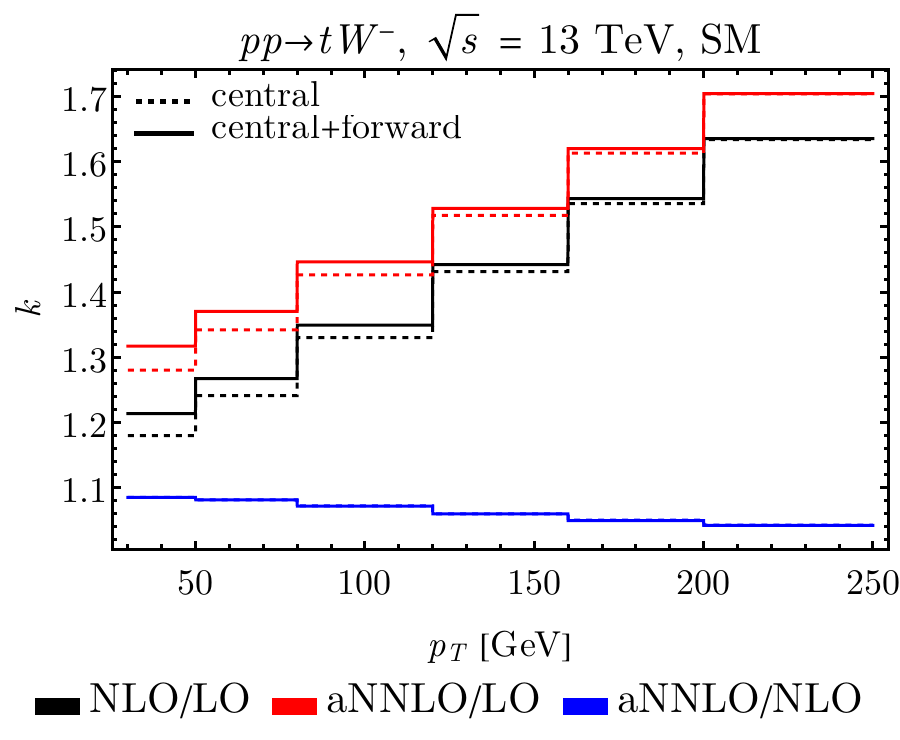}
    \caption{Comparison of LO, NLO, and aNNLO SM distributions (left) at 13 TeV, together with the corresponding $k$ factors (right) across the assumed bins.}
    \label{fig:distributions_sm_all}
\end{figure}

\begin{figure}
    [H]
    \centering
    \includegraphics[width=0.32\linewidth]{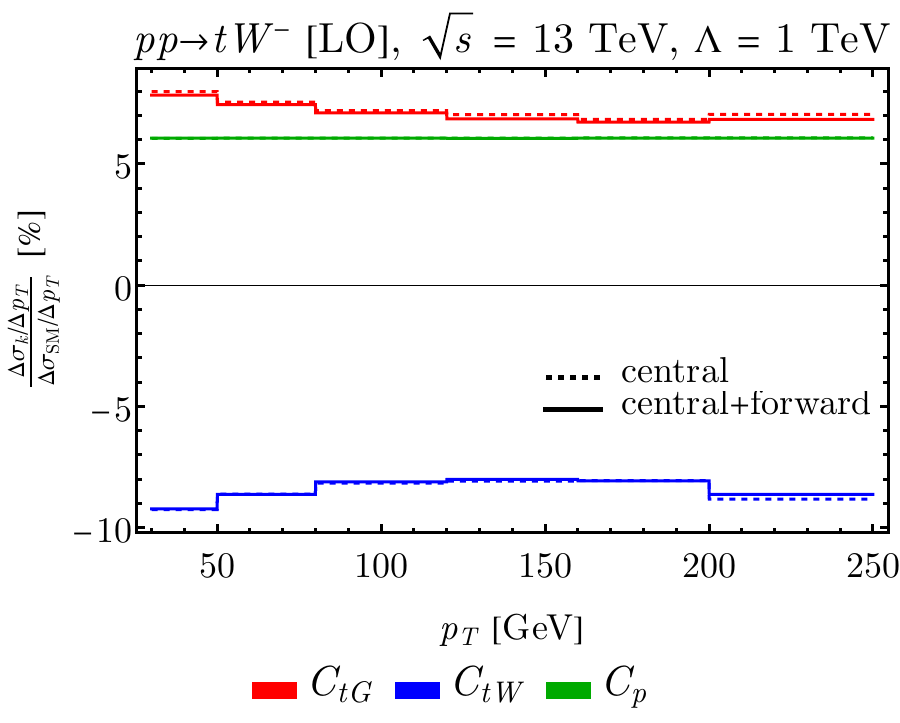}
    \includegraphics[width=0.32\linewidth]{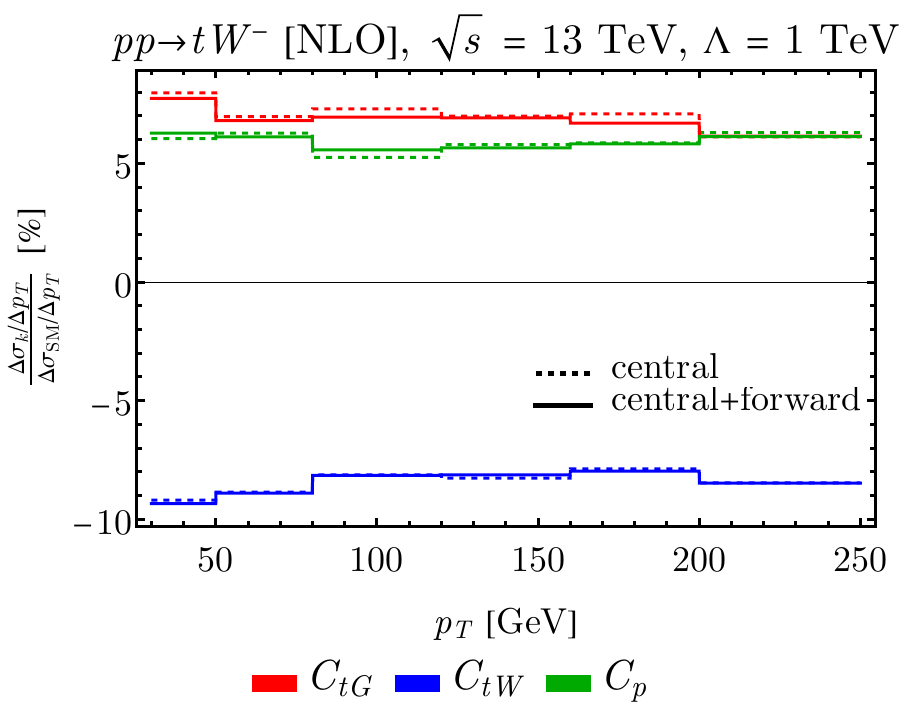}
    \includegraphics[width=0.32\linewidth]{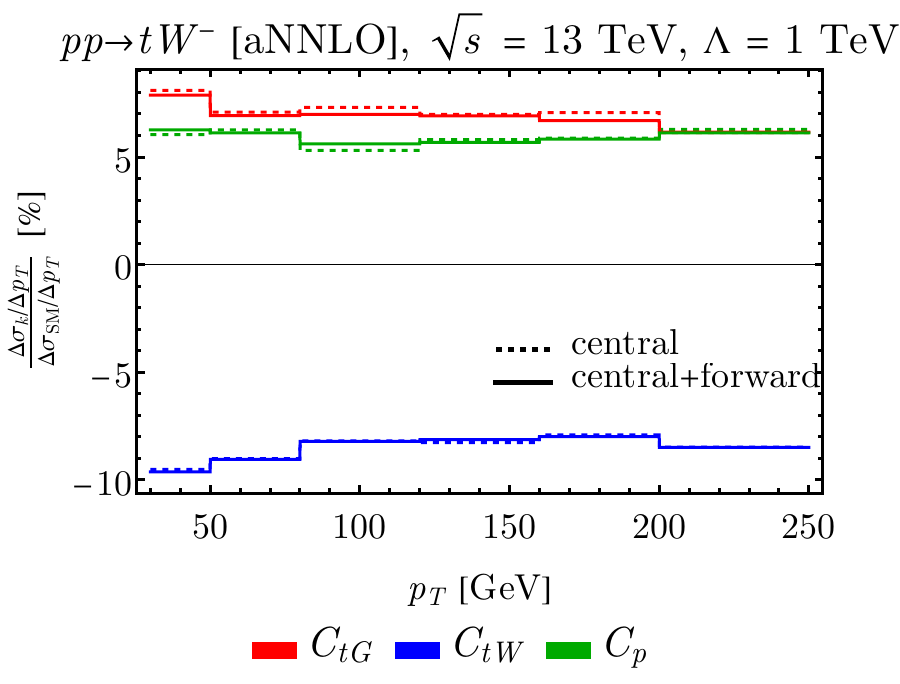}
    \caption{Linear SMEFT corrections to the top distributions characterized by $C_{tG}$, $C_{tW}$, and $C_p$ relative to the SM in percent at LO (left), NLO (center), and aNNLO (right) at 13 TeV.}
    \label{fig:distributions_linear}
\end{figure}

\begin{figure}
    [H]
    \centering
    \includegraphics[width=0.32\linewidth]{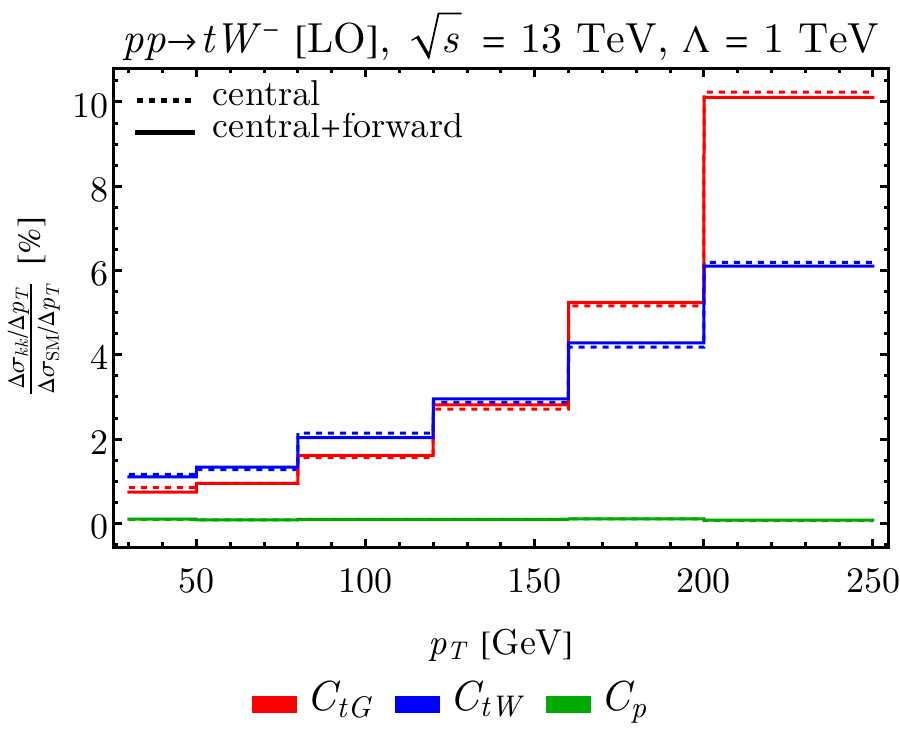}
    \includegraphics[width=0.32\linewidth]{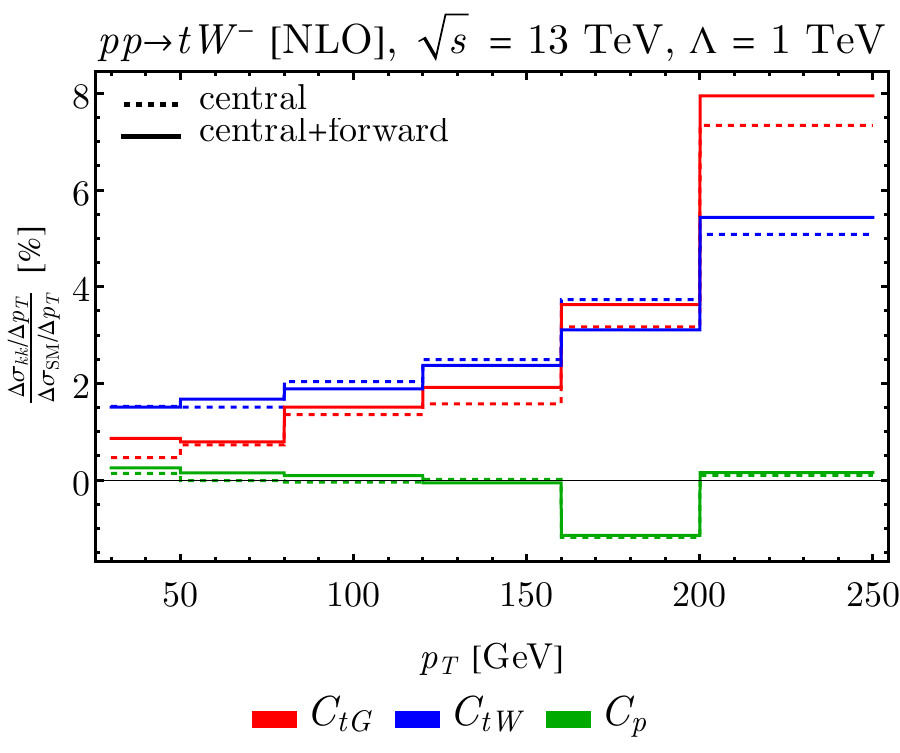}
    \includegraphics[width=0.32\linewidth]{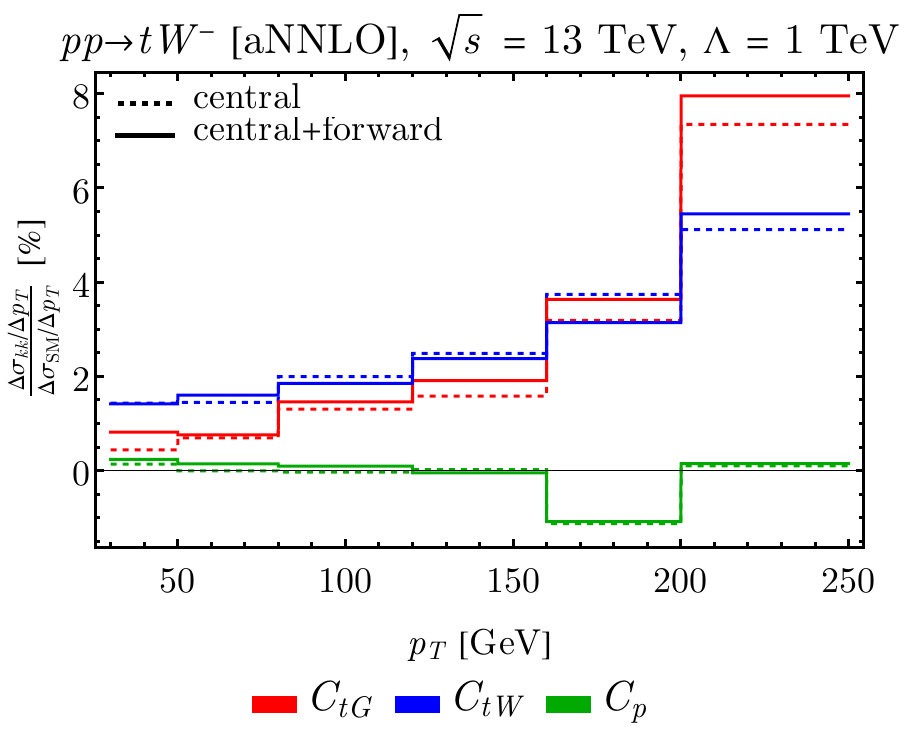}
    \caption{The same as Fig.~\ref{fig:distributions_linear} but for diagonal quadratic corrections characterized by $C_{tG}{}^2$, $C_{tW}{}^2$, and $C_p{}^2$.}
    \label{fig:distributions_quadratic}
\end{figure}

 
In Fig.~\ref{fig:distributions_sm_all}, we compare the LO, NLO, and aNNLO distributions and also present the $k$-factors across the assumed bins at 13 TeV within the SM. On the left panel, the black line indicates the LO, red NLO, and blue aNNLO; on the right, the black line is the $k$-factor for NLO and LO, red for aNNLO and LO, and blue for aNNLO and NLO. On both panels, the dashed line represents the \textit{central} rapidity contributions at a given $p_T$ bin and the solid one the combined \textit{central} and \textit{forward} contributions. The $k$ factor is nearly flat for aNNLO/NLO, varying only mildly around $\sim 1.08$ across the assumed bins. In contrast, the NLO/LO and aNNLO/LO $k$ factors exhibit a clear increase with $p_T$.
In Fig.~\ref{fig:distributions_linear} and \ref{fig:distributions_quadratic}, we present the linear and diagonal quadratic SMEFT contributions, respectively, to the top distributions relative to the SM in units of percent at LO, NLO, and aNNLO at 13 TeV. Here, the red line is for $C_{tG}$, blue for $C_{tW}$, and green for $C_p$. The dashed curve is for the \textit{central} rapidity contributions and solid for the combined \textit{central} and \textit{forward} contributions. We observe that the linear SMEFT correction characterized by $C_p$ is at around 6\% and is essentially flat across bins at LO. Once higher-order effects are included, it develops small but noticeable bin-to-bin variations. The linear corrections associated with $C_{tG}$ and $C_{tW}$ are at around 8\% and exhibit mild shape dependence. A similar pattern holds for the diagonal quadratic terms. The $C_p{}^2$ contribution is around 0.1\% at LO with little to no bin sensitivity, but including running Wilson coefficients at higher orders again induces small yet visible variations. In contrast, the $C_{tG}{}^2$ and $C_{tW}{}^2$ contributions show a pronounced $p_T$ dependence, growing significantly toward the higher end. As anticipated, the sensitivity gets stronger toward larger $p_T$; however, this is the far tail where the $tW$ signal yield is already small. The behavior is similar for 13.6 TeV. 
We now describe our statistical procedure. We perform linear and quadratic SMEFT fits at dimension 6 on the three Wilson coefficients of interest $C_{tG}$, $C_{tW}$, and $C_p$. We choose the doubly differential cross section as our observable. For our fitting, we perform a standard $\chi^2$ test statistics. We define
\begin{gather}
    \chi^2 = (\mathcal O - \hat{\mathcal O})^\top \mathcal H (\mathcal O - \hat{\mathcal O}),
\end{gather}
where $\mathcal O$ is the pseudodata, $\hat{\mathcal O}$ is the model function, namely the observable linear or quadratic in SMEFT parameters, and $\mathcal H = \mathcal E^{-1}$ is the inverse of the uncertainty matrix, $\mathcal E$. We start with the uncertainty matrix. It has two parts, experimental and theoretical:
\begin{gather}
    \mathcal E = \mathcal E^{\rm exp} + \mathcal E^{\rm theo}.
\end{gather}
Here, the experimental part is given by
\begin{gather}
    \mathcal E_{bb'}^{\rm exp}= I_{bb'} (\delta_b^{\rm stat} \oplus \delta_b^{\rm eff})^2 + \rho_{bb'} \delta_b^{\rm corr} \delta_{b'}^{\rm corr},
\end{gather}
where $b,b'=1, \ldots, N_{\rm bin}$ is the bin index, $N_{\rm bin} = 12$ is the number of bins, $I$ is the identity matrix, $\delta_b^{\rm stat/eff/corr}$ is the absolute statistical, efficiency, and correlated uncertainties at the $b$th bin, respectively, given by
\begin{gather}
    \delta_b^{\rm stat} = {1 \over \sqrt{\sigma_{{\rm SM}, b} \mathcal L}} \times \sigma_{{\rm SM}, b}, \quad 
    \delta_b^{\rm eff} = 5\% \times \sigma_{{\rm SM}, b}, \quad 
    \delta_b^{\rm corr} = 1.7\% \times \sigma_{{\rm SM}, b},
\end{gather}
and $\oplus$ indicates sum in quadrature. We assume $\rho _{bb'} = 1$ for all bins. On the theoretical side, we have fully correlated PDF uncertainties, as well as uncorrelated 3-point scale variations. The PDF uncertainty matrix for Hessian sets is given by
\begin{gather}
    \mathcal E_{bb'}^{\rm pdf} = {1\over 4} \sum_{m=1}^{N_{\rm pdf} / 2} \pp{\mathcal O_{2m} - \mathcal O_{2m-1}}_b \pp{\mathcal O_{2m} - \mathcal O_{2m-1}}_{b'},
\end{gather}
and the scale variations are defined as 
\begin{gather}
    \mathcal E_{bb'}^{\rm scale} = I_{bb'} \max\cc{\vv{\mathcal O_b^{(f_R,f_F)} - \mathcal O_b^{(1,1)}}^2}
\end{gather}
where $N_{\rm pdf}$ is the number of PDF members in a given set and we consider $(f_R, f_F) = \{(0.5,0.5), (1,1), (2,2)\}$. We note that the model function is taken without theoretical uncertainties. PDF and scale variations are defined at the SM baseline and propagated consistently. Assigning them separately to SMEFT corrections would double count and introduce spurious $C$-dependence in the uncertainty matrix. To verify that this choice does not materially affect the extracted constraints, we perform a robustness test in which the fractional PDF and scale uncertainties of the SM baseline are used to rescale the theory covariance bin-by-bin by the EFT-corrected prediction, introducing a controlled $C$-dependent proxy for theory errors. Since a $C$-dependent covariance renders the $\chi^2$ non-polynomial if inverted exactly, we restore a well-defined EFT truncation by expanding $\mathcal{E}^{-1}$ consistently in $1/\Lambda^2$ and truncating at $\mathcal{O}(1/\Lambda^2)$ and $\mathcal{O}(1/\Lambda^4)$ for the linear and quadratic analyses, respectively. Repeating the 13~TeV aNNLO analysis under this prescription, we find that the linear Fisher matrix and derived bounds are unchanged, and that the quadratic Fisher matrix changes at below the 1\% level, leading to negligible shifts in the extracted bounds and correlations. We therefore retain the fixed SM-point covariance as our baseline.
\par 
The pseudodata is generated by smearing SM values around using the experimental uncertainties:
\begin{gather}
    \mathcal O_b = \mathcal O_{{\rm SM}, b} + r_b (\delta_b^{\rm stat} \oplus \delta_b^{\rm eff}) + r' \delta_b^{\rm corr},
\end{gather}
where $r_b, r' \in \mathcal N(0, 1)$ unit normal variates. The correlated uncertainty is introduced with the same factor so that all the bins feel the shift by the same amount. For our analysis to make sense statistically, we need to run multiple pseudoexperiments, hence it is now appropriate to introduce the subscript $e$ so we have $\chi^2_e$ denoting the chi-squared function of the $e$th pseudoexperiment. We obtain the best-fit values $\bar{\vec C}_e$ by setting the gradient equal to zero in the parameter space,
\begin{gather}
    \vec \nabla \chi_e^2 (\bar{\vec C}_e) = \vec 0,
\end{gather}
and the Fisher information matrix is just the Hessian evaluated at the best-fit values,
\begin{gather}
    \mathcal F_e = \vec \nabla \vec \nabla \chi_e^2 (\bar{\vec C}_e).
\end{gather}
The best-fit values of SMEFT parameters averaged over pseudoexperiments is given by 
\begin{gather}
    \bar{\vec C} = \pp{\sum_{e=1}^{N_{\rm exp}} \mathcal F_e}^{-1} \sum_{e=1}^{N_{\rm exp}} \mathcal F_e \bar{\vec C}_e,
\end{gather}
and the Fisher matrix averaged over the pseudoexperiments by 
\begin{gather} 
    \mathcal F = {1 \over N_{\rm exp}} \sum_{e=1}^{N_{\rm exp}} \mathcal F_e, 
\end{gather} 
where $N_{\rm exp}$ is the number of pseudoexperiments. We find that with $N_{\rm exp} = 10^4$ pseudoexperiments, the fits stabilize and we attain reliable statistics. For the combined fit of Run II+III, we assume they are independent so we simply add the individual Fisher matrices, $\mathcal F_{\rm II+III} = \mathcal F_{\rm II} + \mathcal F_{\rm III}$. 
\par
\begin{figure}
    [H]
    \centering
    \includegraphics[width=0.4875\linewidth]{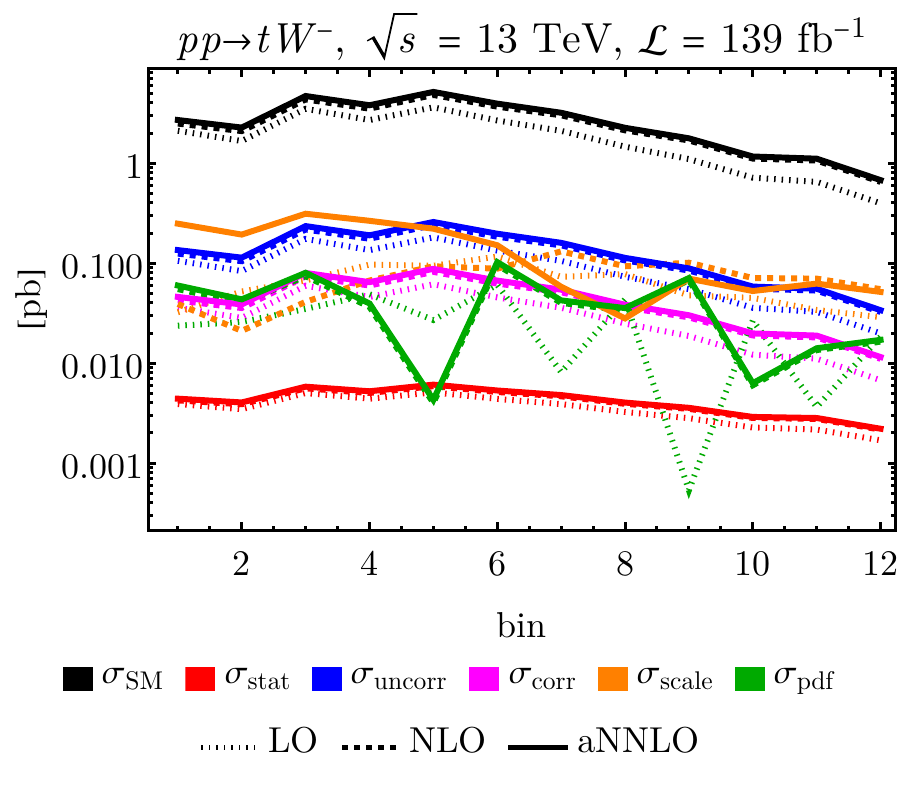}
    \includegraphics[width=0.4875\linewidth]{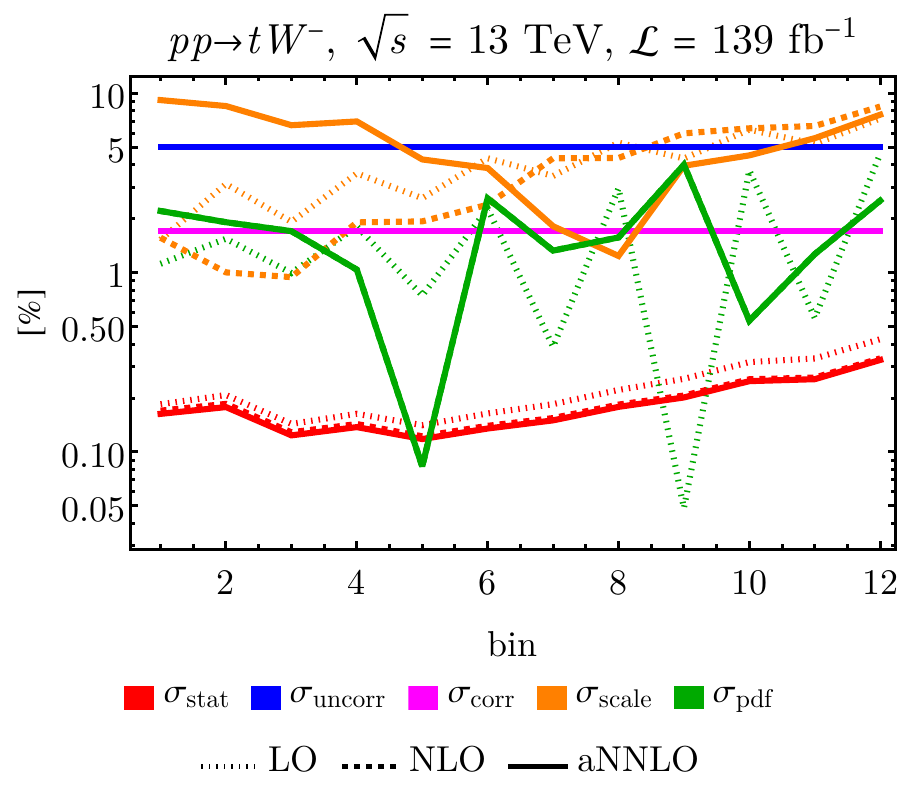}
    \caption{The error budget for the Run II configuration with $\sqrt s = 13 \ \TeV$ and $\mathcal L = 139 \ \fb^{-1}$, showing the LO, NLO, and aNNLO SM predictions and their uncertainty components in absolute terms (left) and relative to the SM in units of percent (right) for each bin.}
    \label{fig:error_budget_13}
\end{figure}

\begin{figure}
    [H]
    \centering
    \includegraphics[width=0.4875\linewidth]{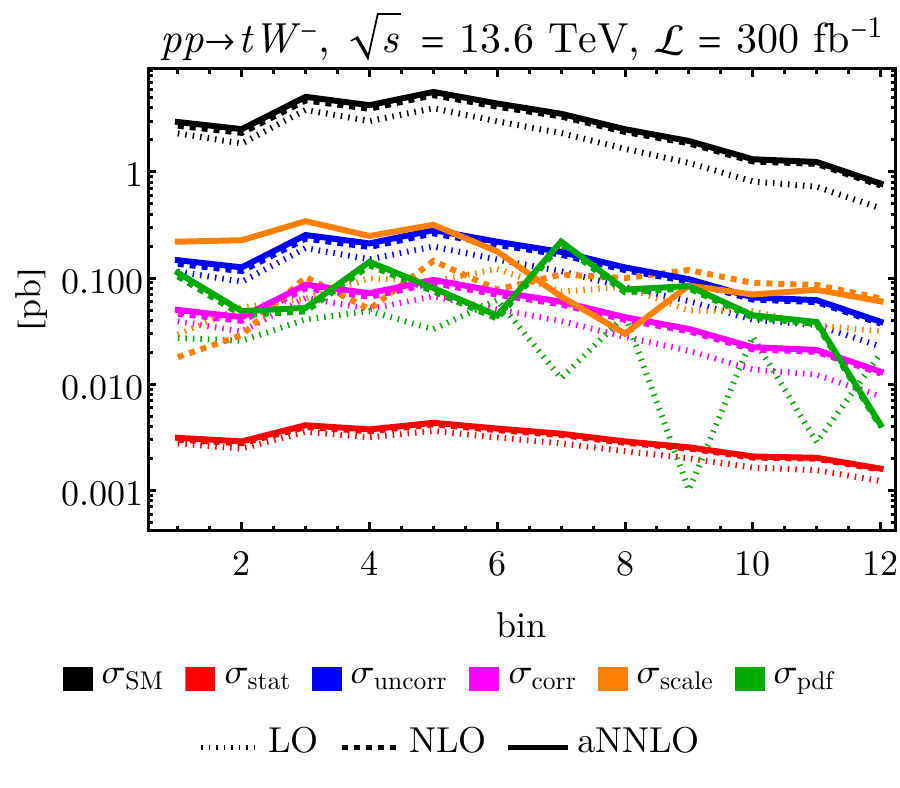}
    \includegraphics[width=0.4875\linewidth]{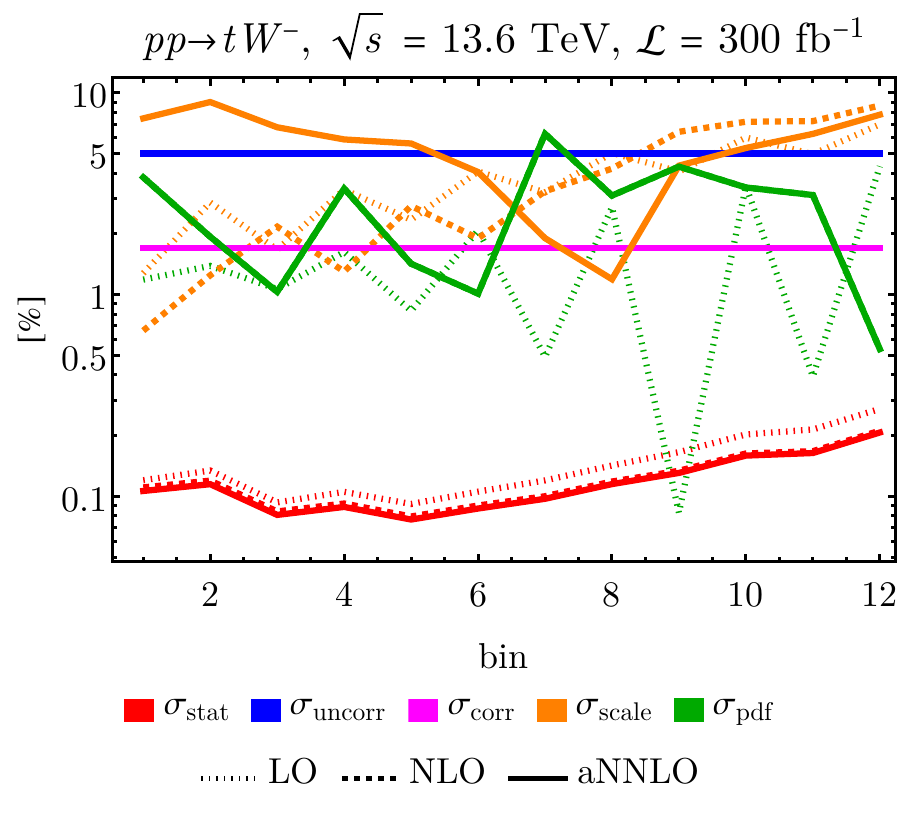}
    \caption{The same as Fig.~\ref{fig:error_budget_13} but for the Run III configuration with $\sqrt s = 13.6 \ \TeV$ and $\mathcal L = 300 \ \fb^{-1}$.}
    \label{fig:error_budget_136}
\end{figure}

 
Before presenting the fit results, we summarize the error budget entering our fits, namely the individual contributions to the diagonal entries of the total uncertainty matrix. In Figs.~\ref{fig:error_budget_13} and \ref{fig:error_budget_136}, we show for each bin the SM predictions at LO, NLO, and aNNLO together with the corresponding uncertainty components in absolute terms on the left panels, as well as the same components relative to the SM prediction expressed in units of percent on the right for Run II and Run III configurations, respectively. The black line is for the SM prediction, red for statistical uncertainties, blue for the assumed 5\% global uncorrelated uncertainty, magenta for the assumed 1.7\% correlated luminosity uncertainty, orange for 3-point scale variations, and green for PDF uncertainties. The dotted line is for LO, dashed for NLO, and solid for aNNLO. 
We find that the dominant uncertainty is set by a competition between the 3-point scale variation and the assumed global uncorrelated uncertainty. The next largest contributions come from the PDF variations and the correlated luminosity uncertainty, which are comparable in size. By contrast, the statistical uncertainties play the weakest role. 

\section{SMEFT fit results\label{sec:results}}
In this section, we present the SMEFT fit results. First, we report the Fisher information matrices for the Run II and Run III configurations for the linear and quadratic fits at LO, NLO, and aNNLO:
\begin{gather}
    \mathcal F_{\rm II}^{\rm LO, linear} = \left(
\begin{array}{ccc}
 16.24 & -19.02 & 13.46 \\
 -19.02 & 22.32 & -15.78 \\
 13.46 & -15.78 & 11.22 \\
\end{array}
\right), \\
    \mathcal F_{\rm III}^{\rm LO, linear} = \left(
\begin{array}{ccc}
 16.92 & -19.75 & 13.99 \\
 -19.75 & 23.08 & -16.35 \\
 13.99 & -16.35 & 11.64 \\
\end{array}
\right), \\
    \mathcal F_{\rm II}^{\rm LO, quadratic} = \left(
\begin{array}{ccc}
 21.34 & -20.21 & 13.49 \\
 -20.21 & 25.94 & -15.22 \\
 13.49 & -15.22 & 11.12 \\
\end{array}
\right), \\
    \mathcal F_{\rm III}^{\rm LO, quadratic} = \left(
\begin{array}{ccc}
 22.15 & -21.00 & 13.95 \\
 -21.00 & 27.57 & -16.07 \\
 13.95 & -16.07 & 11.56 \\
\end{array}
\right), \\
    \mathcal F_{\rm II}^{\rm NLO, linear} = \left(
\begin{array}{ccc}
 13.07 & -15.97 & 10.92 \\
 -15.97 & 19.74 & -13.53 \\
 10.92 & -13.53 & 9.316 \\
\end{array}
\right), \\
    \mathcal F_{\rm III}^{\rm NLO, linear} = \left(
\begin{array}{ccc}
 9.579 & -11.63 & 8.418 \\
 -11.63 & 14.23 & -10.26 \\
 8.418 & -10.26 & 7.551 \\
\end{array}
\right), \\
    \mathcal F_{\rm II}^{\rm NLO, quadratic} = \left(
\begin{array}{ccc}
 18.28 & -15.21 & 10.30 \\
 -15.21 & 25.14 & -13.38 \\
 10.30 & -13.38 & 9.674 \\
\end{array}
\right), \\
    \mathcal F_{\rm III}^{\rm NLO, quadratic} = \left(
\begin{array}{ccc}
 14.57 & -12.24 & 7.425 \\
 -12.24 & 18.42 & -10.59 \\
 7.425 & -10.59 & 8.228 \\
\end{array}
\right), \\
    \mathcal F_{\rm II}^{\rm aNNLO, linear} = \left(
\begin{array}{ccc}
 11.68 & -14.21 & 9.842 \\
 -14.21 & 17.47 & -12.11 \\
 9.842 & -12.11 & 8.425 \\
\end{array}
\right), \\
    \mathcal F_{\rm III}^{\rm aNNLO, linear} = \left(
\begin{array}{ccc}
 7.368 & -9.042 & 6.543 \\
 -9.042 & 11.18 & -8.069 \\
 6.543 & -8.069 & 5.911 \\
\end{array}
\right), \\
    \mathcal F_{\rm II}^{\rm aNNLO, quadratic} = \left(
\begin{array}{ccc}
 15.74 & -13.31 & 9.343 \\
 -13.31 & 21.50 & -11.79 \\
 9.343 & -11.79 & 8.864 \\
\end{array}
\right), \\
    \mathcal F_{\rm III}^{\rm aNNLO, quadratic} = \left(
\begin{array}{ccc}
 11.63 & -9.680 & 5.712 \\
 -9.680 & 15.12 & -8.472 \\
 5.712 & -8.472 & 6.671 \\
\end{array}
\right).
\end{gather}
We emphasize that our results are presented for $pp\to tW^-$ throughout. The charge-conjugate channel $pp\to \bar t W^+$ yields essentially the same kinematic dependence and nearly identical rates, so its impact on the statistical sensitivity is, to an excellent  approximation, an overall doubling of the information content. Consequently, if both channels are included with identical binning and uncertainty treatment, one may approximate the combined constraints by rescaling the Fisher information matrix as $F_{tW^- + \bar t W^+}\simeq 2 F_{tW^-}$, which implies the corresponding marginalized uncertainties shrink by a factor $1/\sqrt{2}$. We have checked that this approximation is sufficient at the level of precision targeted in this work.
\par
\begin{figure}
    [H]
    \centering
    \includegraphics[width=.55\linewidth]{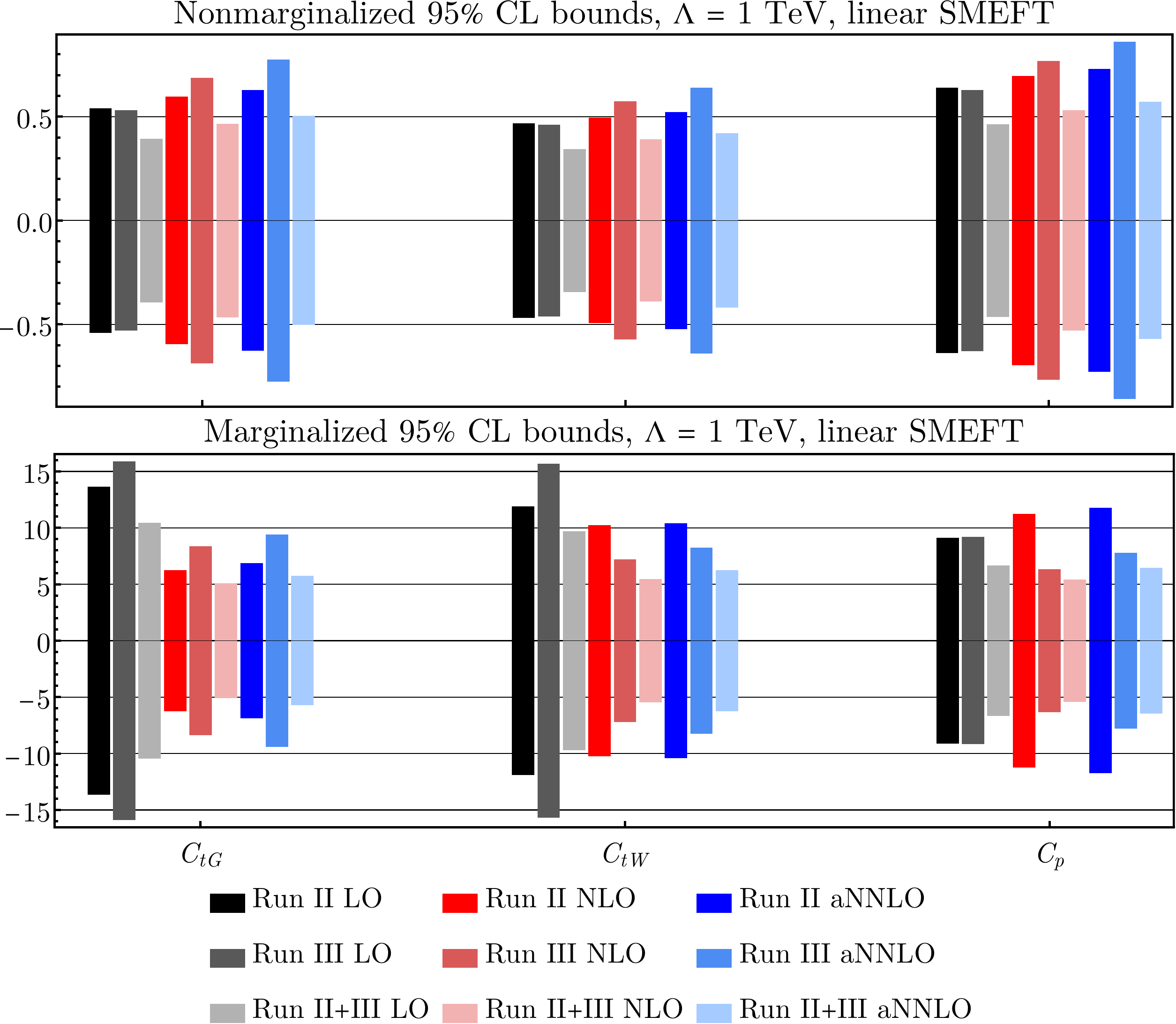}
    \caption{Nonmarginalized (top) and marginalized (bottom) 95\% CL bounds for the linear fit for Run II, Run III, and Run II+III at LO, NLO, and aNNLO.}
    \label{fig:bounds_linear}
\end{figure}

\begin{figure}
    [H]
    \centering
    \includegraphics[width=.55\linewidth]{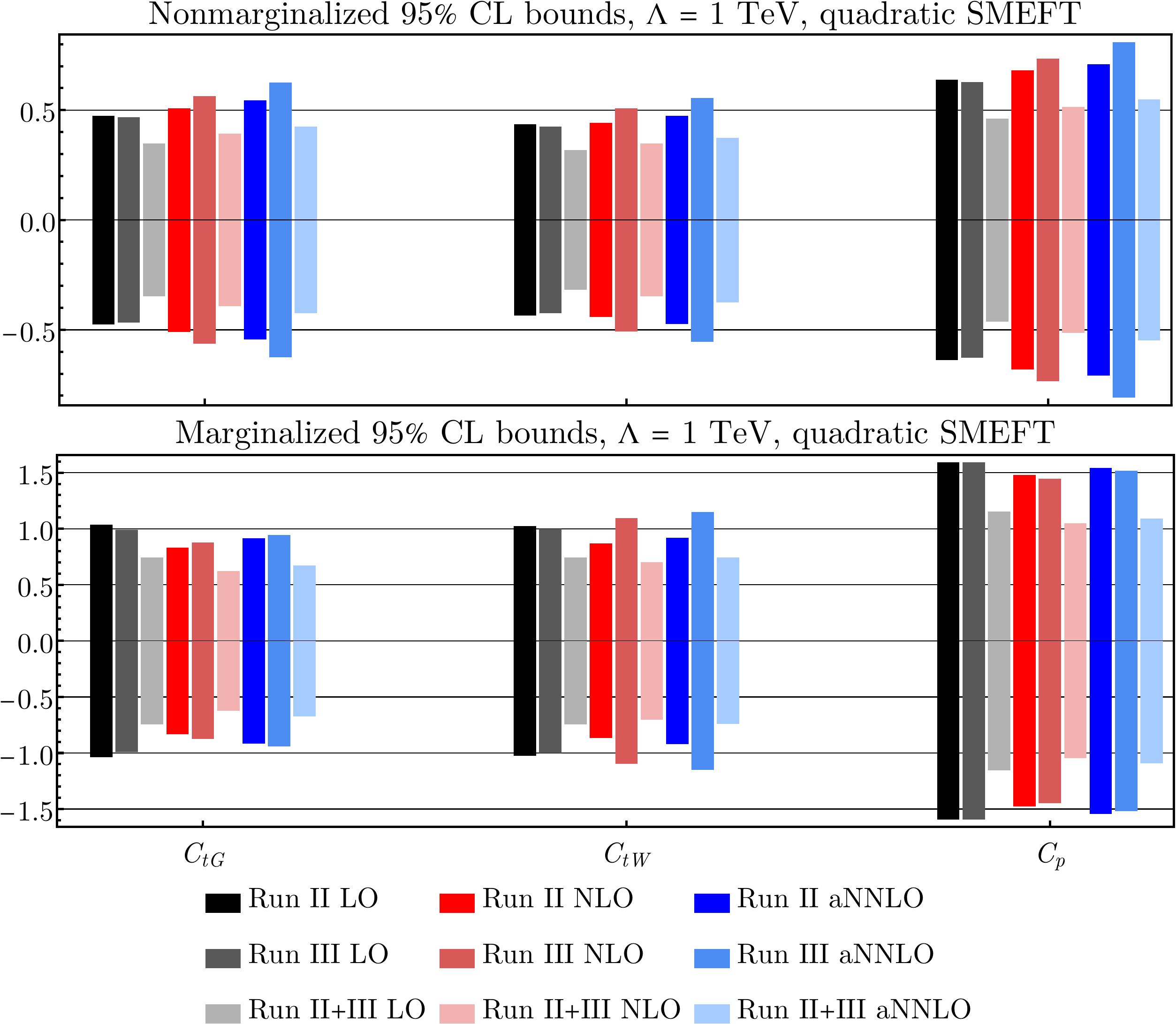}
    \caption{The same as Fig.~\ref{fig:bounds_linear} but for the quadratic fit.} 
    \label{fig:bounds_quadratic}
\end{figure}

 
Next, we present 95\% CL bounds on $C_{tG}$, $C_{tW}$, and $C_p$ for linear and quadratic SMEFT fits. Fig.~\ref{fig:bounds_linear} shows the nonmarginalized and marginalized bounds for the linear fit, and Fig.~\ref{fig:bounds_quadratic} shows the same for the quadratic fit. Here, nonmarginalized refers to fits in which only two Wilson coefficients are allowed to vary at a time, with the remaining coefficient set to zero, while marginalized refers to the full 3-parameter fit projected onto the chosen two-dimensional subspace.  In these figures, the band of colors on the left is for $C_{tG}$, center for $C_{tW}$, and right for $C_p$. The black color family is for LO, red for NLO, and blue for aNNLO. The darkest shade of each color family is for Run II, middle for Run III, and the lightest for the combined fit Run II and Run III. 
At linear order in the Wilson coefficients, the nonmarginalized bounds are at the level of $\mathcal O(0.1)$, while the marginalized bounds degrade to $\mathcal O(10)$. This large weakening indicates strong correlations in the multi-parameter fit. The Run II and Run III bounds are comparable in size because we use the same bins and the collider energy is slightly increased, so the combined Run II+III fit yields only a modest improvement over the individual results.
At quadratic order, the nonmarginalized bounds are comparable to those from the linear fit, while the difference becomes significant after marginalization. The marginalized quadratic bounds are about an order of magnitude stronger than the marginalized linear bounds, indicating that including squared dimension-6 terms improves the constraining power of the fit. This improvement should be interpreted with care, since quadratic dimension-6 effects are parametrically of the same order as linear dimension-8 contributions, and a consistent assessment would require including the dimension-8 operators. 
\par
\begin{figure}
    [H]
    \centering
    \includegraphics[width=.55\linewidth]{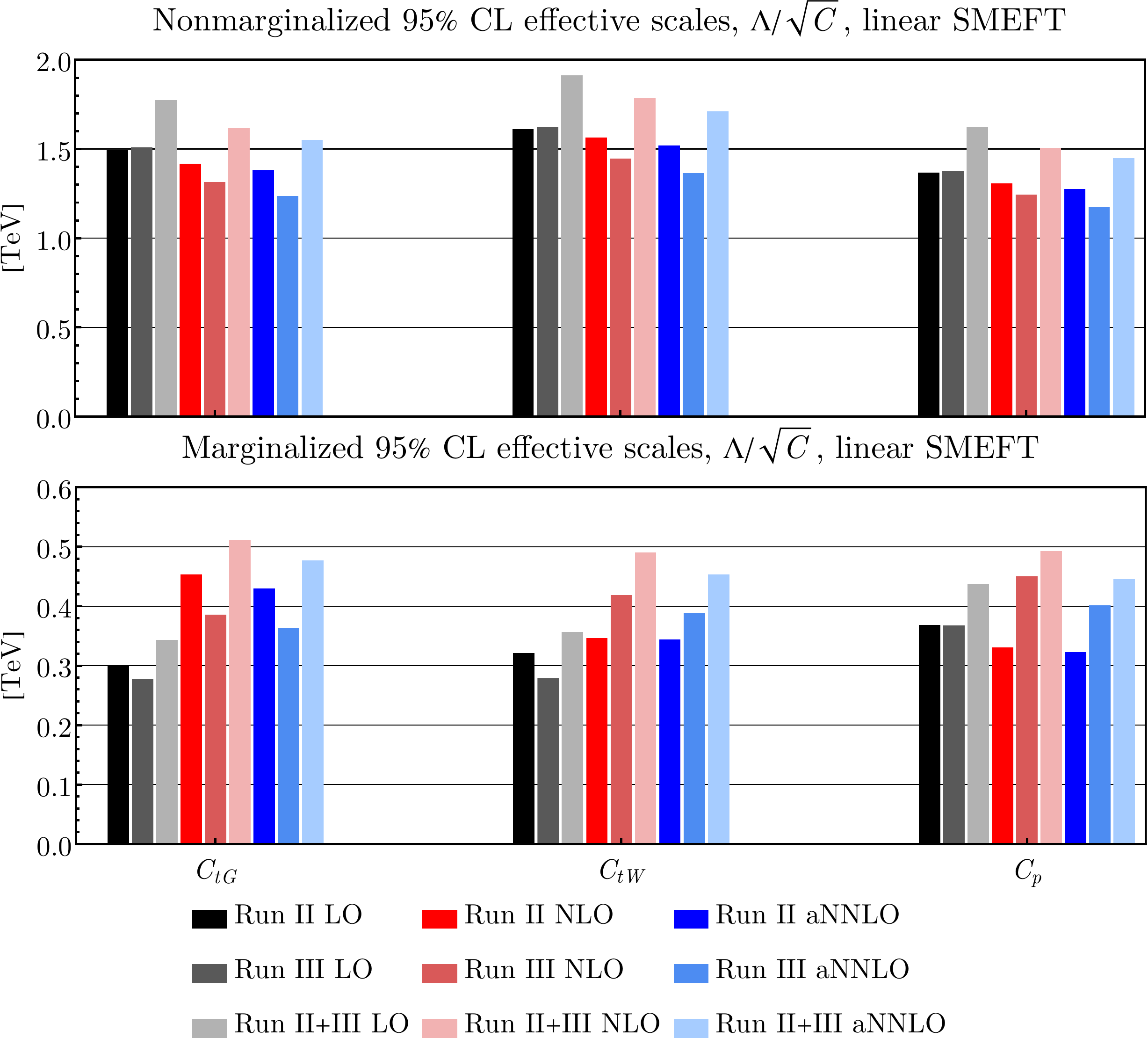}
    \caption{Nonmarginalized (top) and marginalized (marginalized) effective scales $\Lambda/\sqrt{\Delta C_k}$ for the linear fit for Run II, Run III, and Run II+III at LO, NLO, and aNNLO.}
    \label{fig:eff_scales_linear}
\end{figure}

\begin{figure}
    [H]
    \centering
    \includegraphics[width=.55\linewidth]{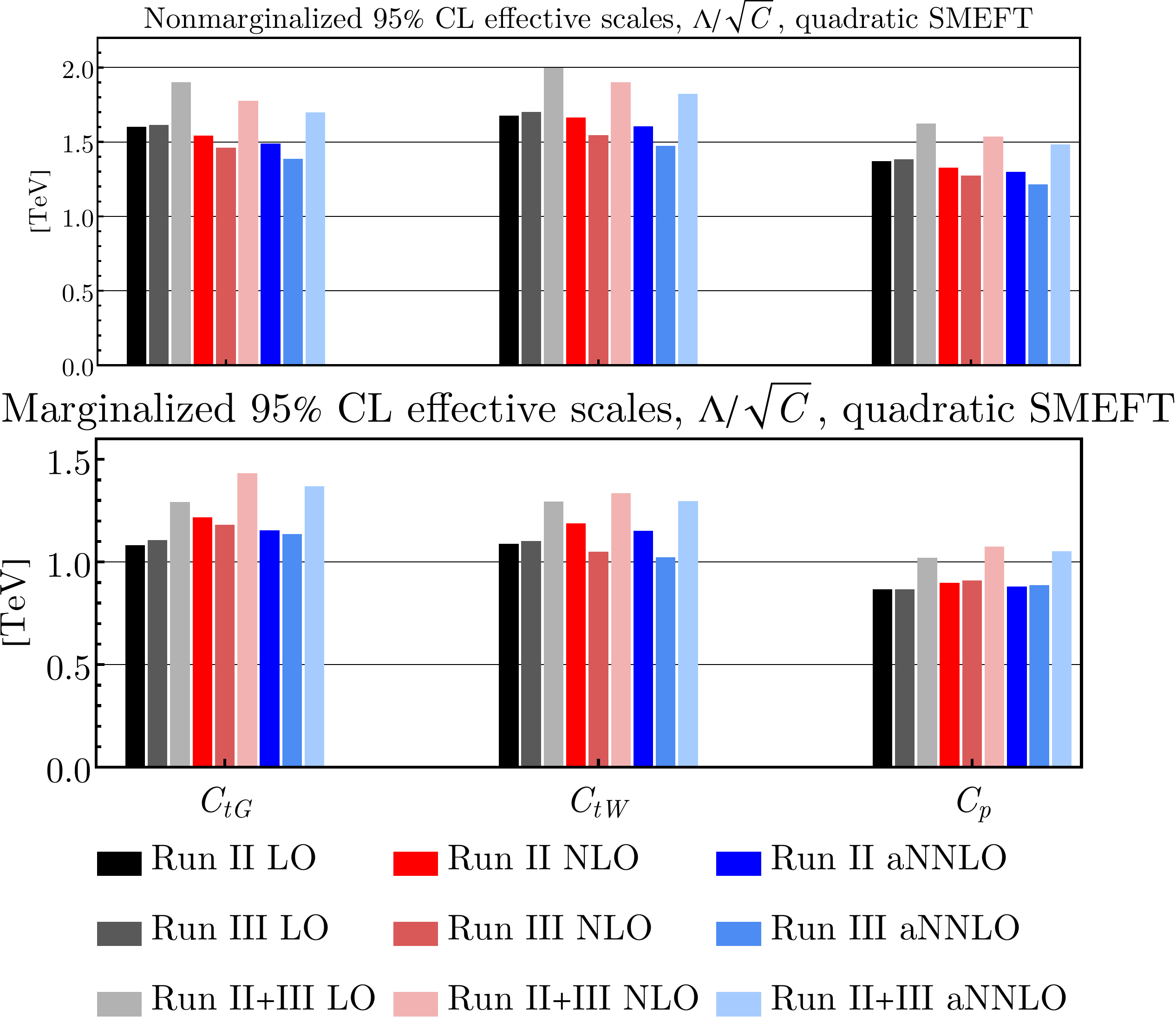}
    \caption{The same as Fig.~\ref{fig:eff_scales_linear} but for the quadratic fit.}
    \label{fig:eff_scales_quadratic}
\end{figure}

 
Next, we quote the corresponding effective scales defined as $\Lambda/\sqrt{\Delta C_k}$ using the 95\% CL bounds. Fig.~\ref{fig:eff_scales_linear} shows the nonmarginalized and marginalized effective scales for the linear fit, and Fig.~\ref{fig:eff_scales_quadratic} shows the same for the quadratic fit. The organization of the plots is the same as the bounds. 
At linear order, the nonmarginalized bounds correspond to effective scales up to about 2 TeV, while the marginalized bounds reduce this reach to roughly 0.5 TeV. For the quadratic fit, the nonmarginalized reach is again around 2 TeV, and the marginalized reach increases to nearly 1.5 TeV. Since the characteristic scale of the process is set by the top quark mass, these numbers suggest that the EFT expansion parameter $m_t/\Lambda$ is better behaved once the squared dimension-6 terms are included. This provides additional motivation to extend the analysis in the future by incorporating dimension-8 effects for a more complete EFT description. 
\par
\begin{figure}
    [H]
    \centering
    \includegraphics[width=0.26\linewidth]{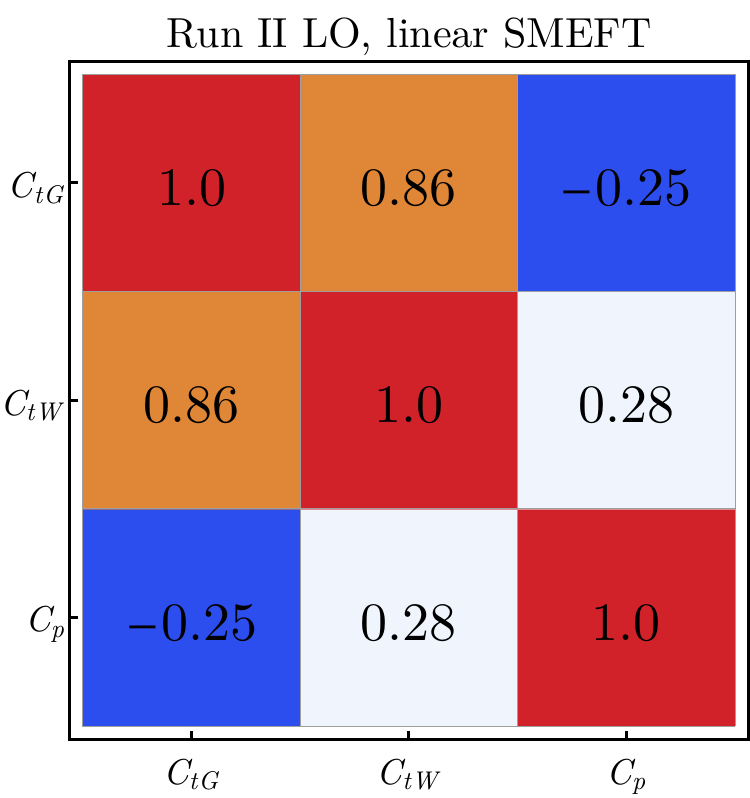}
    \includegraphics[width=0.26\linewidth]{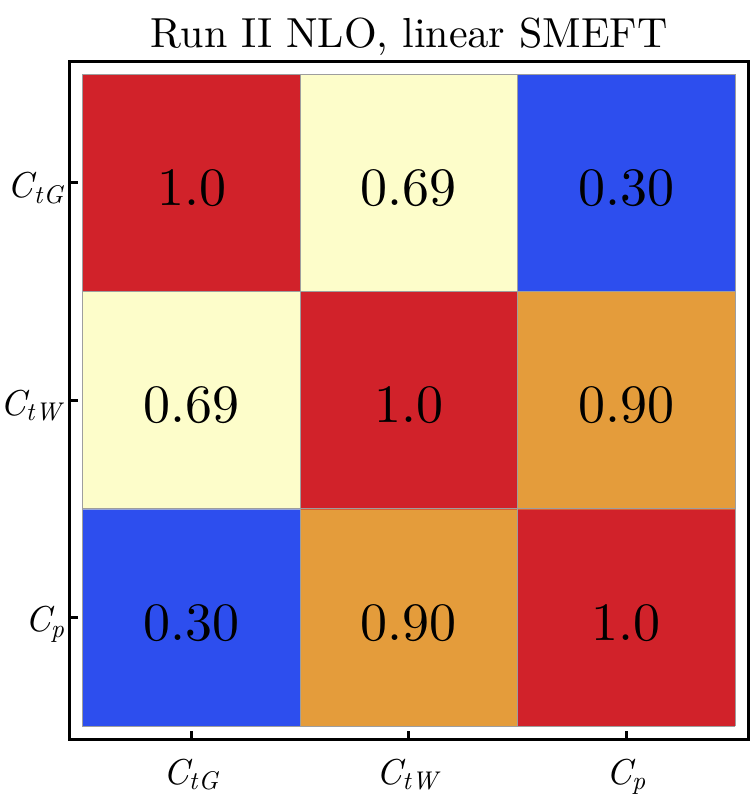}
    \includegraphics[width=0.26\linewidth]{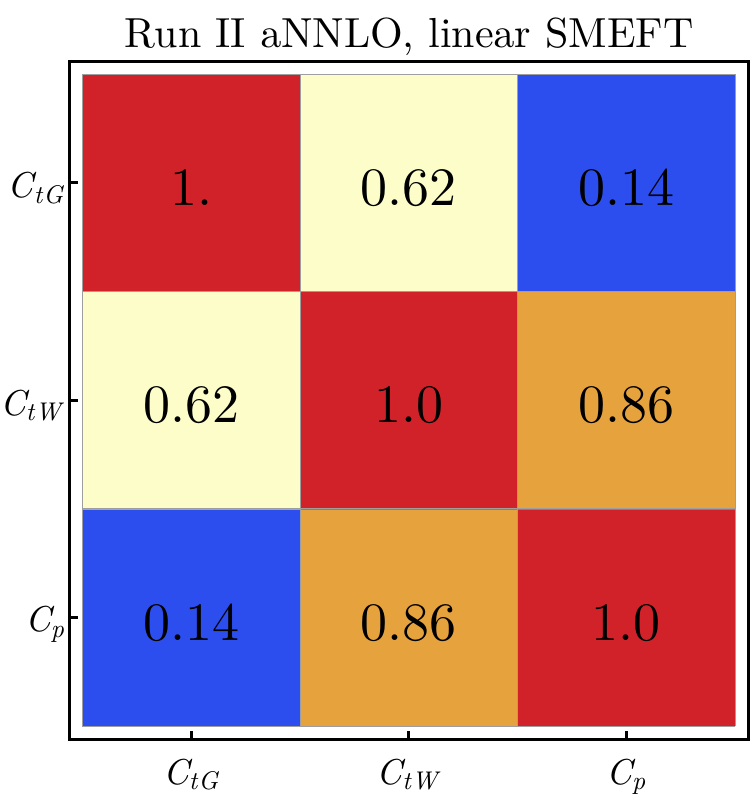}
    \includegraphics[width=0.26\linewidth]{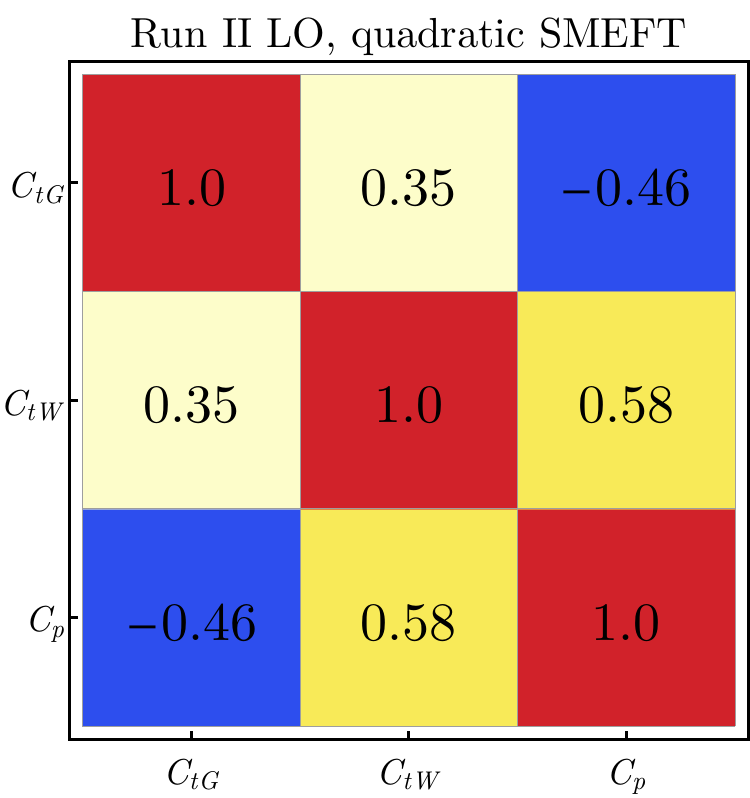}
    \includegraphics[width=0.26\linewidth]{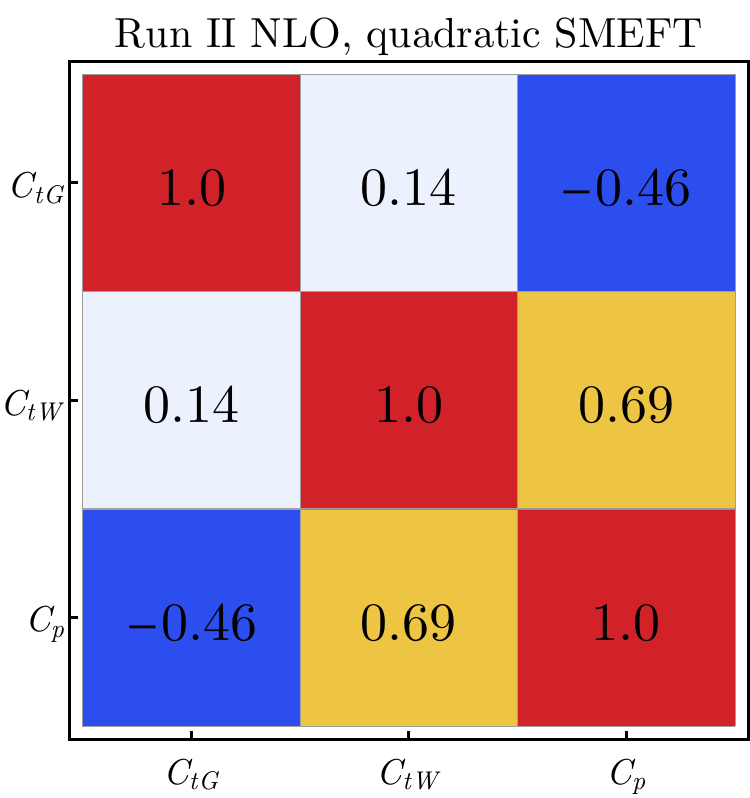}
    \includegraphics[width=0.26\linewidth]{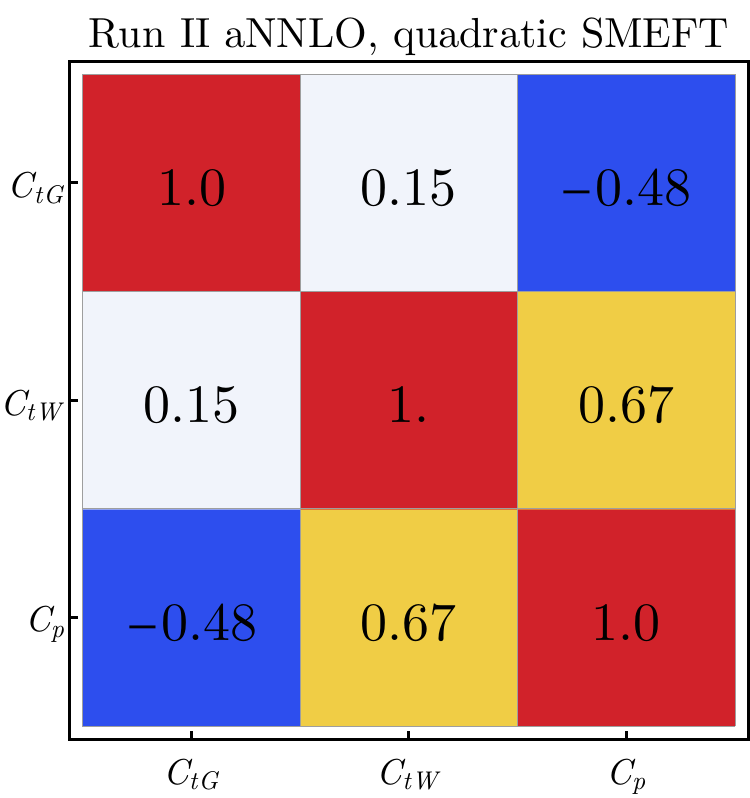}
    \caption{Correlation matrices for the marginalized linear (top) and quadratic (bottom) fits for Run II at LO (left), NLO (center), and aNNLO (right).}
    \label{fig:corr_matrices_13}
\end{figure}

 
We next show correlation matrices for the marginalized fits. Fig.~\ref{fig:corr_matrices_13} shows the Run II correlation matrices for both linear and quadratic fits at LO, NLO, and aNNLO to illustrate the discussion; the complete set including Run III and Run II+III is provided in Appendix~\ref{app:corr}.
The correlation matrices for the 3-parameter fits indicate sizable correlations among the SMEFT coefficients, consistent with the strong degradation observed upon marginalization. At the same time, the matrices remain well behaved, showing that the fits are not dominated by exact flat directions. Moving between perturbative orders can change both the magnitude and the pattern of these correlations, reflecting the modified kinematic dependence and normalization of the SMEFT contributions at each order, but it does not qualitatively alter the structure of the fit. Since we use the same binning for the Run II and Run III configurations, the overall correlation patterns are expected to be similar, with differences driven primarily by the higher collider energy. 
\par
To make the connection between the correlation structure and the underlying operator degeneracies explicit, we diagonalize the Fisher matrix for each fit. The eigenvectors define the principal directions of the likelihood ellipsoid and the eigenvalues quantify how strongly each direction is constrained. At aNNLO as our case example, in the linear fit we find a strong eigenvalue hierarchy with $\lambda_{\min} \sim \mathcal{O}(10^{-2})$, revealing an approximate flat direction well approximated by $-0.8\,C_{tG} + C_{tW} - 0.7\,C_p$, which is responsible for the large off-diagonal correlations and the severe marginalization penalty. The best-constrained direction is dominated by a correlated $(C_{tW}, C_p)$ combination, approximately $0.24\,C_{tG} + 0.89\,C_{tW} + C_p$. In the quadratic fit at aNNLO, the smallest eigenvalue rises to $\mathcal{O}(1)$, indicating that the near-flat direction is lifted by the inclusion of squared dimension-6 terms, consistent with the reduced marginalization penalty and the qualitative change in the correlation pattern.
\par
\begin{figure}
    [H]
    \centering
    \includegraphics[width=0.35\linewidth]{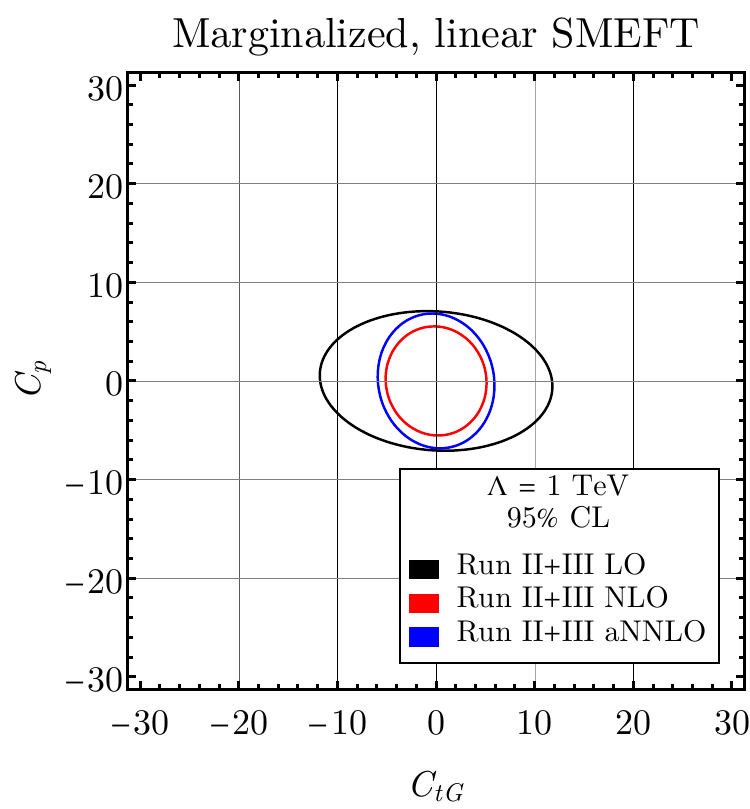}
    \includegraphics[width=0.35\linewidth]{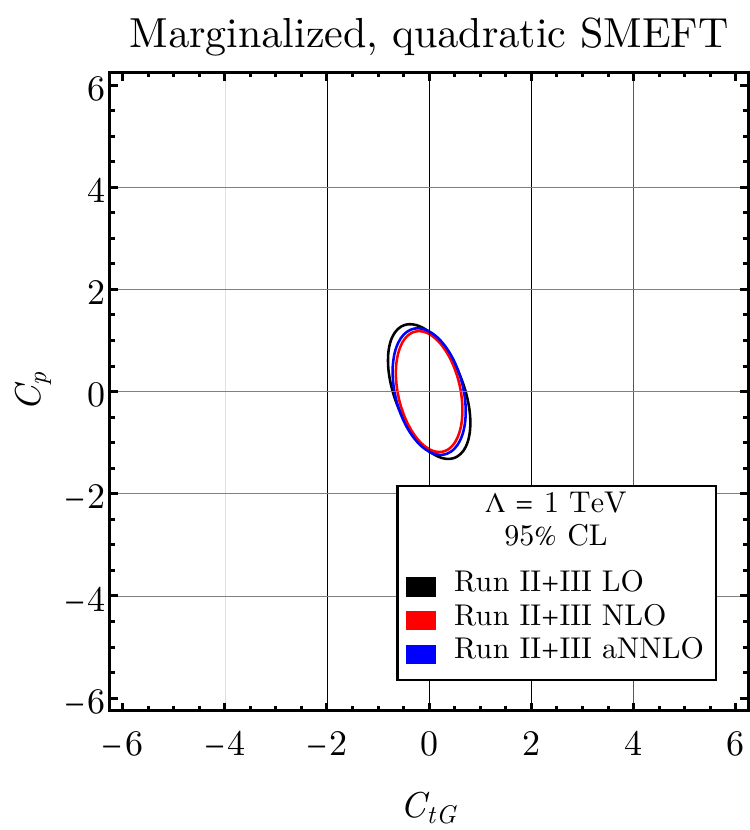}
    \caption{Marginalized confidence ellipses in the parameter subspace $(C_{tG},C_p)$ for the combined Run II+III fit, comparing the linear (left) and quadratic (right) fits at LO, NLO, and aNNLO.}
    \label{fig:ellipses_ctg_cp_linear_vs_quad}
\end{figure}

\begin{figure}
    [H]
    \centering
    \includegraphics[width=0.35\linewidth]{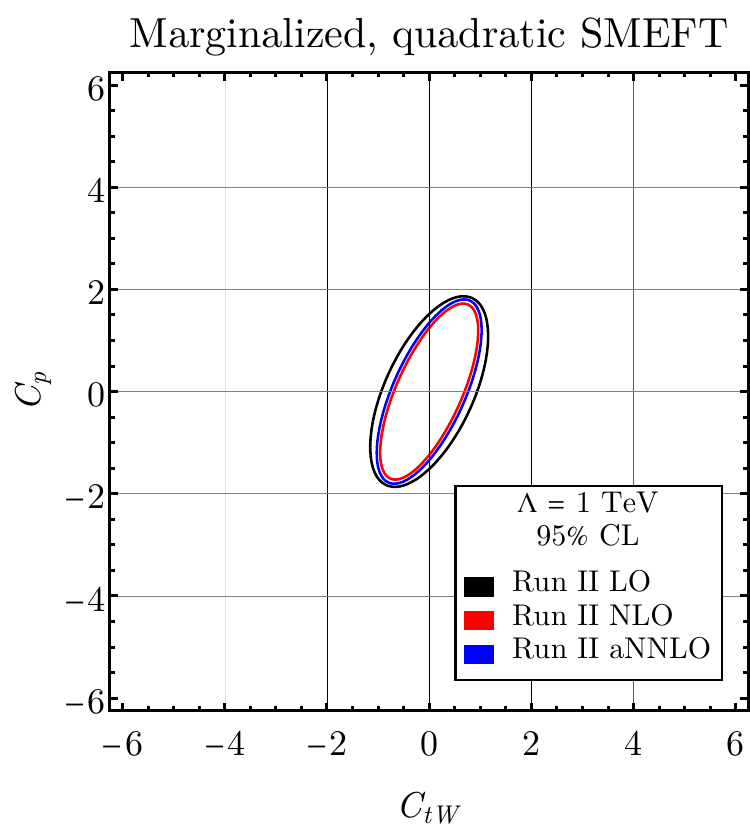}
    \includegraphics[width=0.35\linewidth]{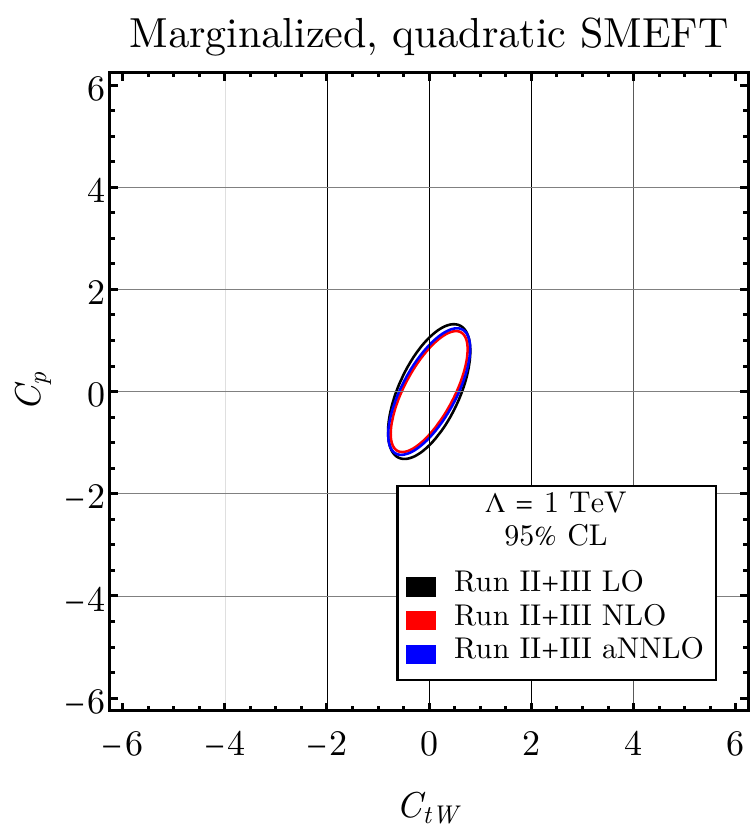}
    \caption{Marginalized confidence ellipses in the parameter subspace $(C_{tW},C_p)$ for the quadratic fit at LO, NLO, and aNNLO, comparing Run II (left) and the combined Run II+III (right) fits.}
    \label{fig:ellipses_ctw_cp_run2_vs_combined}
\end{figure}

 
We now show selected confidence ellipses. The complete set of ellipses is provided in Appendix~\ref{app:ellipses}. Fig.~\ref{fig:ellipses_ctg_ctw_nm_vs_m} compares the nonmarginalized and marginalized constraints in the $(C_{tG},C_{tW})$ parameter subspace for Run II at LO, NLO, and aNNLO. In this figure, the black curve denotes the confidence ellipse at LO, red at NLO, and blue at aNNLO. In the parameter subspace spanned by $C_{tG}$ and $C_{tW}$, the nonmarginalized contours exhibit a pronounced correlation, reflecting the similar imprint of these operators in this channel. After marginalization, the projected constraints take the form of more regular ellipses, indicating that the global fit remains well defined once all coefficients are floated simultaneously. The resulting constraints on $C_{tG}$ and $C_{tW}$ are comparable in size between the 2-parameter and 3-parameter fits, suggesting that the inclusion of the third coefficient primarily stabilizes the fit by lifting potential degeneracies rather than substantially reshaping the allowed region in this projection.
\begin{figure}
    [H]
    \centering
    \includegraphics[width=0.35\linewidth]{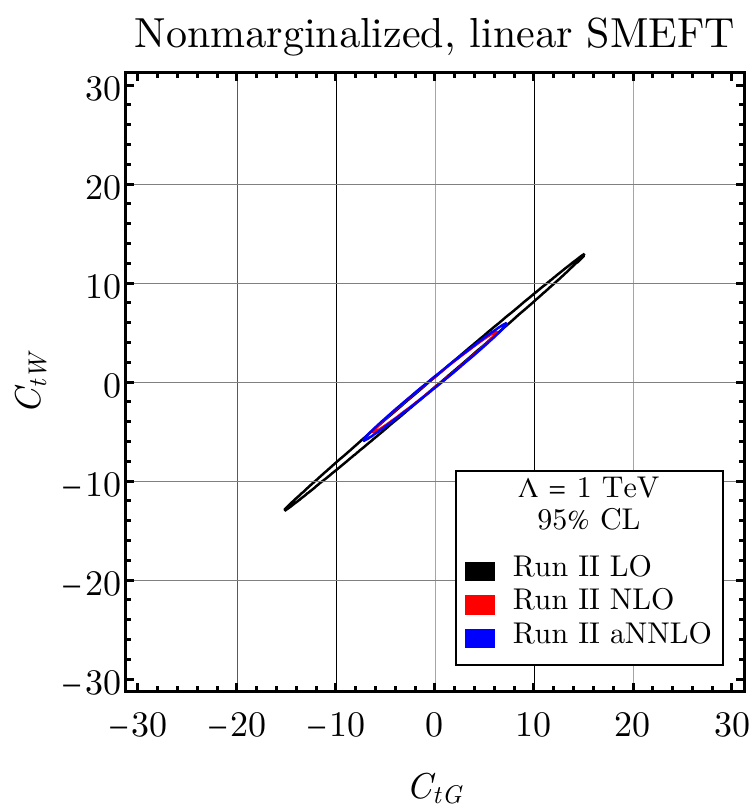}
    \includegraphics[width=0.35\linewidth]{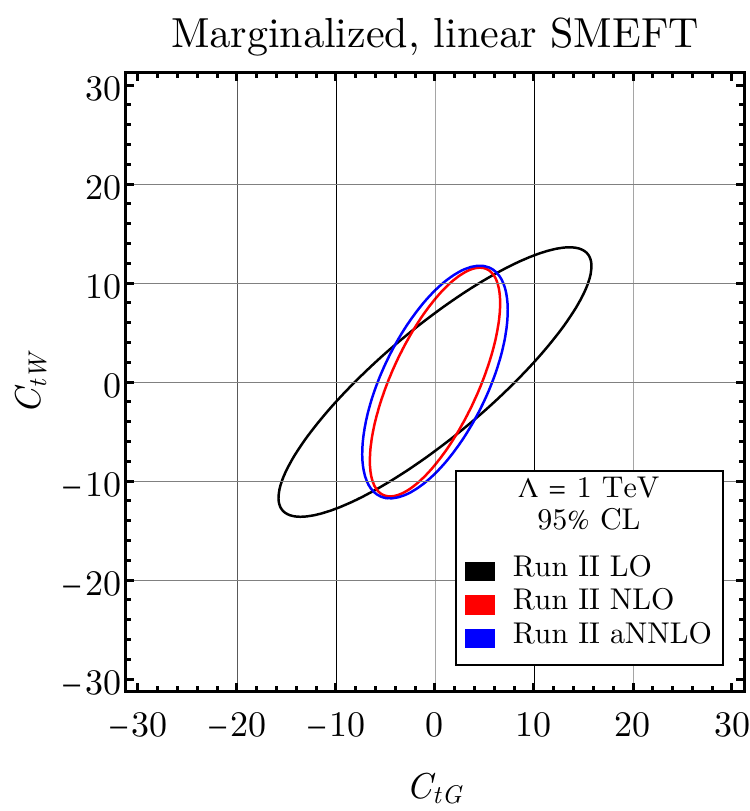}
    \caption{Nonmarginalized (left) and marginalized (right) confidence ellipses in the parameter subspace $(C_{tG},C_{tW})$ for the linear fit for Run II at LO, NLO, and aNNLO.}
    \label{fig:ellipses_ctg_ctw_nm_vs_m}
\end{figure}
Fig.~\ref{fig:ellipses_ctg_cp_linear_vs_quad} compares the marginalized linear and quadratic constraints in the parameter subspace $(C_{tG},C_p)$ for the combined Run II+III fit. The organization of the plots is identical to Fig.~\ref{fig:ellipses_ctg_ctw_nm_vs_m}. Focusing on the comparison between the linear and quadratic fits, we find that the marginalized linear confidence ellipse shows only mild correlations between $C_{tG}$ and $C_p$, but the overall extent of the allowed region remains relatively broad. Once the quadratic dimension-6 terms are included, the constraint tightens substantially, in line with the order-of-magnitude improvement observed in the marginalized bounds. This sensitivity to quadratic dimension-6 effects provides further motivation for extending the EFT description in future work by incorporating dimension-8 operators. 
Fig.~\ref{fig:ellipses_ctw_cp_run2_vs_combined} shows the effect of combining Run II and Run III in the parameter subspace $(C_{tW},C_p)$ for the quadratic fit. The plot layout follows that of Fig.~\ref{fig:ellipses_ctg_ctw_nm_vs_m}, with the same ordering in perturbative accuracy. Since we use the same binning and the same assumptions for the uncorrelated and correlated experimental uncertainties in Run II and Run III, and since the increase in the collider energy is modest, the two runs provide very similar constraint contours. The main difference between the configurations is the larger integrated luminosity in Run III, but because the fits are not statistics-limited, the combined Run II+III ellipses are only moderately tighter than the individual Run II result, at the level expected from effectively adding an additional, comparable dataset. A more pronounced gain would likely require additional differential information, such as finer binning or extended kinematic coverage in Run III, which would come at increased computational cost in the fixed order predictions and their uncertainty evaluation. 
A final point is that $C_p$ represents a specific linear combination of three Wilson coefficients, $C_{\varphi Q}^{(3)}$, $C_{\ell\ell}$, and $C_{\varphi \ell}^{(3)}$. As a result, the $tW$ analysis alone cannot disentangle these three directions, and there will remain an intrinsic degeneracy associated with resolving the individual underlying operators unless additional observables or complementary processes are included. 
\par 
\par 
We now address the interpretation of the quadratic fits in the context of EFT truncation. Since squared dimension-6 contributions are $\mathcal{O}(1/\Lambda^4)$, they are parametrically indistinguishable from linear dimension-8 interference terms, and the quadratic results therefore carry implicit assumptions about the UV completion. To assess this without introducing an explicit dimension-8 model, we perform two diagnostics.
\par 
First, we test the kinematic stability of the fits by systematically removing the highest-$p_T$ bins. We define four datasets A, B, C, and D corresponding to dropping the last $n = 0, 1, 2, 3$ $p_T$ bins (each removal removes both rapidity slices), respectively, and repeat the full analysis on each. We find that the tightening of the marginalized quadratic bounds relative to the linear ones persists across all four datasets, confirming that the improvement is not an artifact of sensitivity in the highest-$p_T$ tail. We use the aNNLO prediction at $\sqrt{s}=13~\TeV$ as the representative case for this test.
\par 
Second, we note that the nonmarginalized bounds are similar between the linear and quadratic fits across all datasets A--D, whereas the marginalized bounds differ substantially. This pattern shows that the gain from including squared dimension-6 terms is primarily a consequence of lifting approximate flat directions in the 3-parameter space rather than a large increase in intrinsic per-coefficient sensitivity, consistent with the sizable correlations seen in the linear Fisher matrices.
\par 
To quantify the regime of applicability, we assign a hardness proxy to each dataset,
\begin{gather}
    Q(p_{T,\max}) = \sqrt{m_t{}^2 + p_{T,\max}{}^2} + \sqrt{m_W{}^2 + p_{T,\max}{}^2},
\end{gather}
and define the expansion parameter $\epsilon \equiv Q / \Lambda_{\rm eff} = Q \sqrt{\Delta C_k} / \Lambda$ (with $Q$ and $\Lambda = 1~\TeV$ in the same units), where $\Lambda_{\rm eff} = \Lambda/\sqrt{\Delta C_k}$ is the effective scale defined above. For the quadratic marginalized bounds we find $\epsilon \simeq 0.34$--$0.68$, while the linear marginalized bounds yield $\epsilon \simeq 1.16$--$1.86$, indicating that the quadratic fit operates in a considerably more controlled expansion regime. The linear results remain the most model-independent, corresponding to a strict $\mathcal{O}(1/\Lambda^2)$ truncation, while the quadratic results should be understood as contingent on the assumption that dimension-8 contributions are subleading.
\par
\begin{figure}
    [H]
    \centering
    \includegraphics[width=0.5\linewidth]{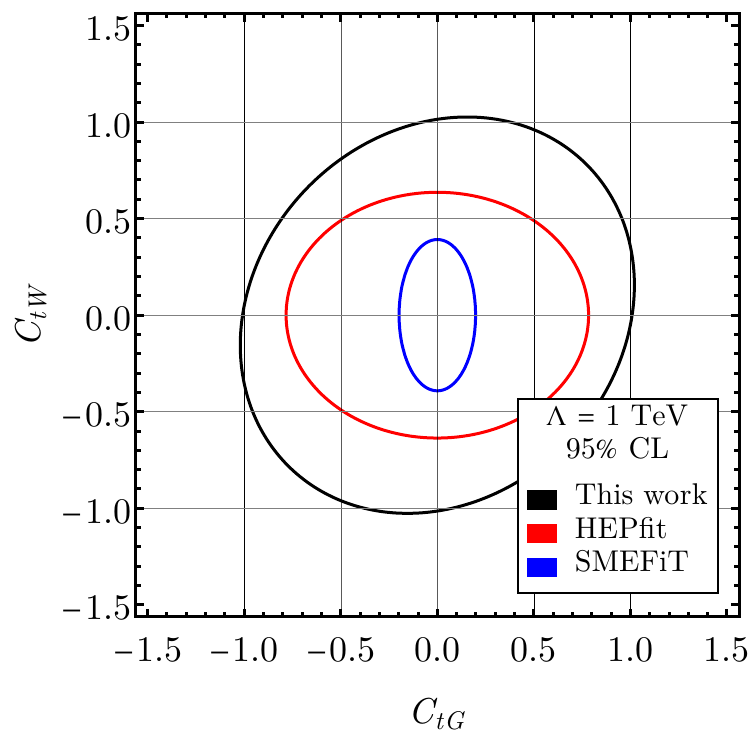}
    \caption{Comparison of the 95\% CL constraints in the parameter subspace $(C_{tG},C_{tW})$ from this work and the global analyses of HEPfit and SMEFiT, shown in the \smeftatnlo~convention for $\Lambda=1$~TeV.}
    \label{fig:lit_comparison_ellipse}
\end{figure}

 
Before we end this section, we compare our fit results with those in the literature in Table~\ref{tab:lit_comparison} and in the parameter subspace spanned by $C_{tG}$ and $C_{tW}$ in Fig.~\ref{fig:lit_comparison_ellipse}. The results summarized here are not strictly equivalent in scope. Our fit is performed in the $tW$ channel using differential information and a three-parameter SMEFT model, and is therefore sensitive to degeneracies internal to this process. By contrast, the HEPfit global analysis~\cite{Cornet-Gomez:2025jot} combines a broad set of legacy and LHC measurements (including electroweak precision observables and multiple collider processes) in a 22-parameter SMEFT fit, which helps break correlations among operators through complementary sensitivity. Similarly, the SMEFiT global fit~\cite{Celada:2024mcf} incorporates a larger operator set together with a wide collection of datasets spanning precision electroweak, Higgs, top, and diboson measurements, such that its constraints reflect the combined information content of many channels. Finally, the single-parameter constraint labeled ``$t\bar t$ cross section"~\cite{Kidonakis:2023htm} is obtained from a fit to an inclusive $t\bar t$ cross-section measurement alone~\cite{ATLAS:2023gsl, CMS:2021vhb}, effectively a single experimental input, and should be interpreted as a dedicated one-operator extraction rather than a global analysis. 
\par 
With these differences in mind, Table~\ref{tab:lit_comparison} indicates that our extracted one-dimensional limits on $C_{tG}$ and $C_{tW}$ are weaker than those obtained in recent global fits, corresponding to smaller effective scales. This trend is expected; namely, global analyses leverage multiple processes with distinct operator dependence, thereby reducing degeneracies between SMEFT corrections characterized by $C_{tG}$ and $C_{tW}$ and typically yielding small correlations in the $(C_{tG},C_{tW})$ subspace after combining complementary inputs. In contrast, a channel-restricted fit in $tW$ can exhibit stronger internal correlations once experimental and theoretical systematics are included, since different operators may induce similar deformations in the available distributions. It is also instructive to compare to the dedicated one-parameter extraction from the inclusive $t\bar t$ cross section, which yields a comparatively strong constraint on $C_{tG}$ despite using only a single observable; this reflects the high intrinsic sensitivity of inclusive $t\bar t$ production to the chromomagnetic dipole operator.
\begin{table}
    [H]
    \centering
    \caption{Comparison of representative constraints on the top dipole operators $C_{tG}$ and $C_{tW}$ from this work and selected recent analyses in the literature. We also report the corresponding effective scales. All coefficients are given in the \smeftatnlo~convention.}
    \label{tab:lit_comparison}
    \begin{tabular}{|l|c|c|c|c|c|c|c|}
        \hline
        & \shortstack{\rm SMEFT\\ \rm order} 
        & Accuracy 
        & \shortstack{\rm Number of\\ \rm parameters} 
        & $C_{tG}$ 
        & \shortstack{$\Lambda_{\rm eff}(tG)$\\ \rm [TeV]} 
        & $C_{tW}$ 
        & \shortstack{$\Lambda_{\rm eff}(tW)$\\ \rm [TeV]} \\
        \hline
        This work & Quadratic & aNNLO & 3  & 0.60  & 1.3 & 0.60 & 1.3 \\
        \hline
        Global fit with HEPfit~\cite{Cornet-Gomez:2025jot} & Linear & NLO & 22 & 0.32 & 1.8 & 0.26 & 2.0 \\
        \hline
        Global fit with SMEFiT~\cite{Celada:2024mcf} & Quadratic & NLO & 50 & 0.081 & 3.5 & 0.16 & 2.5 \\
        \hline
        $t\bar t$ cross section~\cite{Kidonakis:2023htm} & Quadratic & aNNLO & 1 & 0.25 & 2.0 & -- & -- \\
        \hline
    \end{tabular}
\end{table}
The corresponding two-dimensional comparison is shown in Fig.~\ref{fig:lit_comparison_ellipse}. We display the 95\% CL contours in the parameter subspace spanned by $C_{tG}$ and $C_{tW}$ at $\Lambda=1\ \TeV$. The entries ``This work'', ``HEPfit'', and ``SMEFiT'' are the same as in Table~\ref{tab:lit_comparison}. The HEPfit and SMEFiT contours are reported with negligible $C_{tG}$-$C_{tW}$ correlation in their summaries; in particular, HEPfit (SMEFiT) sets $\rho=0$ whenever $|\rho|<0.05$ ($|\rho|<0.2$), and we adopt the same convention in the figure. Although our confidence ellipse is broader and exhibits a sizable correlation, all three ellipses lie within a compact region of the parameter plane, indicating that our $tW$-based determination is broadly comparable in scale to existing global constraints while providing a complementary, process-specific result at aNNLO accuracy.
To quantify the incremental impact of the $tW$ channel more concretely, we perform an explicit Fisher-matrix combination in the $(C_{tG}, C_{tW})$ subspace. We reconstruct the $2\times 2$ covariance matrices for the published global constraints from their reported marginalized 95\% CL bounds and correlation coefficients, invert to obtain the corresponding Fisher matrices $\mathcal{F}_{\rm glob}$, and form the combined constraint by adding information, $\mathcal{F}_{\rm comb} = \mathcal{F}_{\rm glob} + \mathcal{F}_{tW}$, where $\mathcal{F}_{tW}$ is our $tW$ Fisher matrix projected onto the same subspace.
\par 
Using HEPfit as a representative global analysis, we find that adding differential $tW$ information tightens the marginalized 95\% CL bounds by about 14\% for $C_{tG}$ and 9.6\% for $C_{tW}$, and reduces the confidence ellipse area by about 25\%. The combined correlation coefficient shifts from $\rho = 0$ in the global-only fit and $\rho = 0.29$ in our $tW$-only fit to $\rho \simeq 0.058$ in the combination, indicating that the correlated directions of the inference are only mildly affected. Using SMEFiT as a representative state-of-the-art global constraint, the incremental impact is correspondingly smaller, with marginalized bounds improving by about 1.0\% for $C_{tG}$ and 3.8\% for $C_{tW}$, a 2D ellipse area reduction of about 4.8\%, and a negligible combined correlation $\rho \simeq 0.011$. These results confirm that differential $tW$ information is broadly consistent with existing global constraints and provides complementary sensitivity whose practical impact depends on the baseline global dataset.

\section{Conclusion\label{sec:conclusion}}
In this work, we studied $pp\to tW^-$ production at the LHC within the dimension-6 SMEFT and presented predictions at LO, NLO QCD, and aNNLO QCD accuracy, with the NLO calculation defined using a diagram-removal prescription to avoid overlap with resonant $t\bar t$ contributions, with a focus on establishing perturbative-QCD control and quantifying how higher-order corrections affect EFT sensitivity and the stability of the extracted bounds. We analyzed double differential top quark distributions in transverse momentum and rapidity and provided a complete account of the uncertainty budget entering our fits. 
\par 
We performed linear and quadratic SMEFT fits to $C_{tG}$, $C_{tW}$, and $C_p$ for Run II and Run III configurations. At linear order in SMEFT parameters, the nonmarginalized 95\% CL bounds are $\mathcal O(0.1)$, while the marginalized bounds weaken to $\mathcal O(10)$, reflecting strong correlations in the 3-parameter fit; Run II and Run III constraints are similar, so the combined Run II+III fit improves only modestly. Including quadratic dimension-6 terms leaves the nonmarginalized bounds at a similar level but strengthens the marginalized bounds by about an order of magnitude relative to the marginalized linear fit. This behavior is also visible in the effective scale reach. The nonmarginalized fits probe scales up to about 2 TeV, while the marginalized reach is roughly 0.5 TeV in the linear fit and around 1.5 TeV in the quadratic fit. These improvements motivate a more complete EFT treatment, since the quadratic dimension-6 effects are parametrically comparable to linear dimension-8 contributions, and a consistent interpretation would benefit from including dimension-8 operators. In addition, $C_p$ is a fixed linear combination of $C_{\varphi Q}^{(3)}$, $C_{\ell\ell}$, and $C_{\varphi \ell}^{(3)}$, so resolving the underlying operator directions will require additional observables or complementary processes beyond $tW$ alone.
\par 
Finally, we compared our constraints on $C_{tG}$ and $C_{tW}$ to recent results in the literature, including global SMEFT fits and a dedicated one-parameter extraction from the inclusive $t\bar t$ cross section. While global analyses typically yield stronger one-dimensional limits by combining complementary information from many channels, our $tW$-based determination is broadly comparable in scale in the parameter subspace $(C_{tG},C_{tW})$ and provides a complementary, process-specific constraint at aNNLO accuracy. A Fisher-matrix combination in this subspace shows that adding differential $tW$ information yields a non-negligible improvement relative to more conservative global analyses, tightening marginalized bounds by up to 14\% and reducing the confidence ellipse area by up to 25\%, while the impact relative to the most comprehensive current global fits is modest.

\textbf{Acknowledgments:} This material is based upon work supported by the National Science Foundation under Grant No. PHY 2412071. K\c{S} is supported by the Kennesaw State University Office of Research Postdoctoral Fellowship Program. This work was supported in part by research computing resources and technical expertise via a partnership between Kennesaw State University's Office of the Vice President for Research and the Office of the CIO and Vice President for Information Technology.  

\appendix 
\section{Complete set of confidence ellipses\label{app:ellipses}}
\begin{figure}
    [H]
    \centering
    \includegraphics[width=0.24\linewidth]{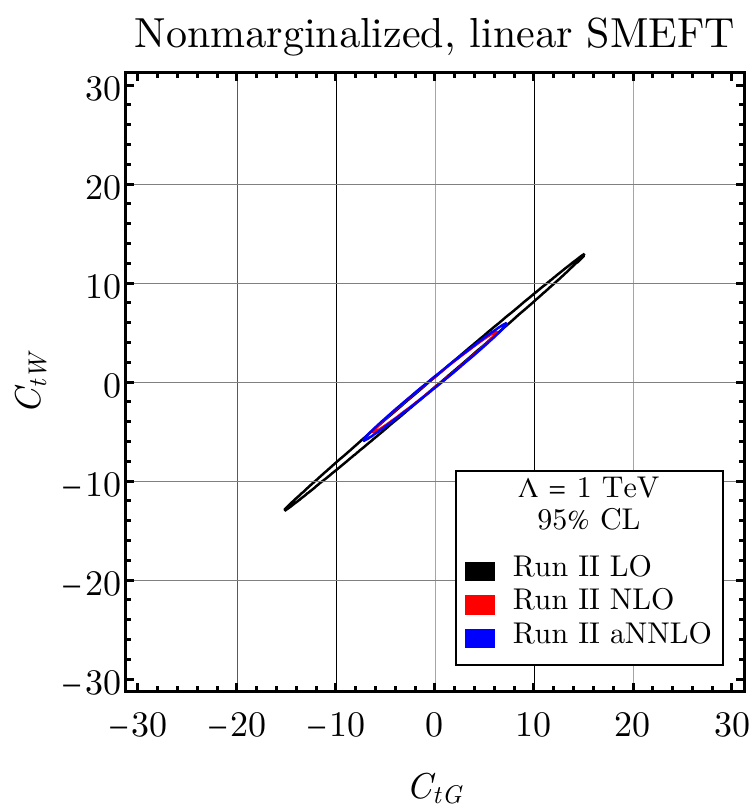}
    \includegraphics[width=0.24\linewidth]{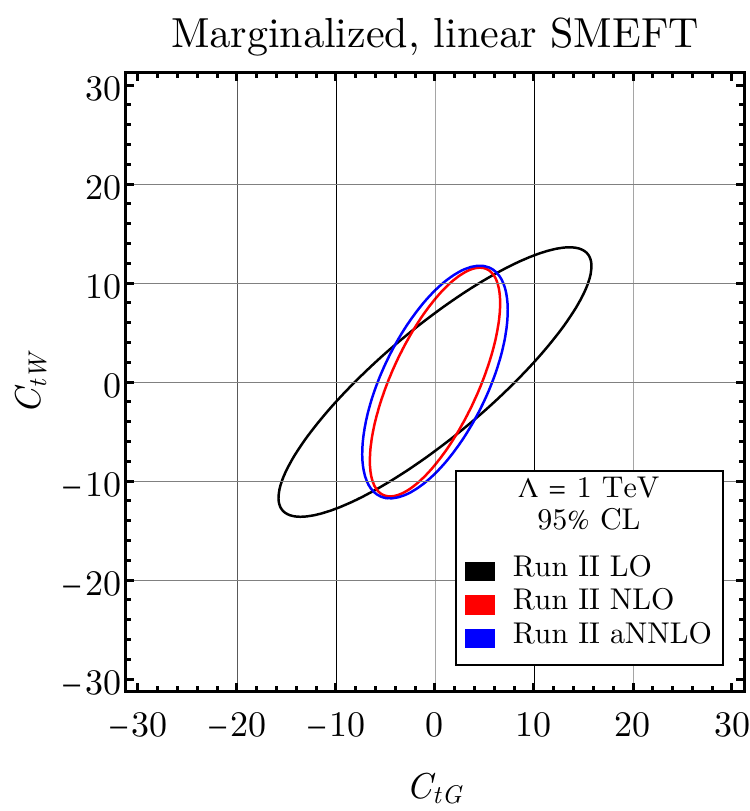}
    \includegraphics[width=0.24\linewidth]{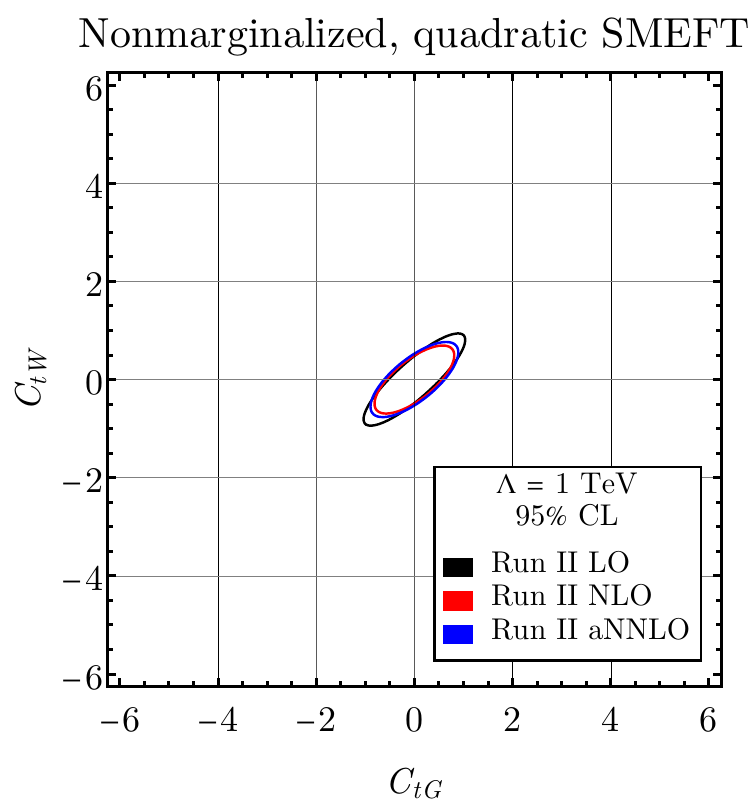}
    \includegraphics[width=0.24\linewidth]{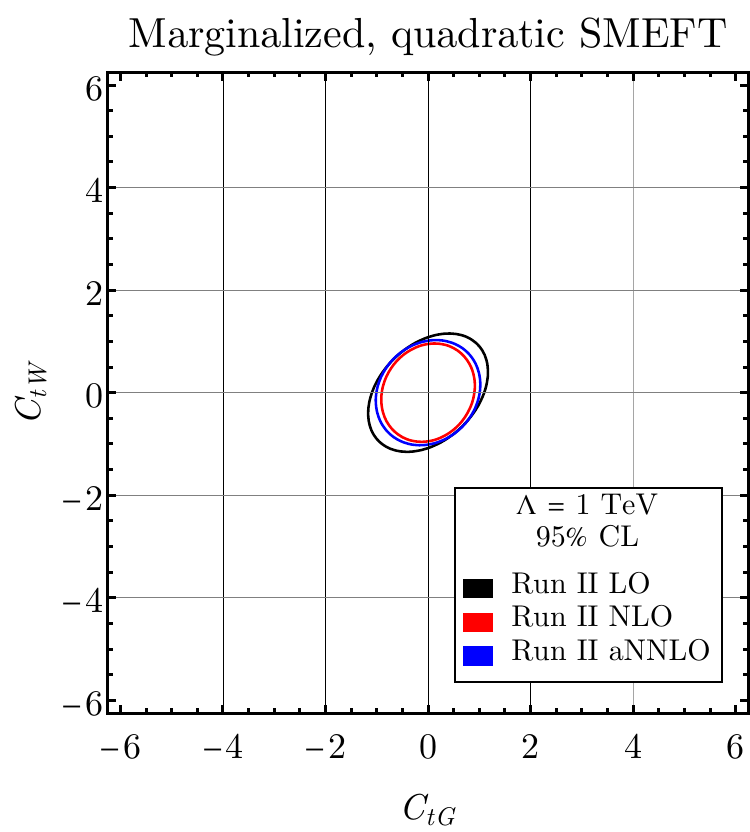}
    \caption{Confidence ellipses in the parameter subspace $(C_{tG},C_{tW})$ for Run II comparing the nonmarginalized linear fit (first), marginalized linear fit (second), nonmarginalized quadratic fit (third), and marginalized quadratic fit (fourth), each shown at LO, NLO, and aNNLO.}
    \label{fig:ellipses_ctg_ctw_13_nm_m_lin_quad}
\end{figure}

\begin{figure}
    [H]
    \centering
    \includegraphics[width=0.24\linewidth]{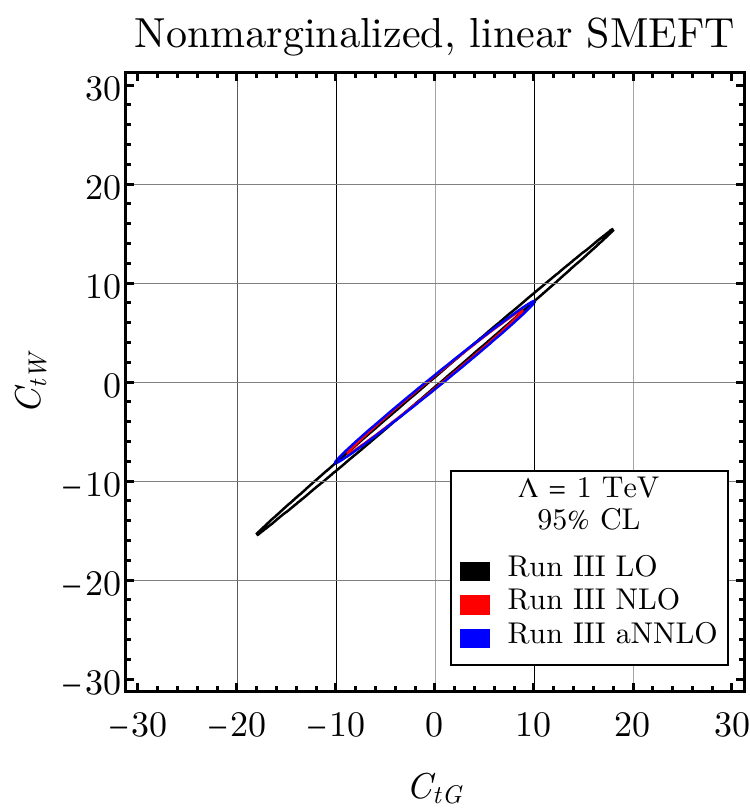}
    \includegraphics[width=0.24\linewidth]{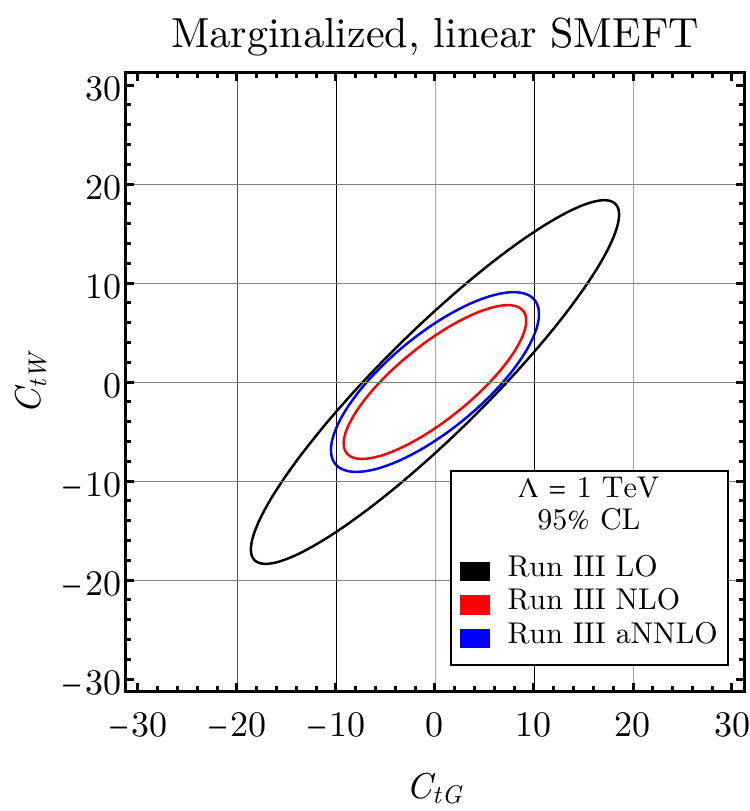}
    \includegraphics[width=0.24\linewidth]{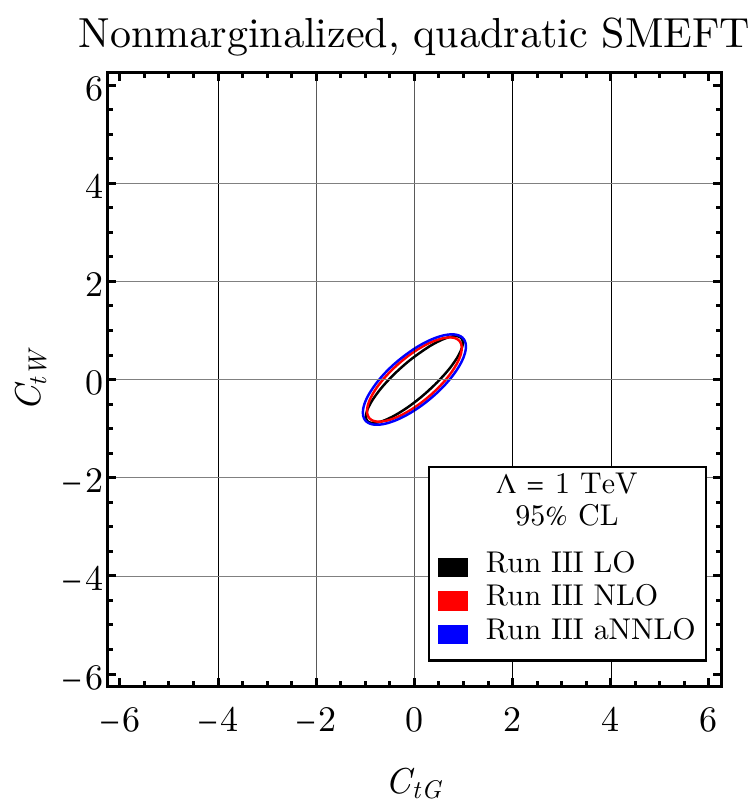}
    \includegraphics[width=0.24\linewidth]{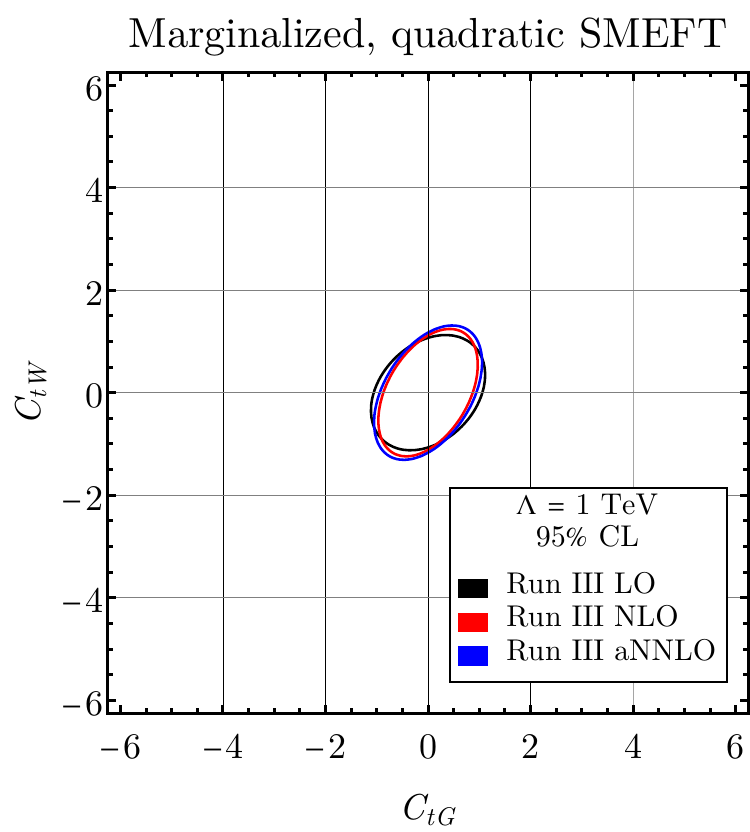}
    \caption{The same as Fig.~\ref{fig:ellipses_ctg_ctw_13_nm_m_lin_quad} but for Run III.}
    \label{fig:ellipses_ctg_ctw_136_nm_m_lin_quad}
\end{figure}

\begin{figure}
    [H]
    \centering
    \includegraphics[width=0.24\linewidth]{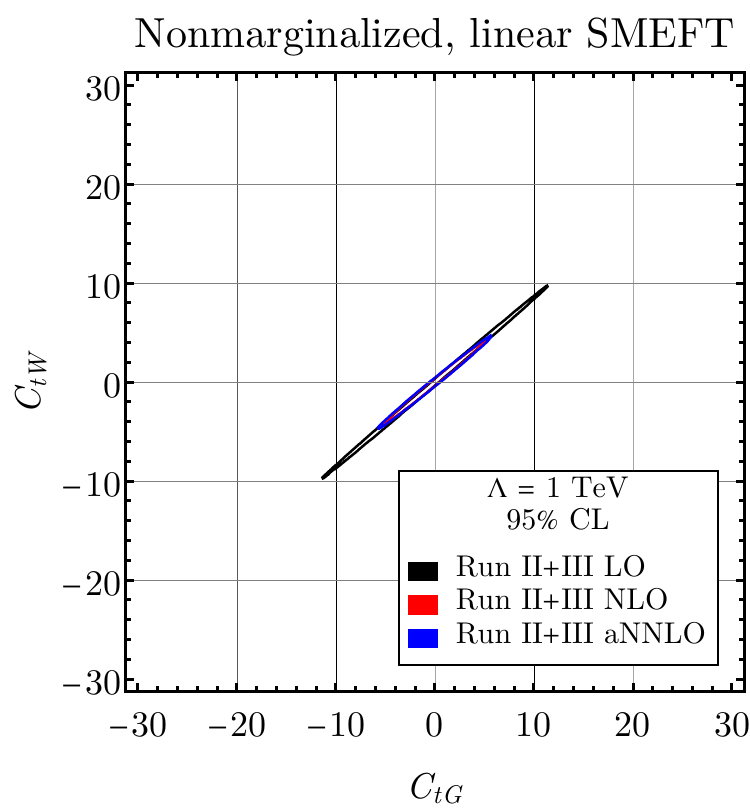}
    \includegraphics[width=0.24\linewidth]{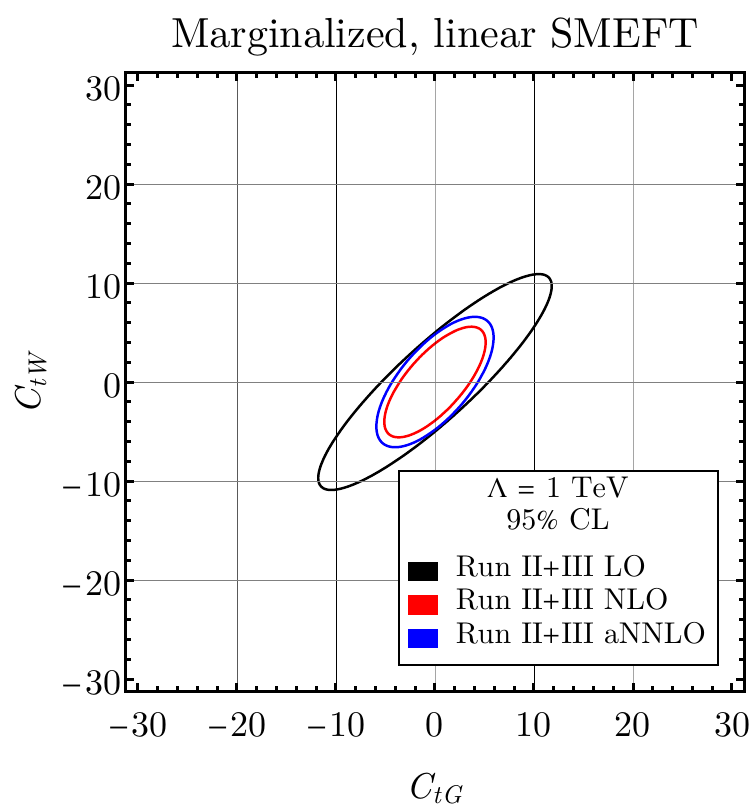}
    \includegraphics[width=0.24\linewidth]{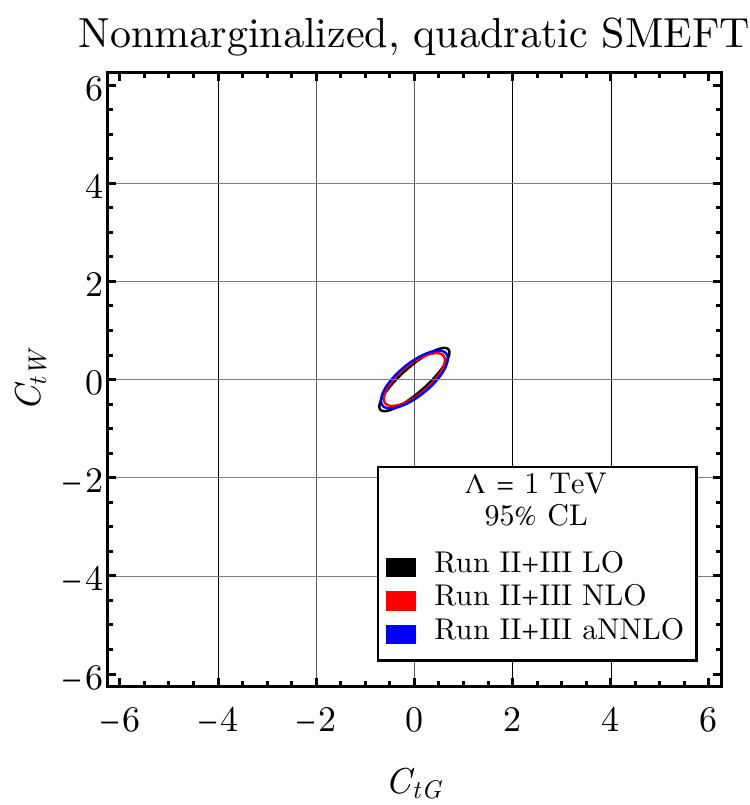}
    \includegraphics[width=0.24\linewidth]{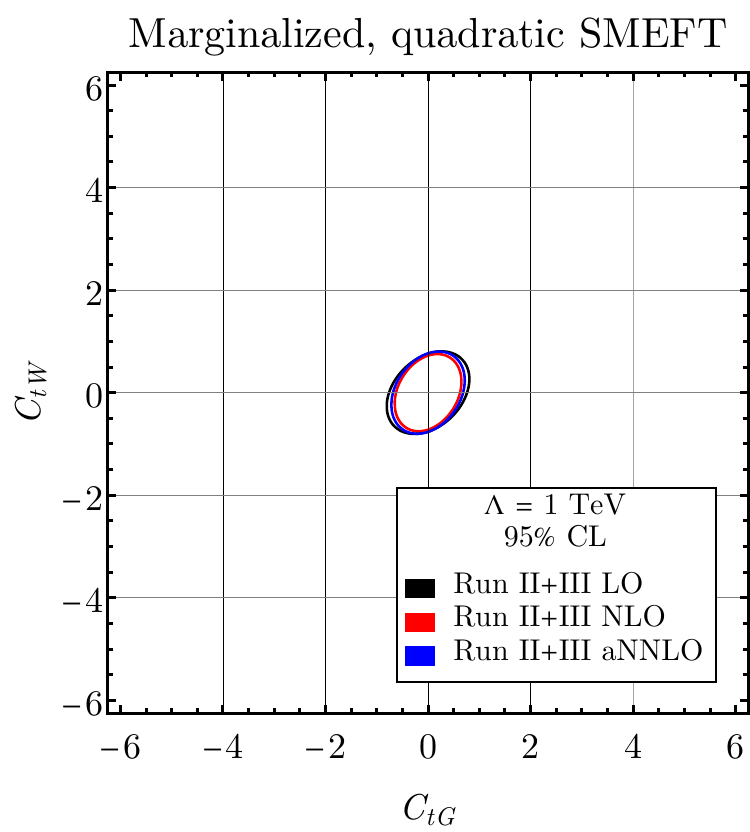}
    \caption{The same as Fig.~\ref{fig:ellipses_ctg_ctw_13_nm_m_lin_quad} but for Run II+III.}
    \label{fig:ellipses_ctg_ctw_combined_nm_m_lin_quad}
\end{figure}

\begin{figure}
    [H]
    \centering
    \includegraphics[width=0.24\linewidth]{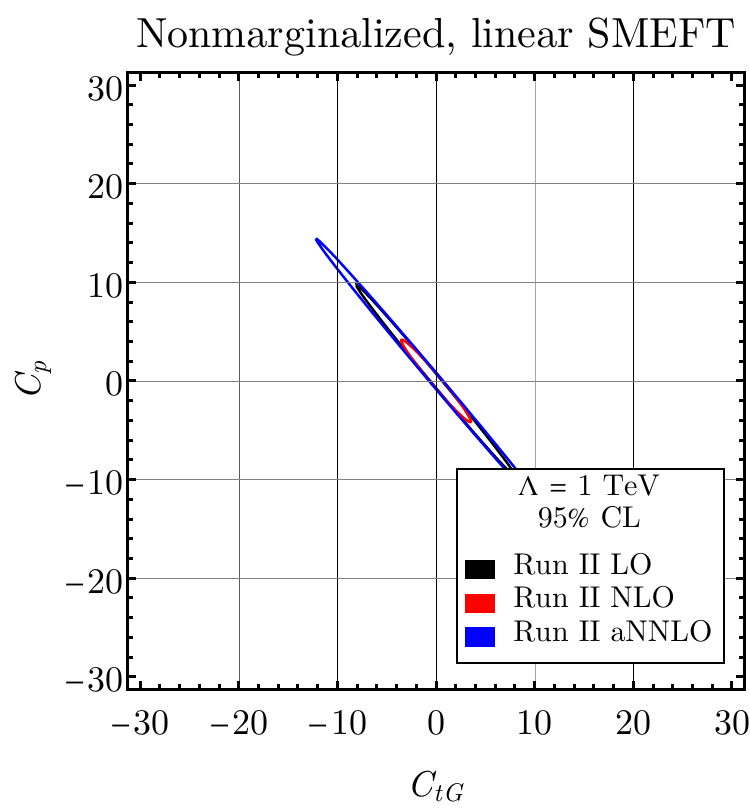}
    \includegraphics[width=0.24\linewidth]{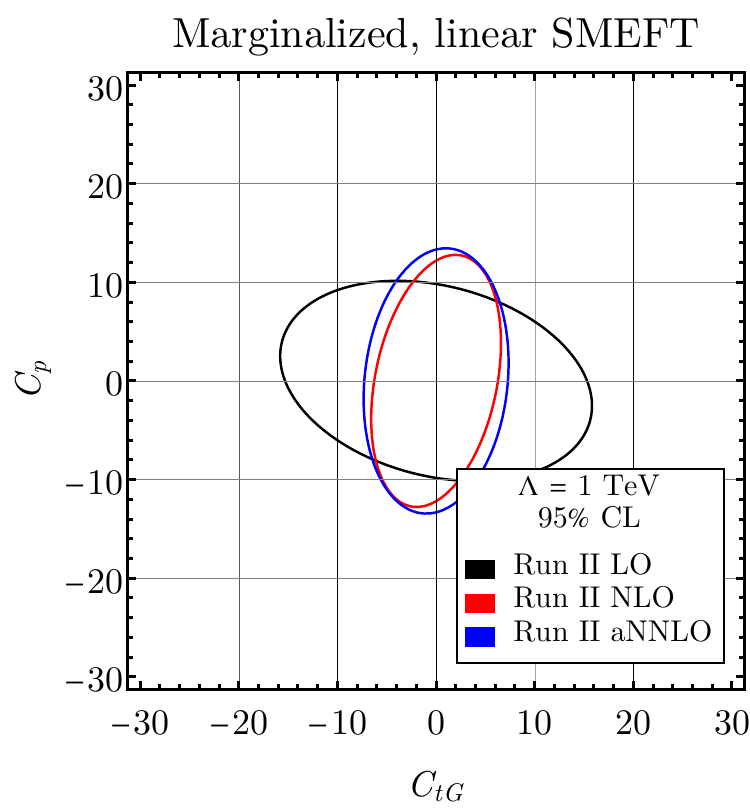}
    \includegraphics[width=0.24\linewidth]{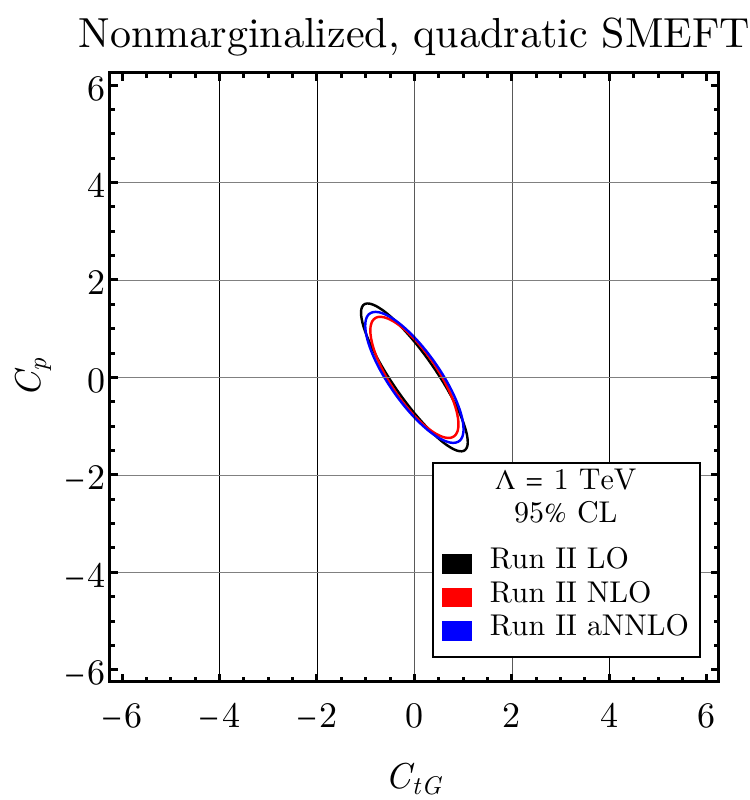}
    \includegraphics[width=0.24\linewidth]{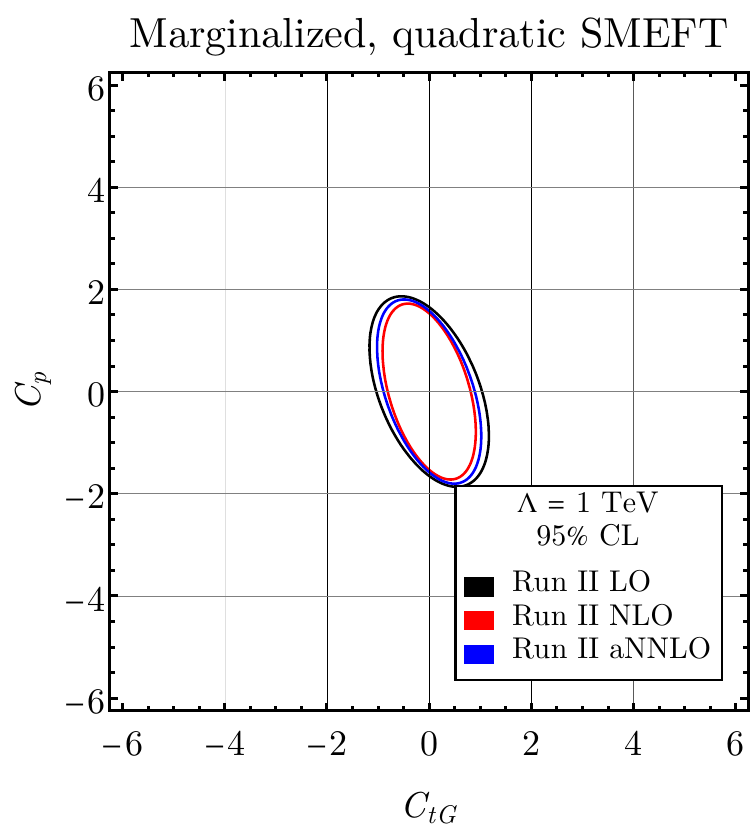}
    \caption{The same as Fig.~\ref{fig:ellipses_ctg_ctw_13_nm_m_lin_quad} but for $(C_{tG}, C_p)$.}
    \label{fig:ellipses_ctg_cp_13_nm_m_lin_quad}
\end{figure}

\begin{figure}
    [H]
    \centering
    \includegraphics[width=0.24\linewidth]{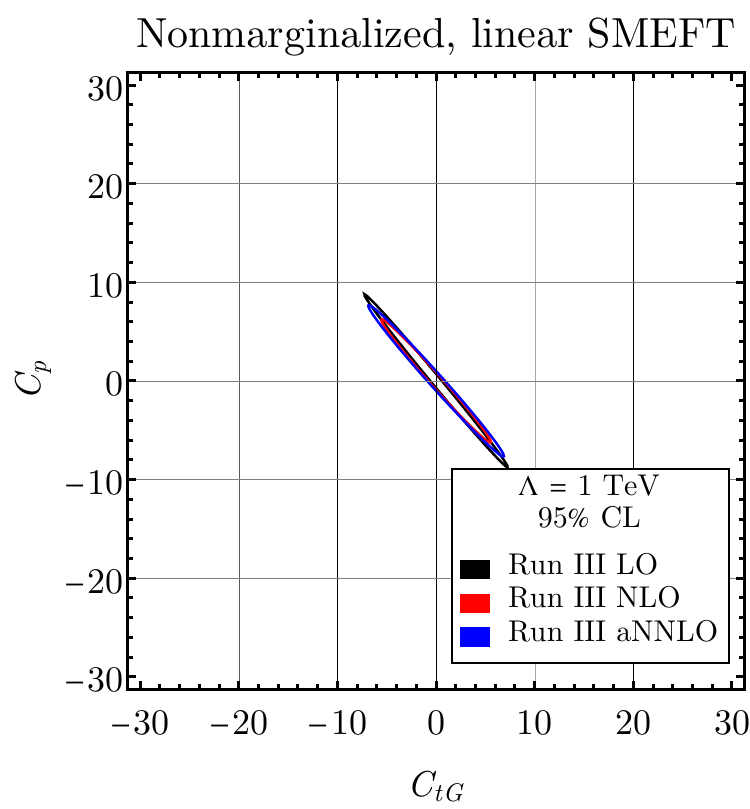}
    \includegraphics[width=0.24\linewidth]{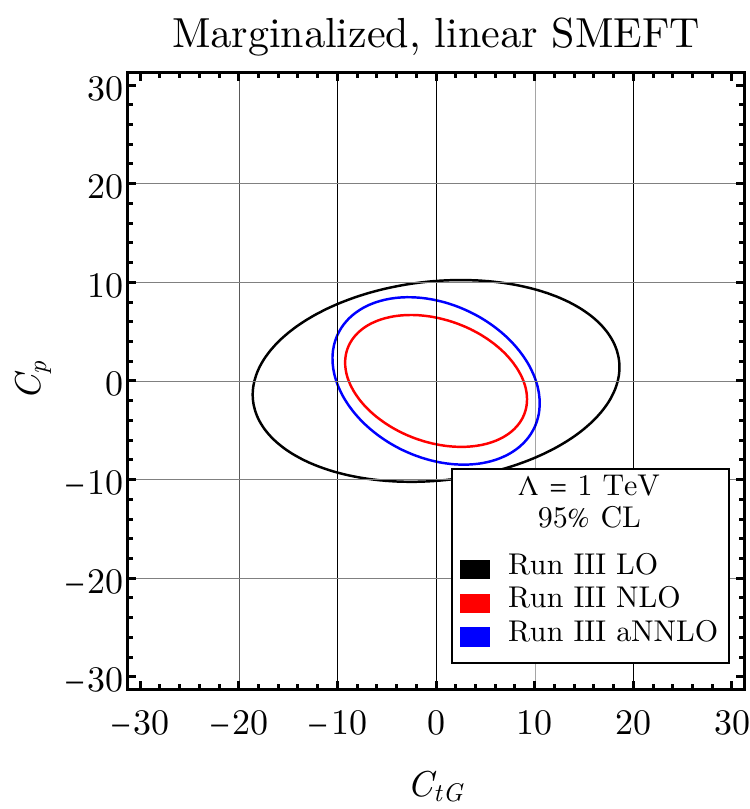}
    \includegraphics[width=0.24\linewidth]{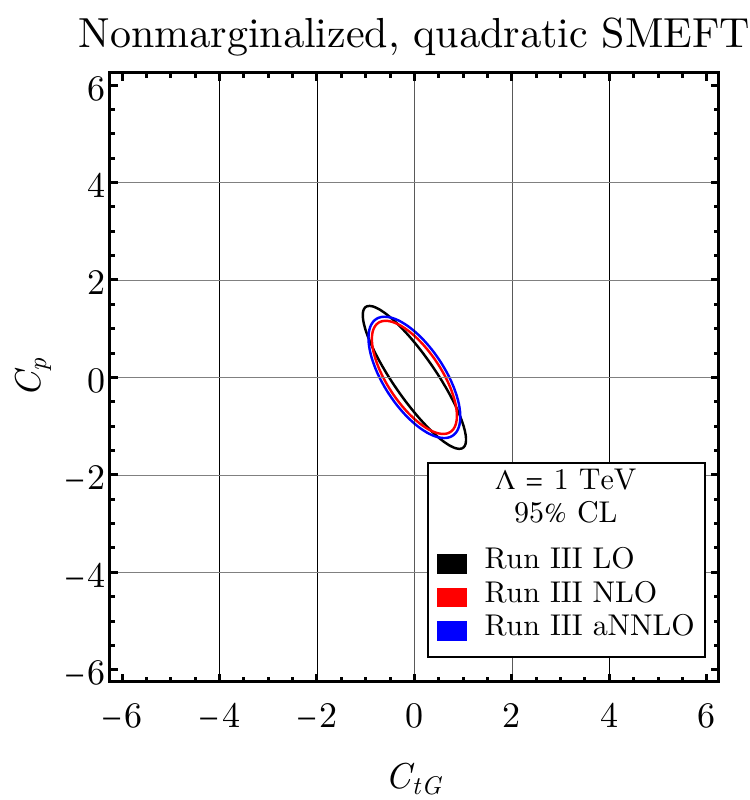}
    \includegraphics[width=0.24\linewidth]{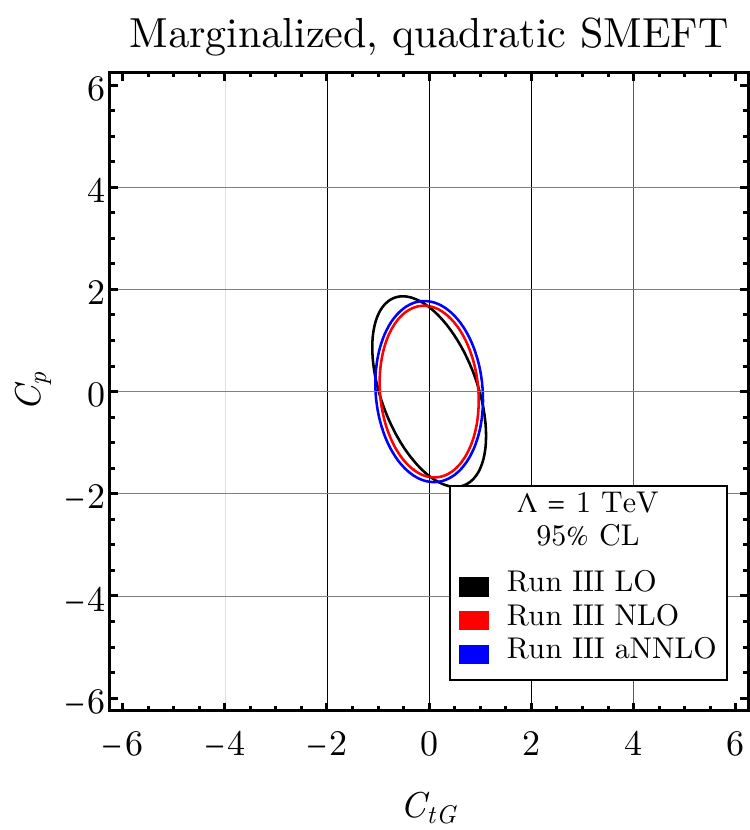}
    \caption{The same as Fig.~\ref{fig:ellipses_ctg_ctw_136_nm_m_lin_quad} but for $(C_{tG}, C_p)$.}
    \label{fig:ellipses_ctg_cp_136_nm_m_lin_quad}
\end{figure}

\begin{figure}
    [H]
    \centering
    \includegraphics[width=0.24\linewidth]{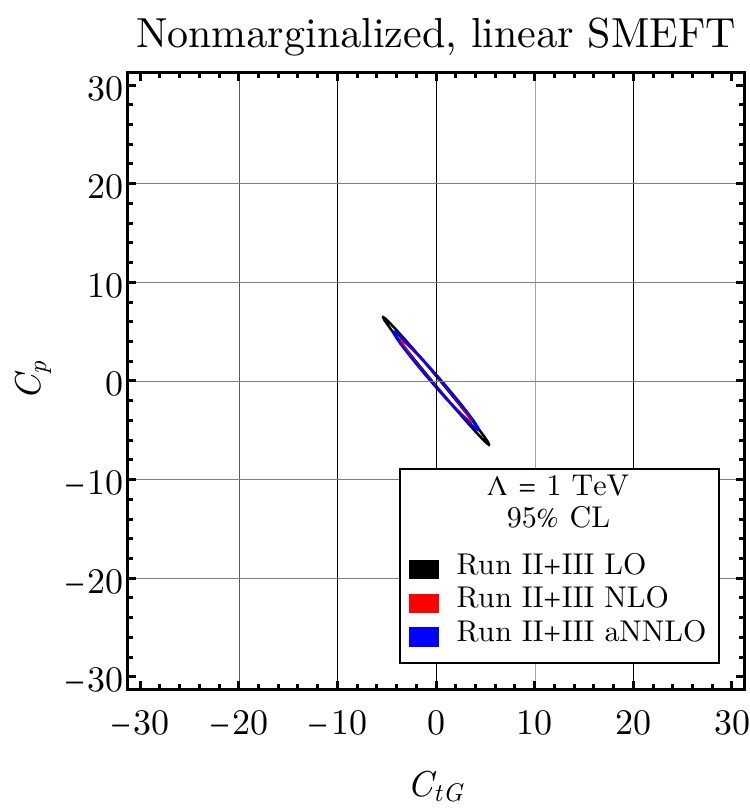}
    \includegraphics[width=0.24\linewidth]{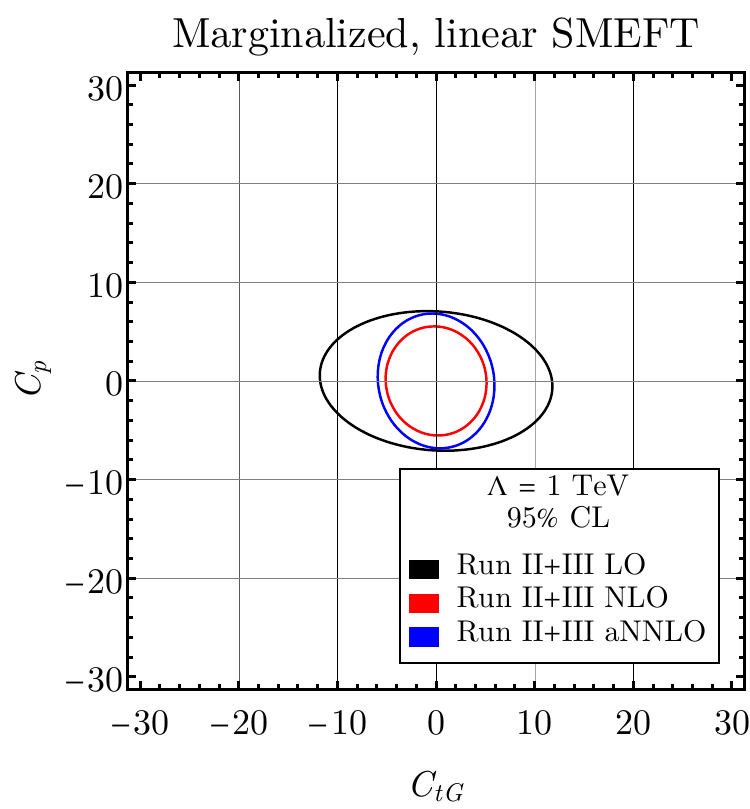}
    \includegraphics[width=0.24\linewidth]{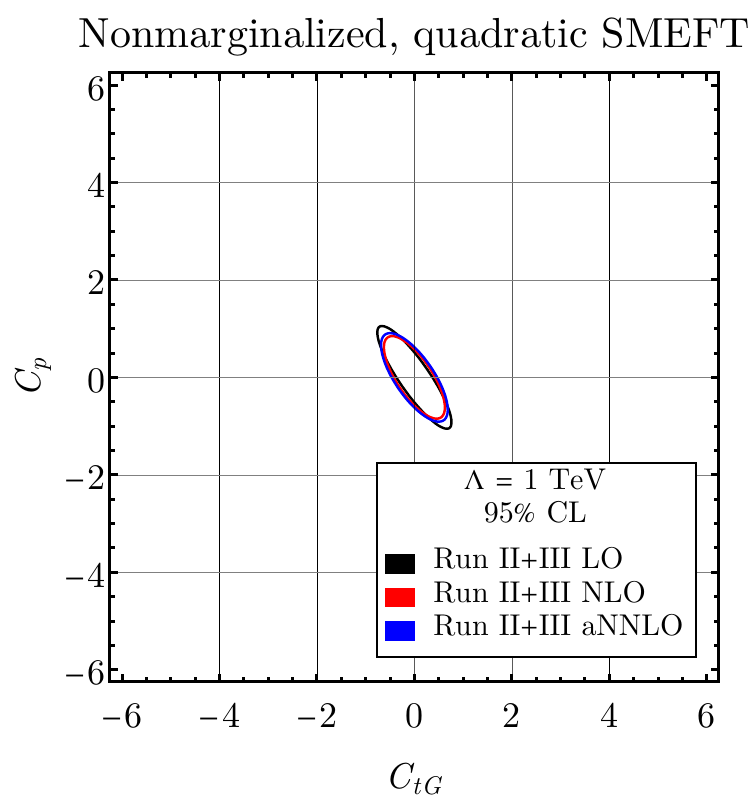}
    \includegraphics[width=0.24\linewidth]{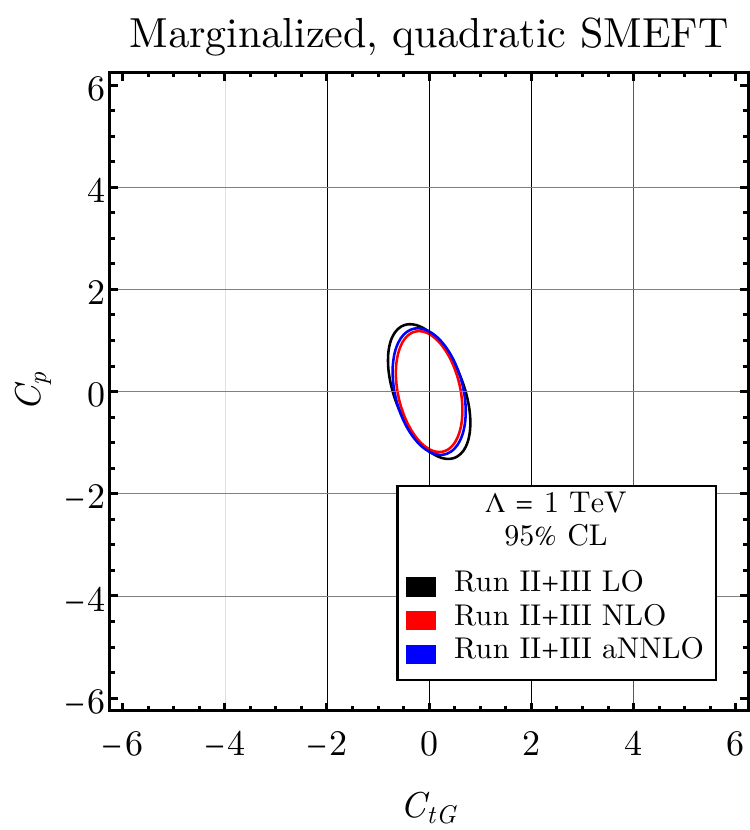}
    \caption{The same as Fig.~\ref{fig:ellipses_ctg_ctw_combined_nm_m_lin_quad} but for $(C_{tG}, C_p)$.}
    \label{fig:ellipses_ctg_cp_combined_nm_m_lin_quad}
\end{figure}

\begin{figure}
    [H]
    \centering
    \includegraphics[width=0.24\linewidth]{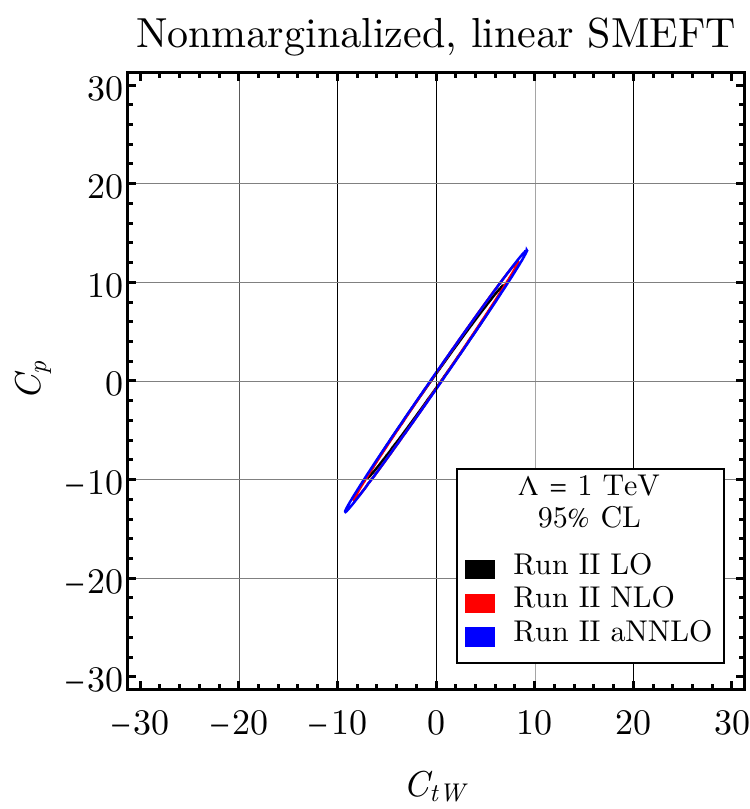}
    \includegraphics[width=0.24\linewidth]{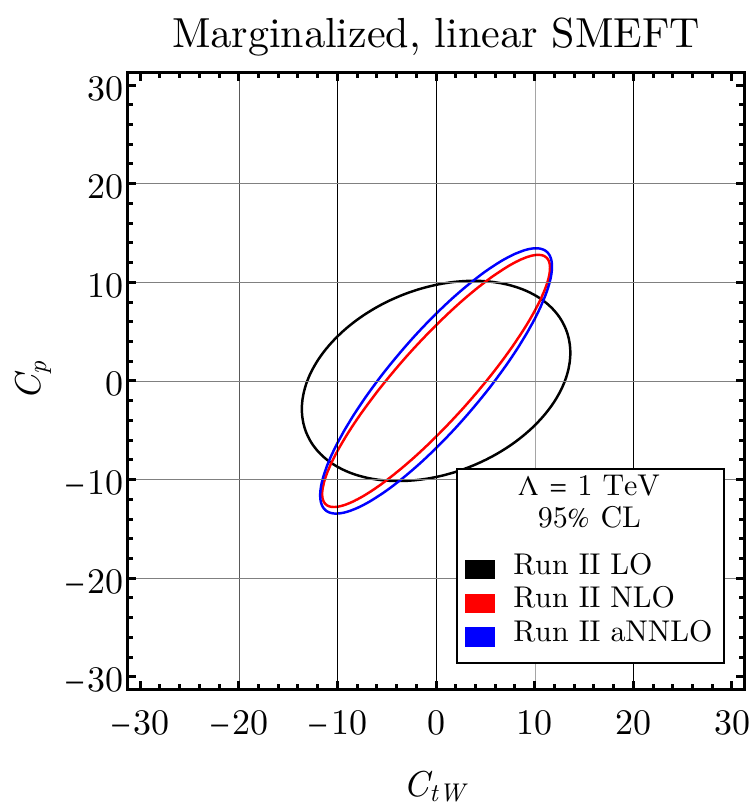}
    \includegraphics[width=0.24\linewidth]{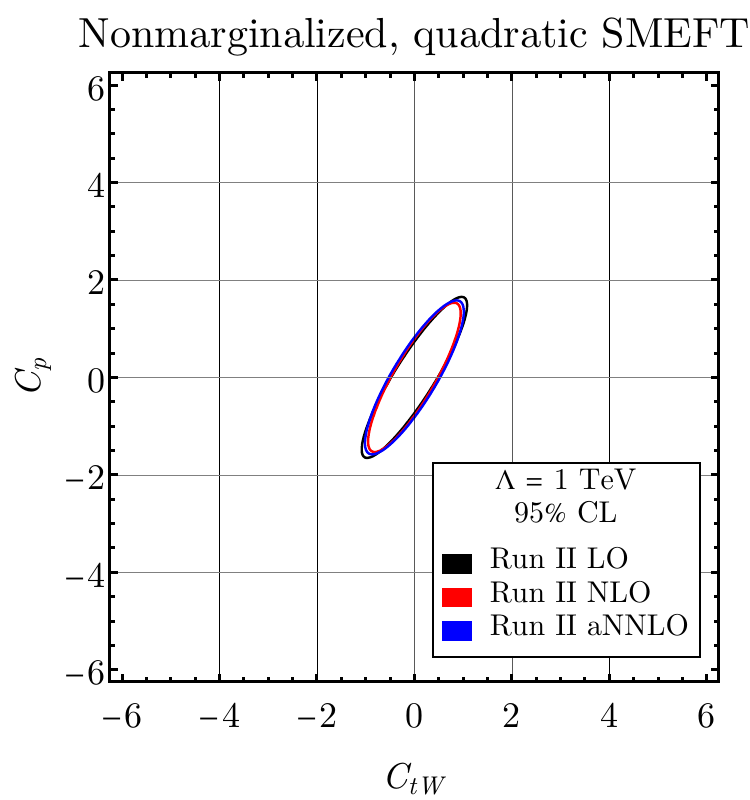}
    \includegraphics[width=0.24\linewidth]{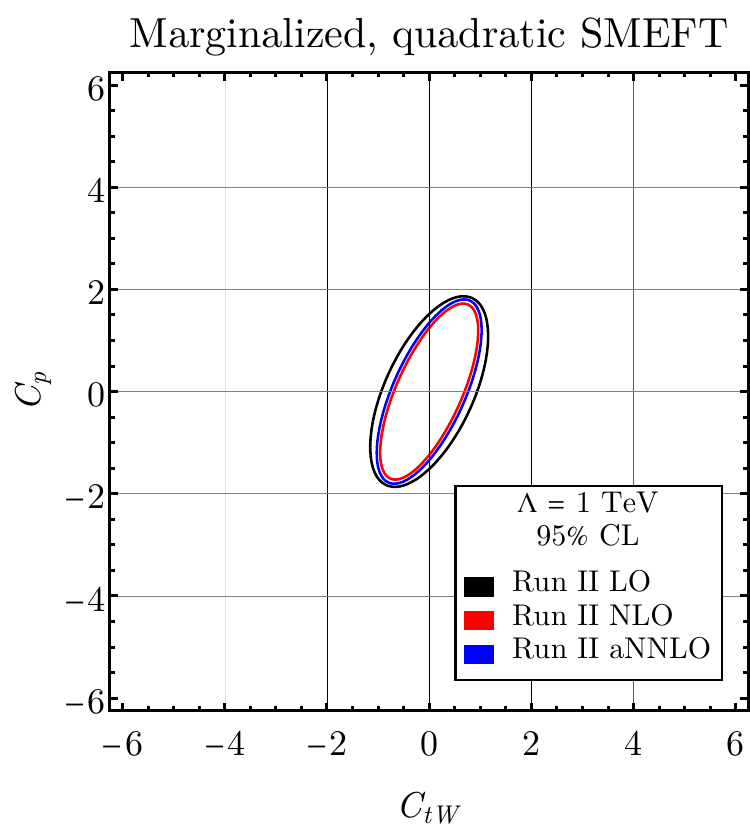}
    \caption{The same as Fig.~\ref{fig:ellipses_ctg_ctw_13_nm_m_lin_quad} but for $(C_{tW}, C_p)$.}
    \label{fig:ellipses_ctw_cp_13_nm_m_lin_quad}
\end{figure}

\begin{figure}
    [H]
    \centering
    \includegraphics[width=0.24\linewidth]{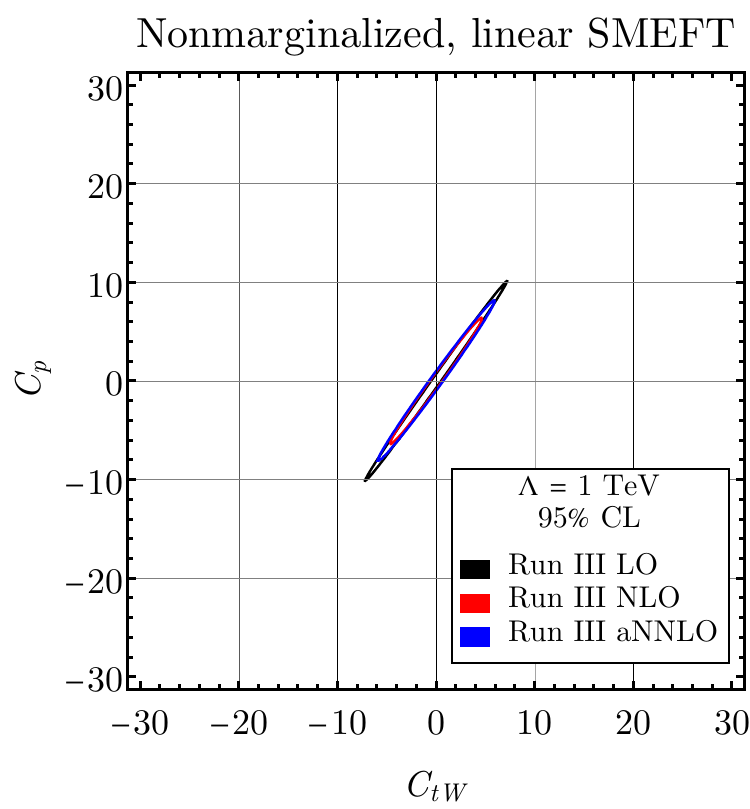}
    \includegraphics[width=0.24\linewidth]{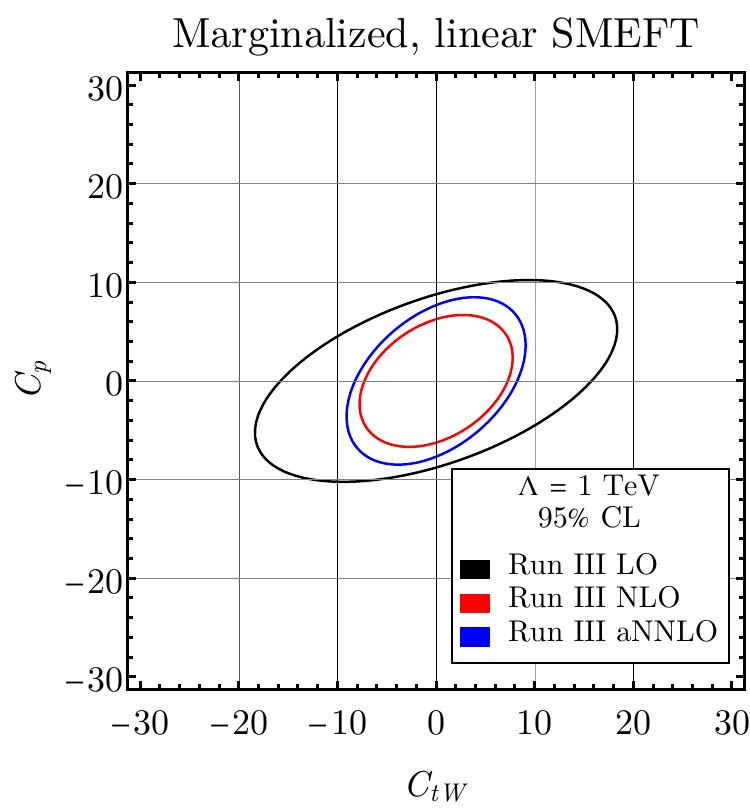}
    \includegraphics[width=0.24\linewidth]{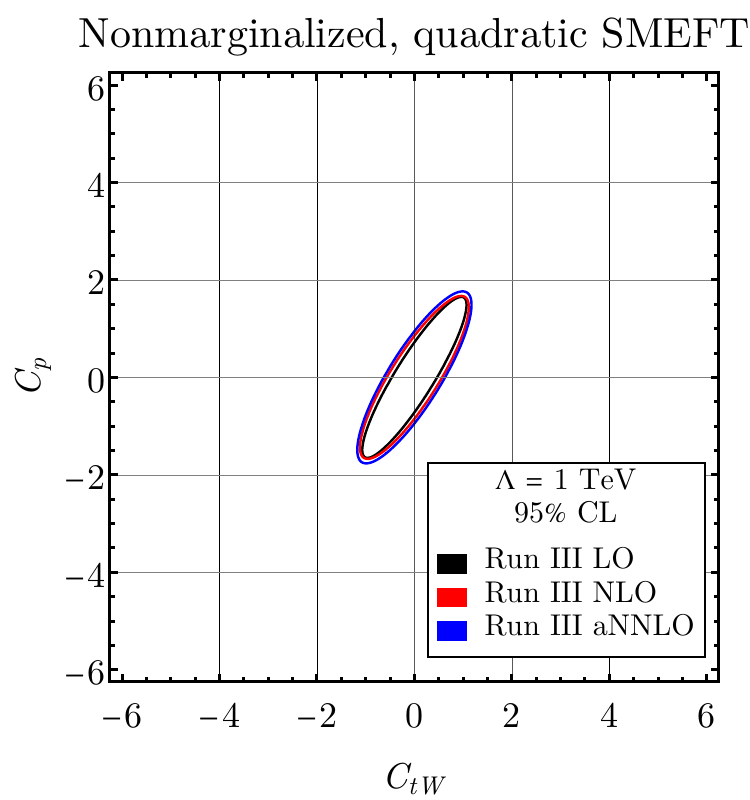}
    \includegraphics[width=0.24\linewidth]{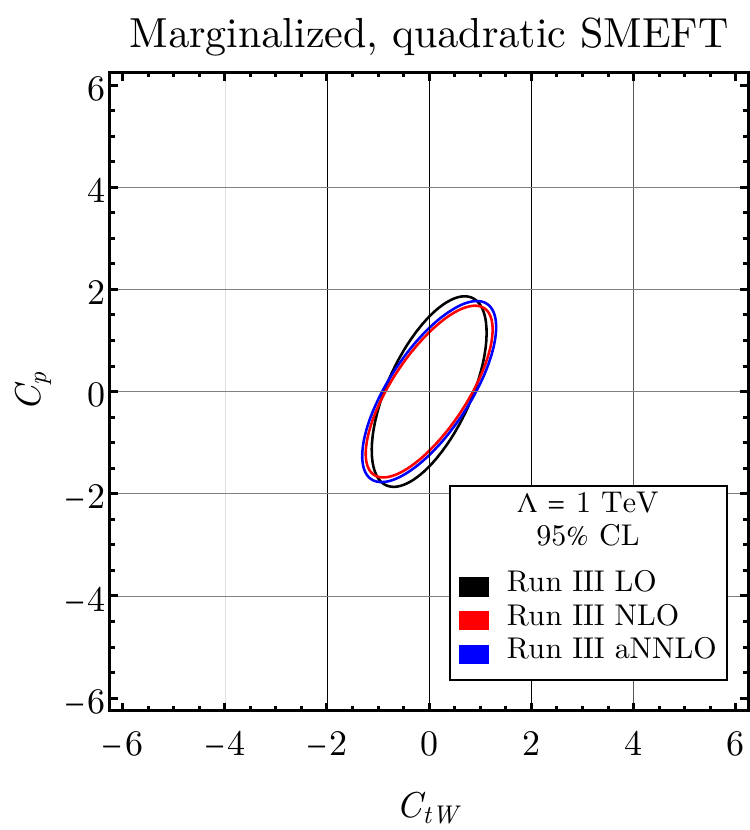}
    \caption{The same as Fig.~\ref{fig:ellipses_ctg_ctw_136_nm_m_lin_quad} but for $(C_{tW}, C_p)$.}
    \label{fig:ellipses_ctw_cp_136_nm_m_lin_quad}
\end{figure}

\begin{figure}
    [H]
    \centering
    \includegraphics[width=0.24\linewidth]{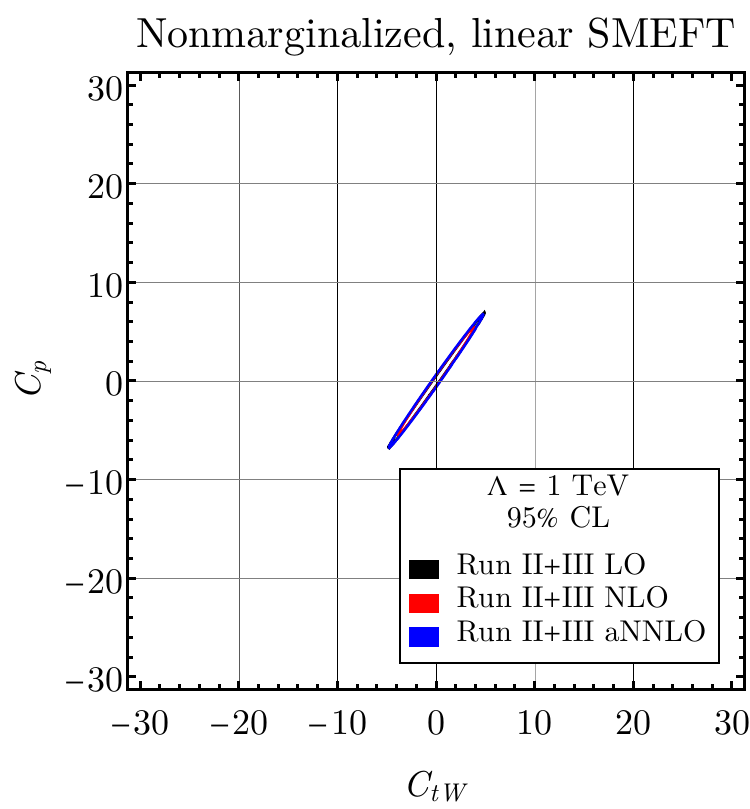}
    \includegraphics[width=0.24\linewidth]{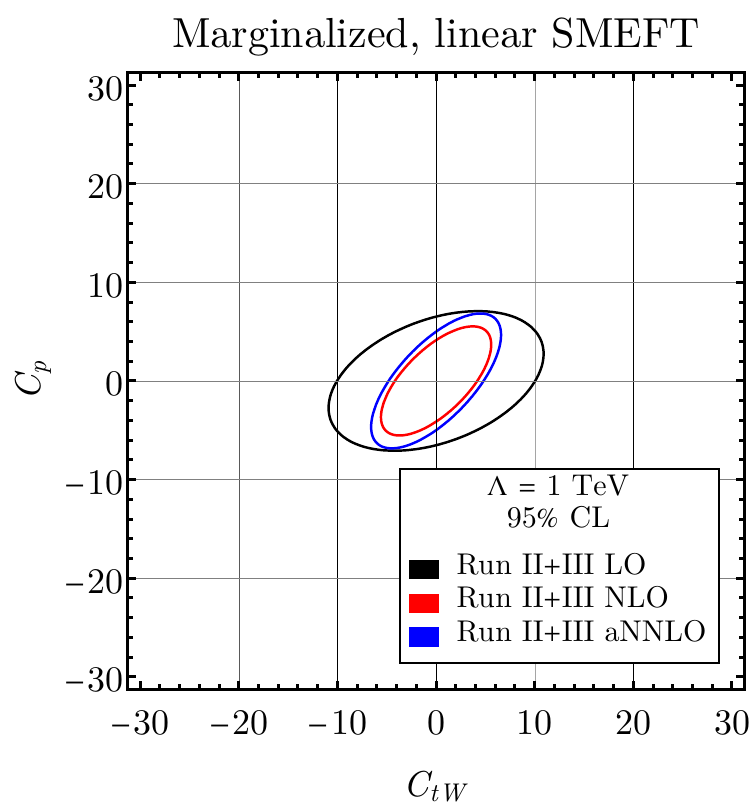}
    \includegraphics[width=0.24\linewidth]{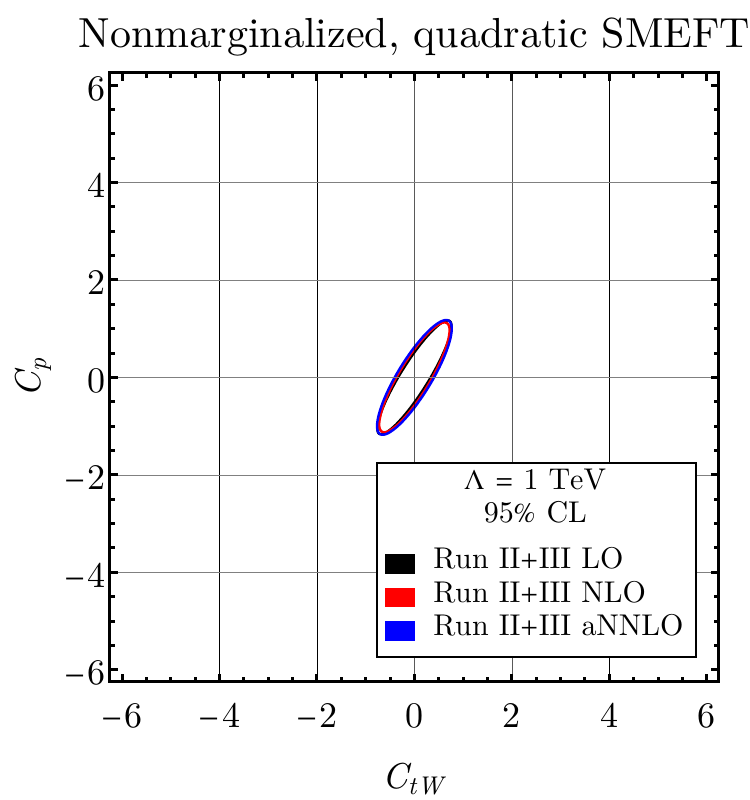}
    \includegraphics[width=0.24\linewidth]{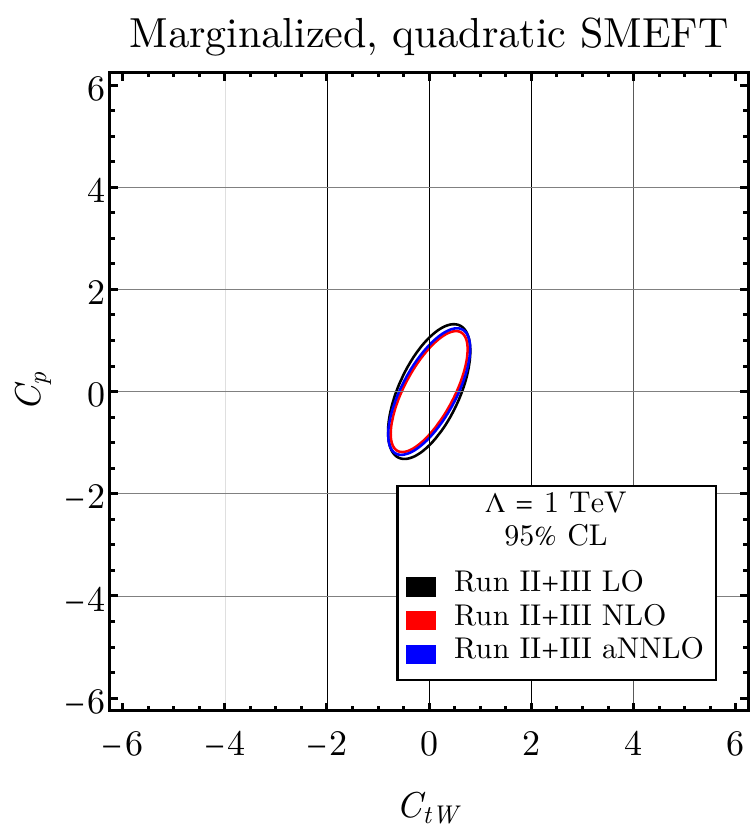}
    \caption{The same as Fig.~\ref{fig:ellipses_ctg_ctw_combined_nm_m_lin_quad} but for $(C_{tW}, C_p)$.}
    \label{fig:ellipses_ctw_cp_combined_nm_m_lin_quad}
\end{figure}

In this section, we present the complete set of confidence ellipses for the SMEFT fit results in Figs.~\ref{fig:ellipses_ctg_ctw_13_nm_m_lin_quad}--\ref{fig:ellipses_ctw_cp_combined_nm_m_lin_quad}.

\section{Complete set of correlation matrices\label{app:corr}}
\begin{figure}
    [H]
    \centering
    \includegraphics[width=0.26\linewidth]{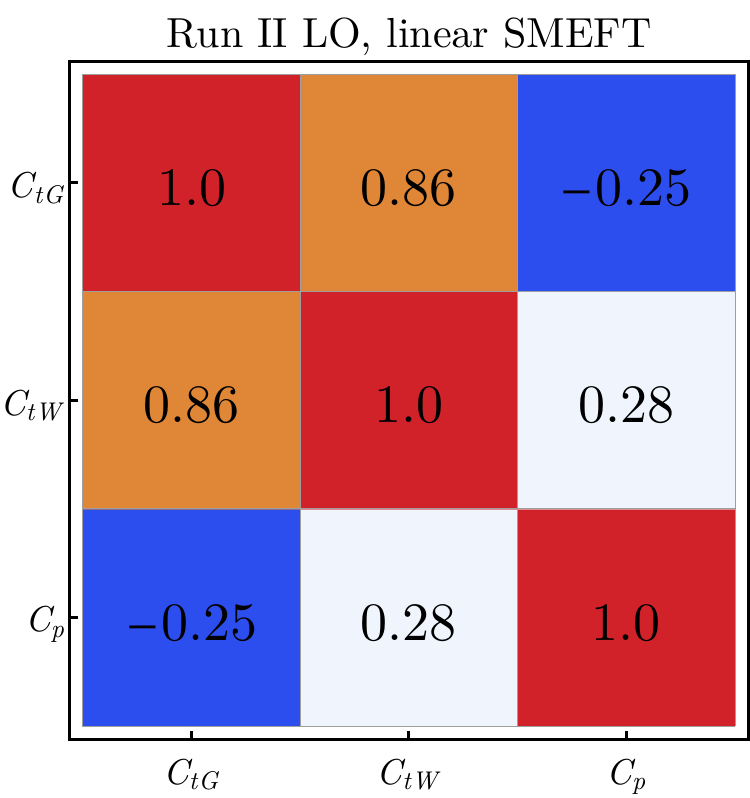}
    \includegraphics[width=0.26\linewidth]{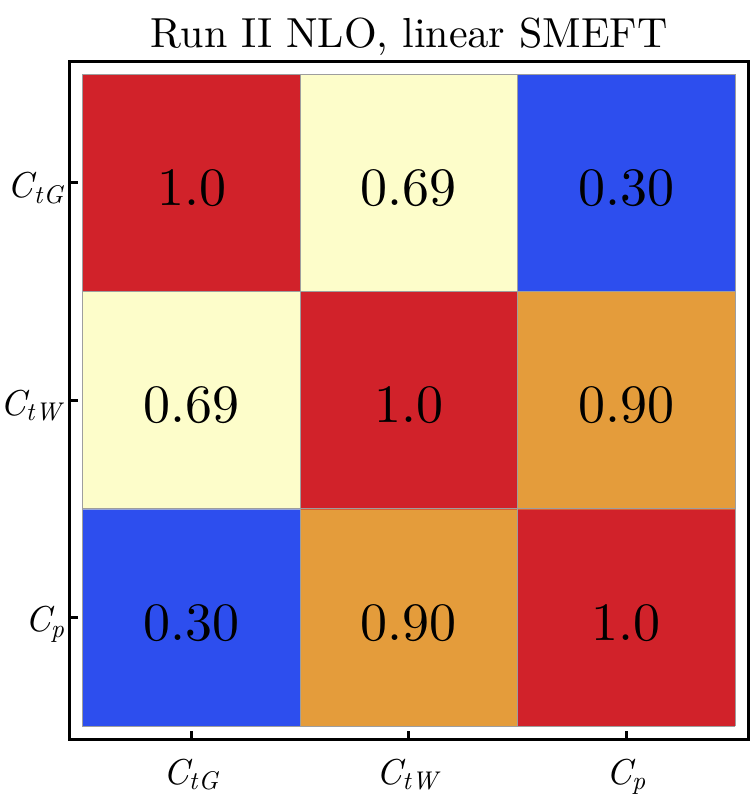}
    \includegraphics[width=0.26\linewidth]{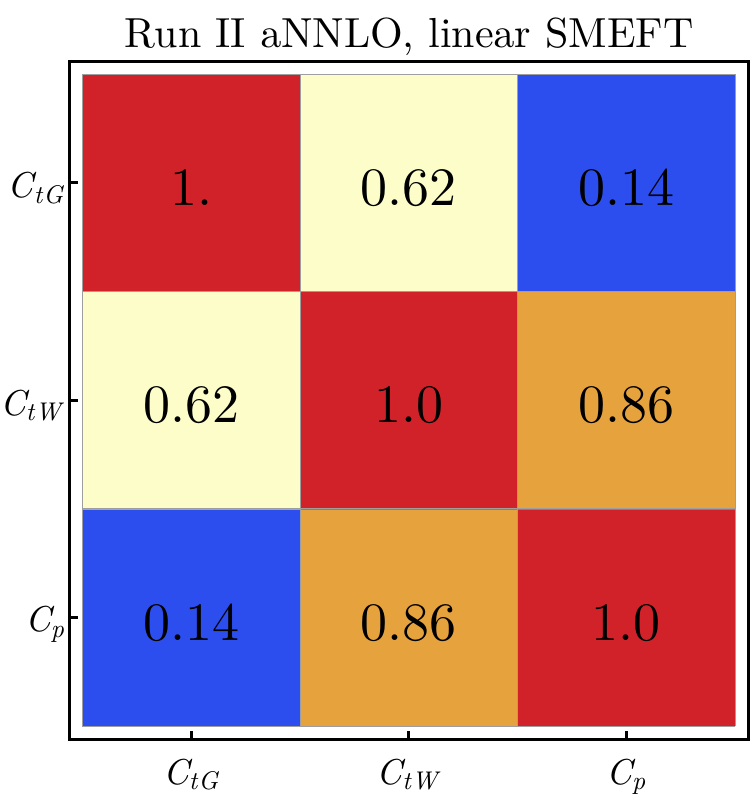}
    \includegraphics[width=0.26\linewidth]{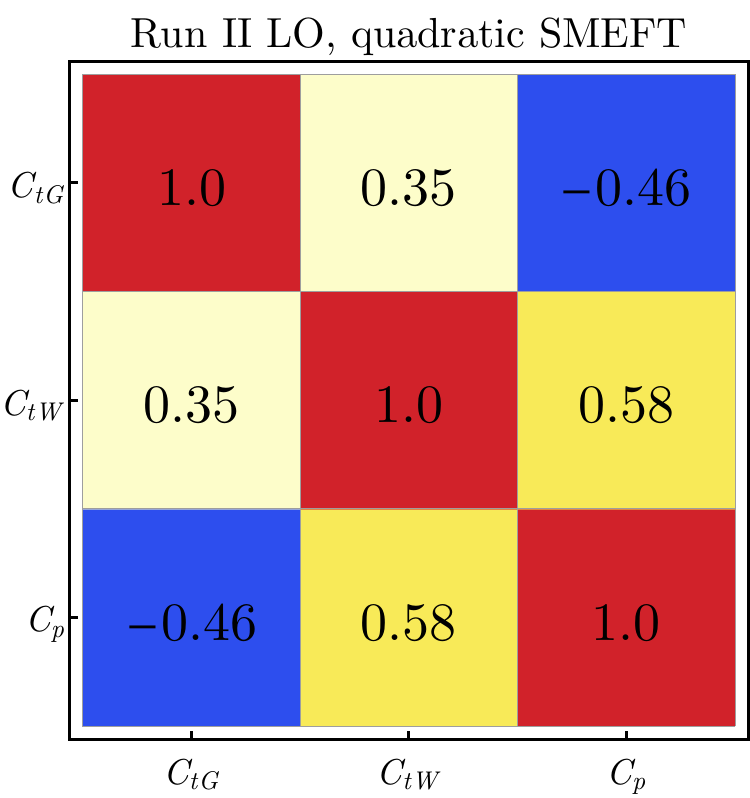}
    \includegraphics[width=0.26\linewidth]{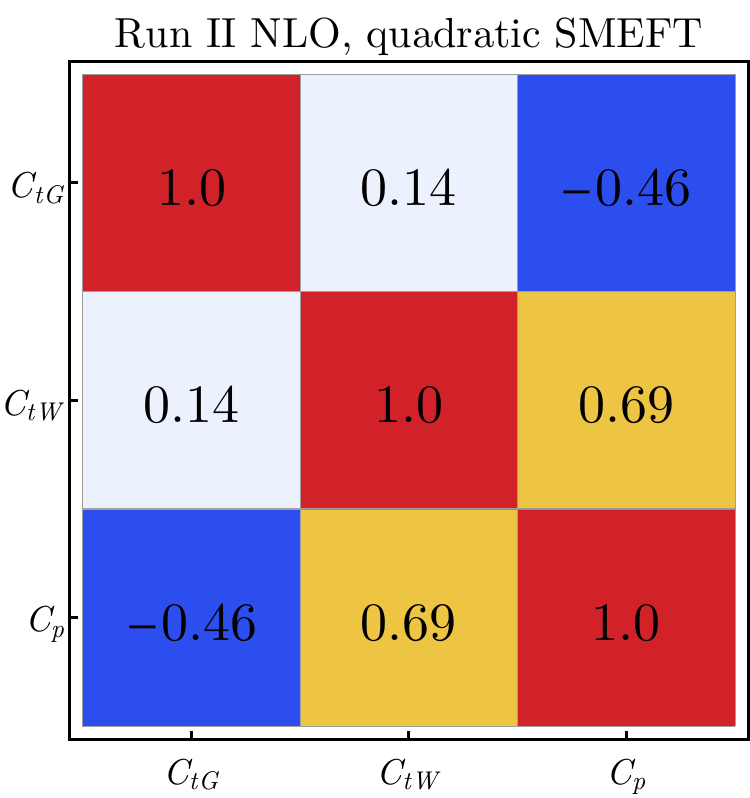}
    \includegraphics[width=0.26\linewidth]{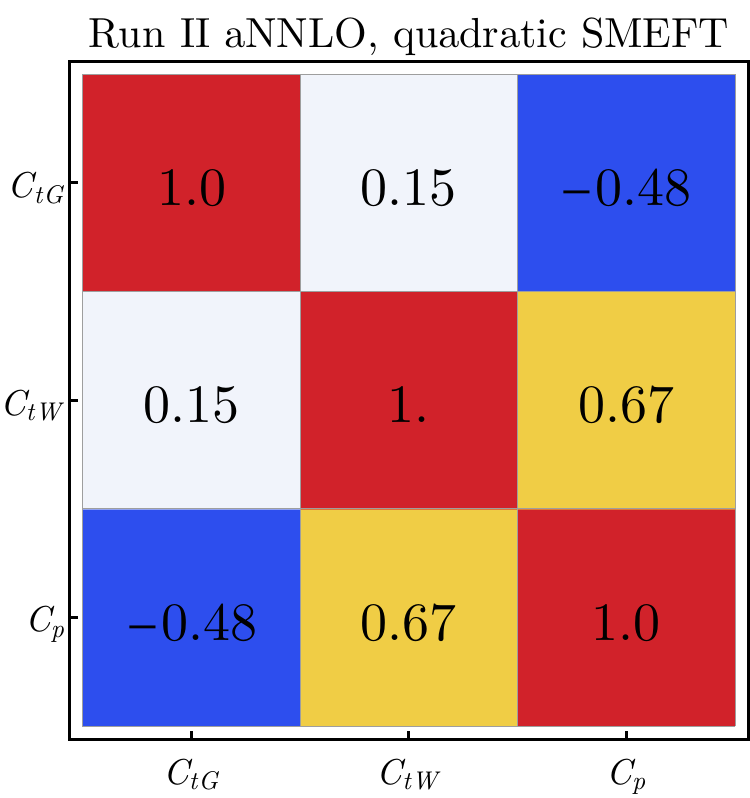}
    \caption{Correlation matrices for the marginalized linear (top) and quadratic (bottom) fits for Run II at LO (left), NLO (center), and aNNLO (right).}
    \label{fig:corr_matrices_13_app}
\end{figure}

\begin{figure}
    [H]
    \centering
    \includegraphics[width=0.26\linewidth]{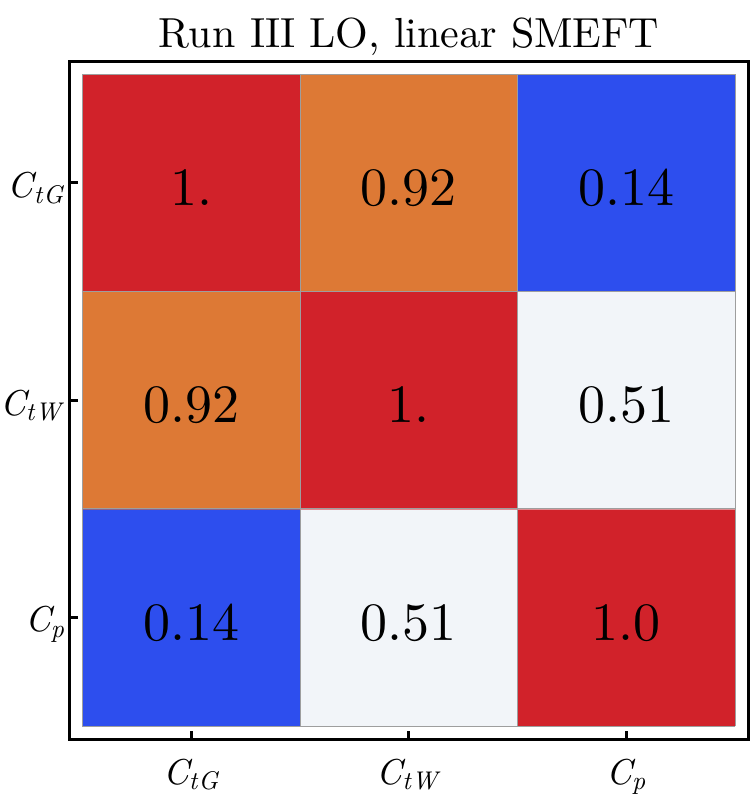}
    \includegraphics[width=0.26\linewidth]{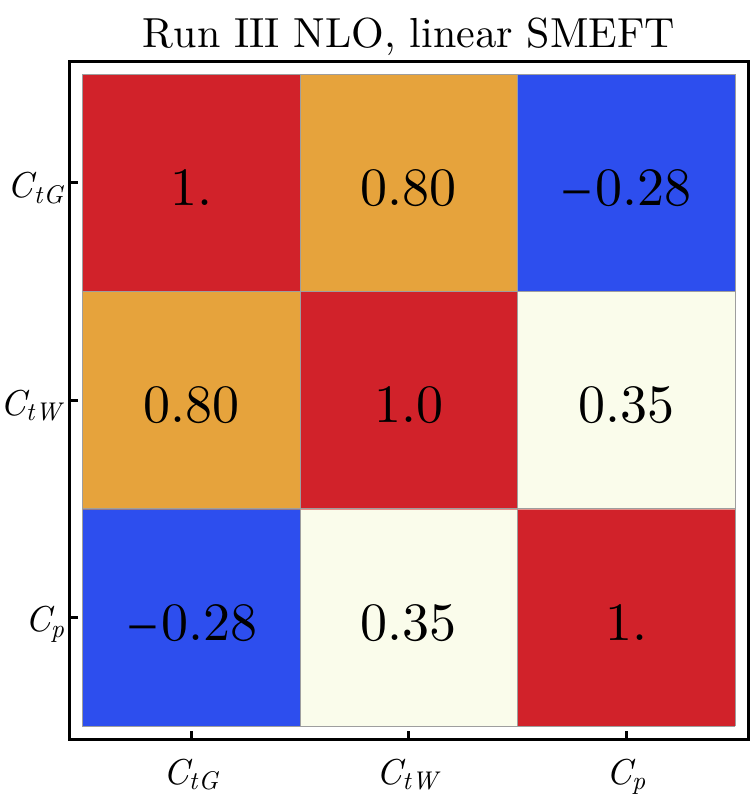}
    \includegraphics[width=0.26\linewidth]{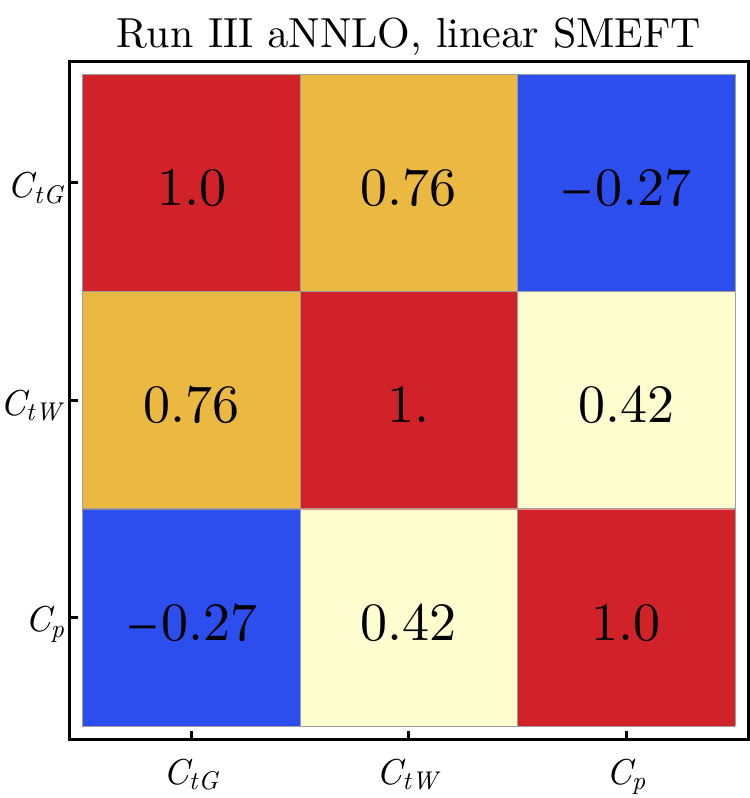}
    \includegraphics[width=0.26\linewidth]{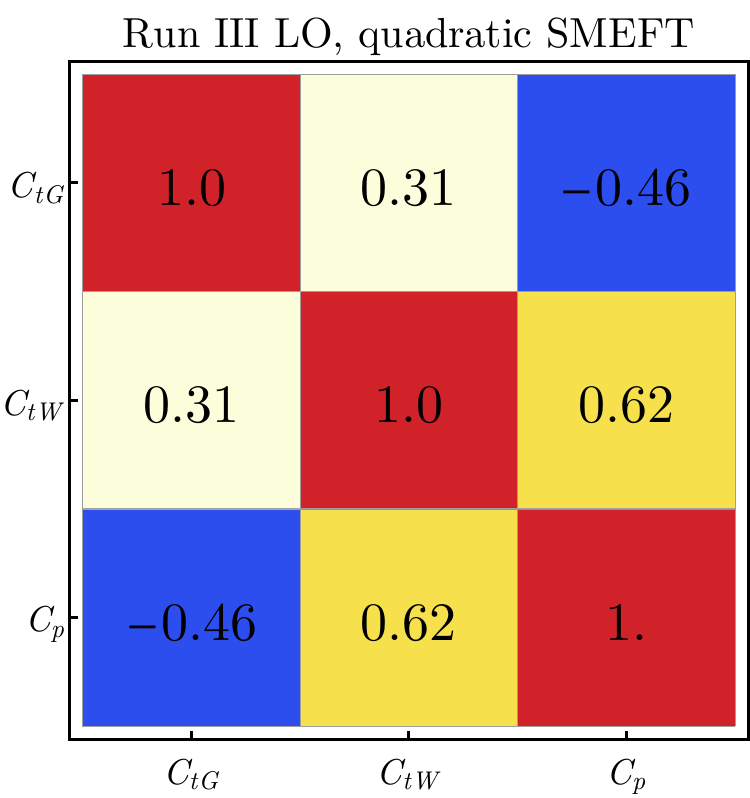}
    \includegraphics[width=0.26\linewidth]{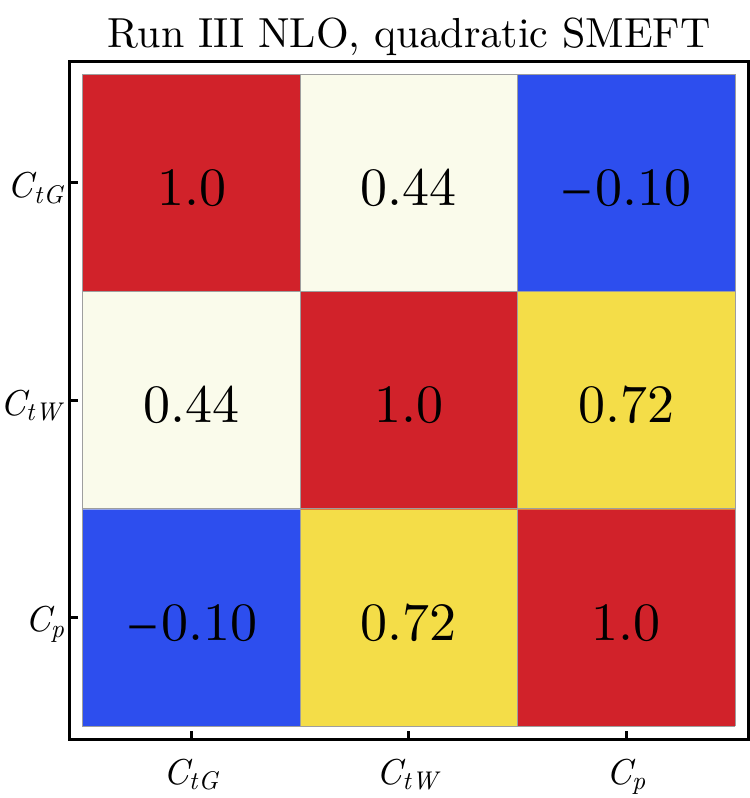}
    \includegraphics[width=0.26\linewidth]{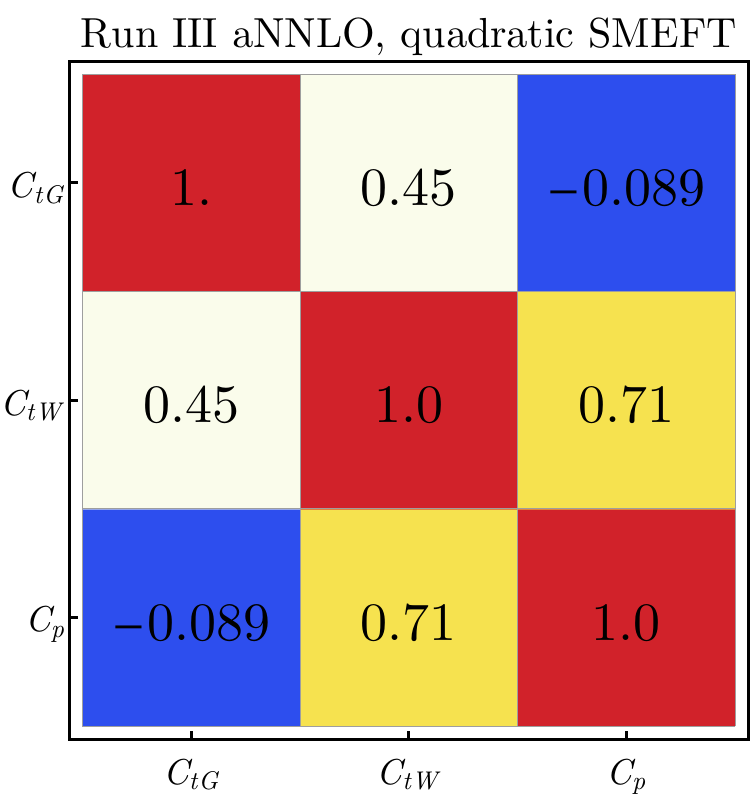}
    \caption{The same as Fig.~\ref{fig:corr_matrices_13_app} but for Run III.}
    \label{fig:corr_matrices_136_app}
\end{figure}

\begin{figure}
    [H]
    \centering
    \includegraphics[width=0.26\linewidth]{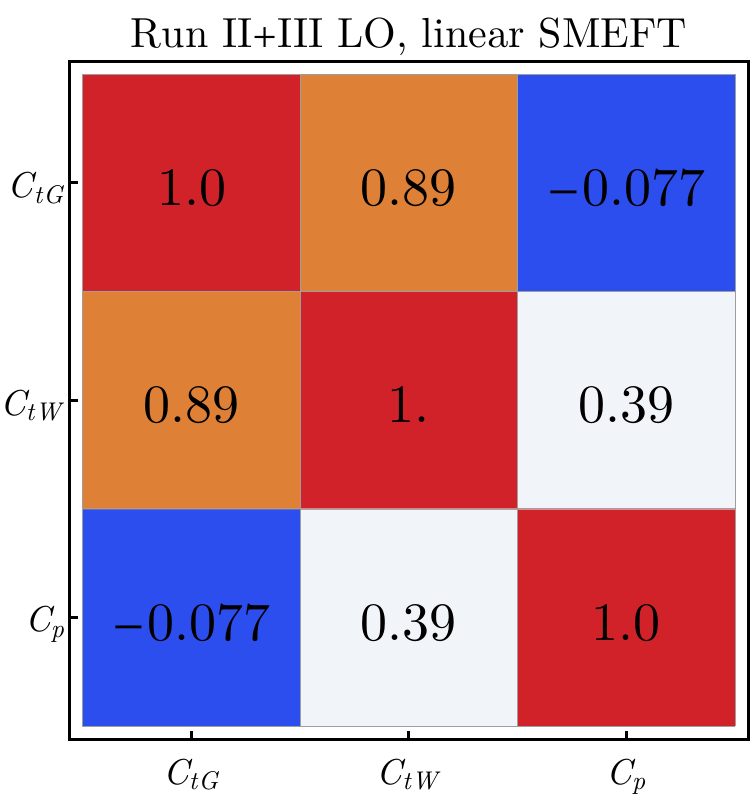}
    \includegraphics[width=0.26\linewidth]{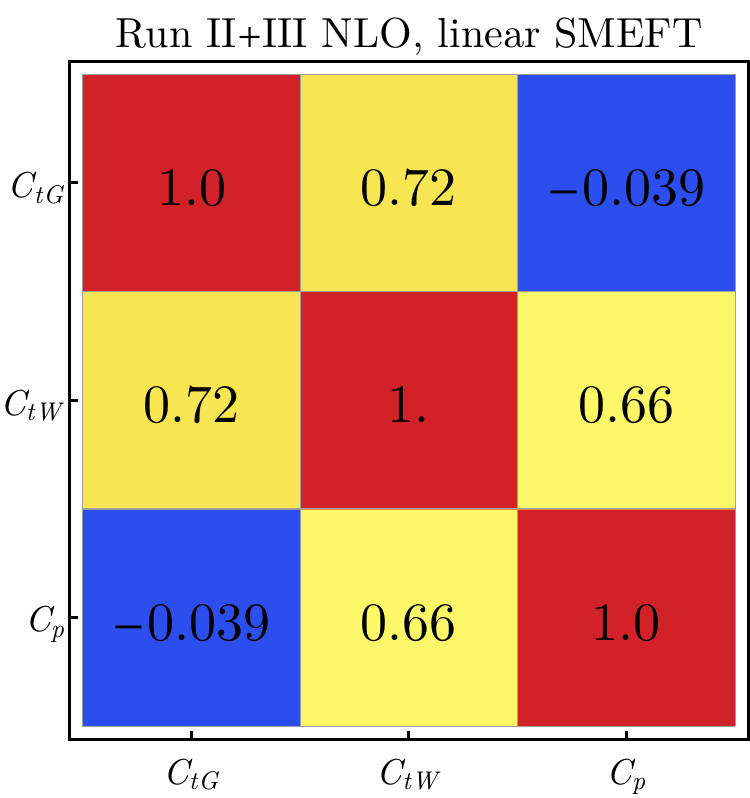}
    \includegraphics[width=0.26\linewidth]{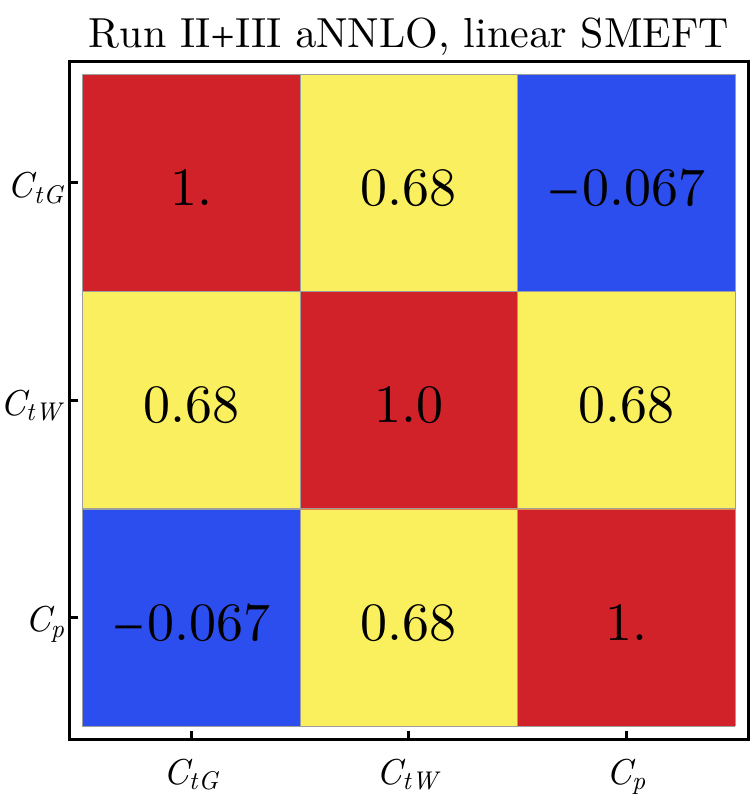}
    \includegraphics[width=0.26\linewidth]{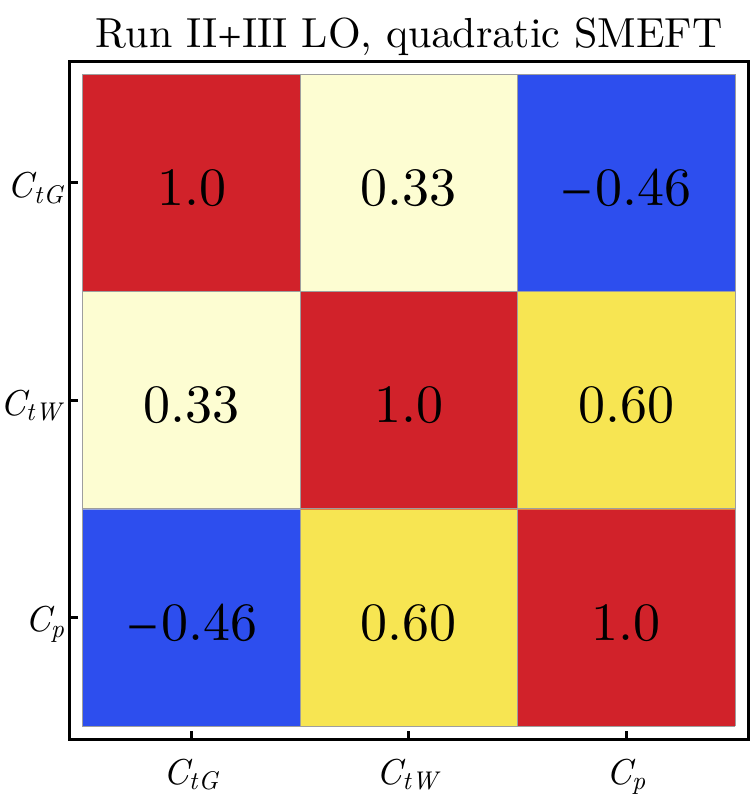}
    \includegraphics[width=0.26\linewidth]{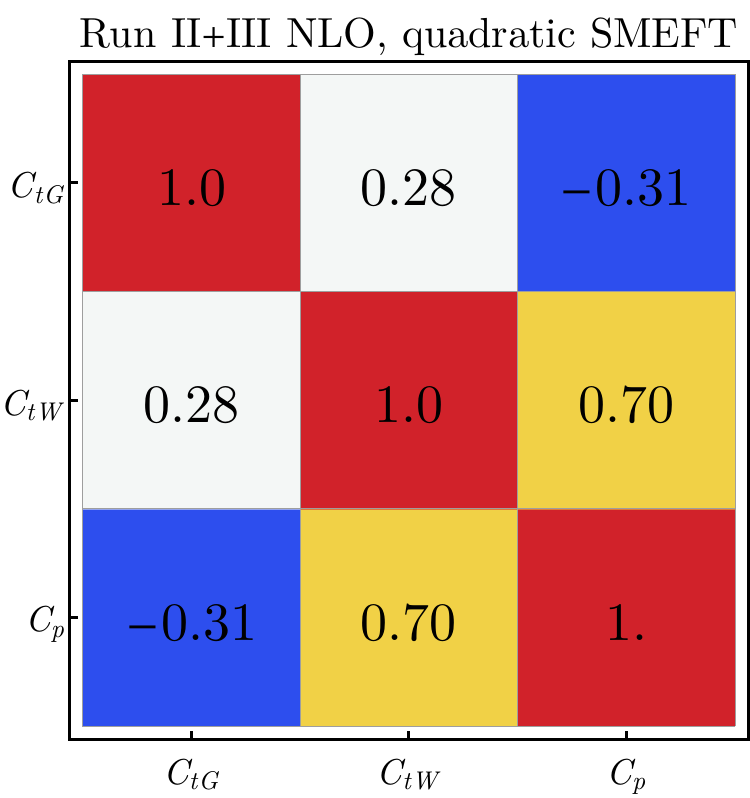}
    \includegraphics[width=0.26\linewidth]{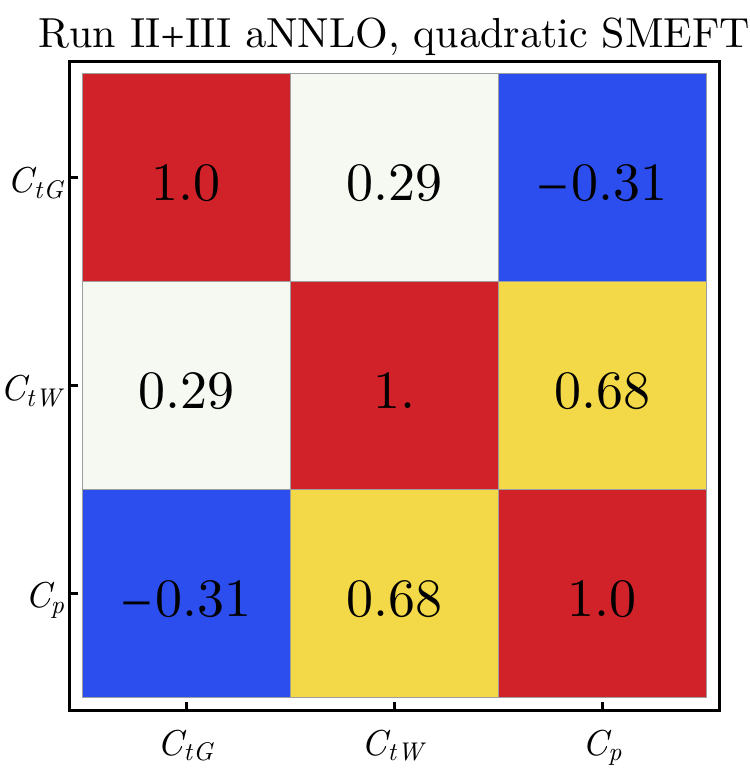}
    \caption{The same as Fig.~\ref{fig:corr_matrices_13_app} but for Run II+III.}
    \label{fig:corr_matrices_combined_app}
\end{figure}

In this section, we present the correlation matrices for the marginalized fits for Run II, Run III, and Run II+III in Figs.~\ref{fig:corr_matrices_13_app}, \ref{fig:corr_matrices_136_app}, and \ref{fig:corr_matrices_combined_app}, respectively. 

\bibliography{refs}

\end{document}